\documentclass[longauth]{aa}  

\usepackage{graphicx}
\usepackage{txfonts}
\usepackage{multirow}
\usepackage{amsmath}
\usepackage{xcolor}
\usepackage[breaklinks, colorlinks, citecolor=blue, linkcolor=blue]{hyperref}
\usepackage{placeins}
\newcommand{\cheops}{{\it CHEOPS}}
\newcommand{\tess}{{\it TESS}}
\newcommand{\starname}{{HD\,108236}}

\begin{document} 

   \title{Characterization of the HD\,108236 system with \cheops{} and \tess. Confirmation of a fifth transiting planet\thanks{CHEOPS detrended light curves are only available in electronic format the CDS via anonymous ftp to cdsarc.u-strasbg.fr (130.79.128.5) or via http://cdsweb.u-strasbg.fr/cgi-bin/qcat?J/A+A/2022/43720 }}
  
   \titlerunning{HD\,108236 system observed with \cheops{} and \tess{}}
   \authorrunning{Hoyer et al.}
   \author{S. Hoyer\inst{1} $^{\href{https://orcid.org/0000-0003-3477-2466}{\includegraphics[scale=0.5]{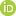}}}$,
A. Bonfanti\inst{2} $^{\href{https://orcid.org/0000-0002-1916-5935}{\includegraphics[scale=0.5]{figures/orcid.jpg}}}$,
A. Leleu\inst{3,4} $^{\href{https://orcid.org/0000-0003-2051-7974}{\includegraphics[scale=0.5]{figures/orcid.jpg}}}$,
L. Acuña\inst{1} $^{\href{https://orcid.org/0000-0002-9147-7925}{\includegraphics[scale=0.5]{figures/orcid.jpg}}}$,
L. M. Serrano\inst{5} $^{\href{https://orcid.org/0000-0001-9211-3691}{\includegraphics[scale=0.5]{figures/orcid.jpg}}}$,
M. Deleuil\inst{1} $^{\href{https://orcid.org/0000-0001-6036-0225}{\includegraphics[scale=0.5]{figures/orcid.jpg}}}$,
A. Bekkelien\inst{3}, 
C. Broeg\inst{4,6} $^{\href{https://orcid.org/0000-0001-5132-2614}{\includegraphics[scale=0.5]{figures/orcid.jpg}}}$,
H.-G. Florén\inst{7,8}, 
D. Queloz\inst{9,10} $^{\href{https://orcid.org/0000-0002-3012-0316}{\includegraphics[scale=0.5]{figures/orcid.jpg}}}$,
T. G. Wilson\inst{11} $^{\href{https://orcid.org/0000-0001-8749-1962}{\includegraphics[scale=0.5]{figures/orcid.jpg}}}$,
S. G. Sousa\inst{12} $^{\href{https://orcid.org/0000-0001-9047-2965}{\includegraphics[scale=0.5]{figures/orcid.jpg}}}$,
M. J. Hooton\inst{10,4} $^{\href{https://orcid.org/0000-0003-0030-332X}{\includegraphics[scale=0.5]{figures/orcid.jpg}}}$,
V. Adibekyan\inst{12} $^{\href{https://orcid.org/0000-0002-0601-6199}{\includegraphics[scale=0.5]{figures/orcid.jpg}}}$,
Y. Alibert\inst{4} $^{\href{https://orcid.org/0000-0002-4644-8818}{\includegraphics[scale=0.5]{figures/orcid.jpg}}}$,
R. Alonso\inst{13,14} $^{\href{https://orcid.org/0000-0001-8462-8126}{\includegraphics[scale=0.5]{figures/orcid.jpg}}}$,
G. Anglada\inst{15,16} $^{\href{https://orcid.org/0000-0002-3645-5977}{\includegraphics[scale=0.5]{figures/orcid.jpg}}}$,
J. Asquier\inst{17}, 
T. Bárczy\inst{18} $^{\href{https://orcid.org/0000-0002-7822-4413}{\includegraphics[scale=0.5]{figures/orcid.jpg}}}$,
D. Barrado\inst{19} $^{\href{https://orcid.org/0000-0002-5971-9242}{\includegraphics[scale=0.5]{figures/orcid.jpg}}}$,
S. C. C. Barros\inst{12,20} $^{\href{https://orcid.org/0000-0003-2434-3625}{\includegraphics[scale=0.5]{figures/orcid.jpg}}}$,
W. Baumjohann\inst{2} $^{\href{https://orcid.org/0000-0001-6271-0110}{\includegraphics[scale=0.5]{figures/orcid.jpg}}}$,
M. Beck\inst{3} $^{\href{https://orcid.org/0000-0003-3926-0275}{\includegraphics[scale=0.5]{figures/orcid.jpg}}}$,
T. Beck\inst{4}, 
W. Benz\inst{4,6} $^{\href{https://orcid.org/0000-0001-7896-6479}{\includegraphics[scale=0.5]{figures/orcid.jpg}}}$,
N. Billot\inst{3} $^{\href{https://orcid.org/0000-0003-3429-3836}{\includegraphics[scale=0.5]{figures/orcid.jpg}}}$,
F. Biondi\inst{21,22}, 
X. Bonfils\inst{23} $^{\href{https://orcid.org/0000-0001-9003-8894}{\includegraphics[scale=0.5]{figures/orcid.jpg}}}$,
A. Brandeker\inst{7} $^{\href{https://orcid.org/0000-0002-7201-7536}{\includegraphics[scale=0.5]{figures/orcid.jpg}}}$,
J. Cabrera\inst{24}, 
S. Charnoz\inst{25} $^{\href{https://orcid.org/0000-0002-7442-491X}{\includegraphics[scale=0.5]{figures/orcid.jpg}}}$,
A. Collier Cameron\inst{11} $^{\href{https://orcid.org/0000-0002-8863-7828}{\includegraphics[scale=0.5]{figures/orcid.jpg}}}$,
Sz. Csizmadia\inst{24} $^{\href{https://orcid.org/0000-0001-6803-9698}{\includegraphics[scale=0.5]{figures/orcid.jpg}}}$,
M. B. Davies\inst{26} $^{\href{https://orcid.org/0000-0001-6080-1190}{\includegraphics[scale=0.5]{figures/orcid.jpg}}}$,
L. Delrez\inst{27,28} $^{\href{https://orcid.org/0000-0001-6108-4808}{\includegraphics[scale=0.5]{figures/orcid.jpg}}}$,
O. D. S. Demangeon\inst{12,20} $^{\href{https://orcid.org/0000-0001-7918-0355}{\includegraphics[scale=0.5]{figures/orcid.jpg}}}$,
B.-O. Demory\inst{6} $^{\href{https://orcid.org/0000-0002-9355-5165}{\includegraphics[scale=0.5]{figures/orcid.jpg}}}$,
D. Ehrenreich\inst{3} $^{\href{https://orcid.org/0000-0001-9704-5405}{\includegraphics[scale=0.5]{figures/orcid.jpg}}}$,
A. Erikson\inst{24}, 
A. Fortier\inst{4,6} $^{\href{https://orcid.org/0000-0001-8450-3374}{\includegraphics[scale=0.5]{figures/orcid.jpg}}}$,
L. Fossati\inst{2} $^{\href{https://orcid.org/0000-0003-4426-9530}{\includegraphics[scale=0.5]{figures/orcid.jpg}}}$,
M. Fridlund\inst{29,30} $^{\href{https://orcid.org/0000-0002-0855-8426}{\includegraphics[scale=0.5]{figures/orcid.jpg}}}$,
D. Gandolfi\inst{5} $^{\href{https://orcid.org/0000-0001-8627-9628}{\includegraphics[scale=0.5]{figures/orcid.jpg}}}$,
M. Gillon\inst{27} $^{\href{https://orcid.org/0000-0003-1462-7739}{\includegraphics[scale=0.5]{figures/orcid.jpg}}}$,
M. Güdel\inst{31}, 
N. Hara\inst{3} $^{\href{https://orcid.org/0000-0001-9232-3314}{\includegraphics[scale=0.5]{figures/orcid.jpg}}}$,
K. Heng\inst{6,32} $^{\href{https://orcid.org/0000-0003-1907-5910}{\includegraphics[scale=0.5]{figures/orcid.jpg}}}$,
K. G. Isaak\inst{33} $^{\href{https://orcid.org/0000-0001-8585-1717}{\includegraphics[scale=0.5]{figures/orcid.jpg}}}$,
J. M. Jenkins\inst{34}, 
L. L. Kiss\inst{35,36}, 
J. Laskar\inst{37} $^{\href{https://orcid.org/0000-0003-2634-789X}{\includegraphics[scale=0.5]{figures/orcid.jpg}}}$,
D. W. Latham\inst{38}, 
A. Lecavelier des Etangs\inst{39} $^{\href{https://orcid.org/0000-0002-5637-5253}{\includegraphics[scale=0.5]{figures/orcid.jpg}}}$,
M. Lendl\inst{3} $^{\href{https://orcid.org/0000-0001-9699-1459}{\includegraphics[scale=0.5]{figures/orcid.jpg}}}$,
C. Lovis\inst{3} $^{\href{https://orcid.org/0000-0001-7120-5837}{\includegraphics[scale=0.5]{figures/orcid.jpg}}}$,
A. Luntzer\inst{40}, 
D. Magrin\inst{22} $^{\href{https://orcid.org/0000-0003-0312-313X}{\includegraphics[scale=0.5]{figures/orcid.jpg}}}$,
P. F. L. Maxted\inst{41} $^{\href{https://orcid.org/0000-0003-3794-1317}{\includegraphics[scale=0.5]{figures/orcid.jpg}}}$,
V. Nascimbeni\inst{22} $^{\href{https://orcid.org/0000-0001-9770-1214}{\includegraphics[scale=0.5]{figures/orcid.jpg}}}$,
G. Olofsson\inst{7} $^{\href{https://orcid.org/0000-0003-3747-7120}{\includegraphics[scale=0.5]{figures/orcid.jpg}}}$,
R. Ottensamer\inst{40}, 
I. Pagano\inst{42} $^{\href{https://orcid.org/0000-0001-9573-4928}{\includegraphics[scale=0.5]{figures/orcid.jpg}}}$,
E. Pallé\inst{13} $^{\href{https://orcid.org/0000-0003-0987-1593}{\includegraphics[scale=0.5]{figures/orcid.jpg}}}$,
C. M. Persson\inst{30}, 
G. Peter\inst{43} $^{\href{https://orcid.org/0000-0001-6101-2513}{\includegraphics[scale=0.5]{figures/orcid.jpg}}}$,
D. Piazza\inst{4}, 
G. Piotto\inst{22,44} $^{\href{https://orcid.org/0000-0002-9937-6387}{\includegraphics[scale=0.5]{figures/orcid.jpg}}}$,
D. Pollacco\inst{32}, 
R. Ragazzoni\inst{22,44} $^{\href{https://orcid.org/0000-0002-7697-5555}{\includegraphics[scale=0.5]{figures/orcid.jpg}}}$,
N. Rando\inst{17}, 
H. Rauer\inst{24,45,46} $^{\href{https://orcid.org/0000-0002-6510-1828}{\includegraphics[scale=0.5]{figures/orcid.jpg}}}$,
I. Ribas\inst{15,16} $^{\href{https://orcid.org/0000-0002-6689-0312}{\includegraphics[scale=0.5]{figures/orcid.jpg}}}$,
G. R. Ricker\inst{47} $^{\href{https://orcid.org/0000-0003-2058-6662}{\includegraphics[scale=0.5]{figures/orcid.jpg}}}$,
S. Salmon\inst{3} $^{\href{https://orcid.org/0000-0002-1714-3513}{\includegraphics[scale=0.5]{figures/orcid.jpg}}}$,
N. C. Santos\inst{12,20} $^{\href{https://orcid.org/0000-0003-4422-2919}{\includegraphics[scale=0.5]{figures/orcid.jpg}}}$,
G. Scandariato\inst{42} $^{\href{https://orcid.org/0000-0003-2029-0626}{\includegraphics[scale=0.5]{figures/orcid.jpg}}}$,
S. Seager\inst{47,48,49} $^{\href{https://orcid.org/0000-0002-6892-6948}{\includegraphics[scale=0.5]{figures/orcid.jpg}}}$,
D. Ségransan\inst{3} $^{\href{https://orcid.org/0000-0003-2355-8034}{\includegraphics[scale=0.5]{figures/orcid.jpg}}}$,
A. E. Simon\inst{4} $^{\href{https://orcid.org/0000-0001-9773-2600}{\includegraphics[scale=0.5]{figures/orcid.jpg}}}$,
A. M. S. Smith\inst{24} $^{\href{https://orcid.org/0000-0002-2386-4341}{\includegraphics[scale=0.5]{figures/orcid.jpg}}}$,
M. Steller\inst{2} $^{\href{https://orcid.org/0000-0003-2459-6155}{\includegraphics[scale=0.5]{figures/orcid.jpg}}}$,
Gy. M. Szabó\inst{50,51}, 
N. Thomas\inst{4}, 
J. D. Twicken\inst{52,34} $^{\href{https://orcid.org/0000-0002-6778-7552  }{\includegraphics[scale=0.5]{figures/orcid.jpg}}}$,
S. Udry\inst{3} $^{\href{https://orcid.org/0000-0001-7576-6236}{\includegraphics[scale=0.5]{figures/orcid.jpg}}}$,
V. Van Grootel\inst{28} $^{\href{https://orcid.org/0000-0003-2144-4316}{\includegraphics[scale=0.5]{figures/orcid.jpg}}}$,
R. K. Vanderspek\inst{47} $^{\href{https://orcid.org/0000-0001-6763-6562}{\includegraphics[scale=0.5]{figures/orcid.jpg}}}$,
N. A. Walton\inst{53} $^{\href{https://orcid.org/0000-0003-3983-8778}{\includegraphics[scale=0.5]{figures/orcid.jpg}}}$,
K. Westerdorff\inst{43}, 
J. N. Winn\inst{54} $^{\href{https://orcid.org/0000-0002-4265-047X}{\includegraphics[scale=0.5]{figures/orcid.jpg}}}$}

  \institute{\label{inst:1} Aix Marseille Univ, CNRS, CNES, LAM, 38 rue Frédéric Joliot-Curie, 13388 Marseille, France \and
\label{inst:2} Space Research Institute, Austrian Academy of Sciences, Schmiedlstrasse 6, A-8042 Graz, Austria \and
\label{inst:3} Observatoire Astronomique de l'Université de Genève, Chemin Pegasi 51, Versoix, Switzerland \and
\label{inst:4} Physikalisches Institut, University of Bern, Gesellsschaftstrasse 6, 3012 Bern, Switzerland \and
\label{inst:5} Dipartimento di Fisica, Universita degli Studi di Torino, via Pietro Giuria 1, I-10125, Torino, Italy \and
\label{inst:6} Center for Space and Habitability, Gesellsschaftstrasse 6, 3012 Bern, Switzerland \and
\label{inst:7} Department of Astronomy, Stockholm University, AlbaNova University Center, 10691 Stockholm, Sweden \and
\label{inst:8} Department of Astronomy, Stockholm University, SE-106 91 Stockholm, Sweden \and
\label{inst:9} ETH Zurich, Department of Physics, Wolfgang-Pauli-Strasse 2, CH-8093 Zurich, Switzerland \and
\label{inst:10} Cavendish Laboratory, JJ Thomson Avenue, Cambridge CB3 0HE, UK \and
\label{inst:11} Centre for Exoplanet Science, SUPA School of Physics and Astronomy, University of St Andrews, North Haugh, St Andrews KY16 9SS, UK \and
\label{inst:12} Instituto de Astrofisica e Ciencias do Espaco, Universidade do Porto, CAUP, Rua das Estrelas, 4150-762 Porto, Portugal \and
\label{inst:13} Instituto de Astrofisica de Canarias, 38200 La Laguna, Tenerife, Spain \and
\label{inst:14} Departamento de Astrofisica, Universidad de La Laguna, 38206 La Laguna, Tenerife, Spain \and
\label{inst:15} Institut de Ciencies de l'Espai (ICE, CSIC), Campus UAB, Can Magrans s/n, 08193 Bellaterra, Spain \and
\label{inst:16} Institut d'Estudis Espacials de Catalunya (IEEC), 08034 Barcelona, Spain \and
\label{inst:17} ESTEC, European Space Agency, 2201AZ, Noordwijk, NL \and
\label{inst:18} Admatis, 5. Kandó Kálmán Street, 3534 Miskolc, Hungary \and
\label{inst:19} Depto. de Astrofisica, Centro de Astrobiologia (CSIC-INTA), ESAC campus, 28692 Villanueva de la Cañada (Madrid), Spain \and
\label{inst:20} Departamento de Fisica e Astronomia, Faculdade de Ciencias, Universidade do Porto, Rua do Campo Alegre, 4169-007 Porto, Portugal \and
\label{inst:21} Max Planck Institute for Extraterrestrial Physics, Gießenbachstraße 1, 85748 Garching bei München, Germany \and
\label{inst:22} INAF, Osservatorio Astronomico di Padova, Vicolo dell'Osservatorio 5, 35122 Padova, Italy \and
\label{inst:23} Université Grenoble Alpes, CNRS, IPAG, 38000 Grenoble, France \and
\label{inst:24} Institute of Planetary Research, German Aerospace Center (DLR), Rutherfordstrasse 2, 12489 Berlin, Germany \and
\label{inst:25} Université de Paris, Institut de physique du globe de Paris, CNRS, F-75005 Paris, France \and
\label{inst:26} Centre for Mathematical Sciences, Lund University, Box 118, 221 00 Lund, Sweden \and
\label{inst:27} Astrobiology Research Unit, Université de Liège, Allée du 6 Août 19C, B-4000 Liège, Belgium \and
\label{inst:28} Space sciences, Technologies and Astrophysics Research (STAR) Institute, Université de Liège, Allée du 6 Août 19C, 4000 Liège, Belgium \and
\label{inst:29} Leiden Observatory, University of Leiden, PO Box 9513, 2300 RA Leiden, The Netherlands \and
\label{inst:30} Department of Space, Earth and Environment, Chalmers University of Technology, Onsala Space Observatory, 43992 Onsala, Sweden \and
\label{inst:31} University of Vienna, Department of Astrophysics, Türkenschanzstrasse 17, 1180 Vienna, Austria \and
\label{inst:32} Department of Physics, University of Warwick, Gibbet Hill Road, Coventry CV4 7AL, United Kingdom \and
\label{inst:33} Science and Operations Department - Science Division (SCI-SC), Directorate of Science, European Space Agency (ESA), European Space Research and Technology Centre (ESTEC),
Keplerlaan 1, 2201-AZ Noordwijk, The Netherlands \and
\label{inst:34} NASA Ames Research Center, Moffett Field, CA, 94035, USA \and
\label{inst:35} Konkoly Observatory, Research Centre for Astronomy and Earth Sciences, 1121 Budapest, Konkoly Thege Miklós út 15-17, Hungary \and
\label{inst:36} ELTE E\"otv\"os Lor\'and University, Institute of Physics, P\'azm\'any P\'eter s\'et\'any 1/A, 1117 \and
\label{inst:37} IMCCE, UMR8028 CNRS, Observatoire de Paris, PSL Univ., Sorbonne Univ., 77 av. Denfert-Rochereau, 75014 Paris, France \and
\label{inst:38} Center for Astrophysics \textbar \ Harvard \& Smithsonian, 60 Garden Street, Cambridge, MA, 02138, USA \and
\label{inst:39} Institut d’astrophysique de Paris, CNRS, UMR 7095, Sorbonne Université, 98 bis bd Arago, 75014 Paris, France \and
\label{inst:40} Department of Astrophysics, University of Vienna, Tuerkenschanzstrasse 17, 1180 Vienna, Austria \and
\label{inst:41} Astrophysics Group, Keele University, Staffordshire, ST5 5BG, United Kingdom \and
\label{inst:42} INAF, Osservatorio Astrofisico di Catania, Via S. Sofia 78, 95123 Catania, Italy \and
\label{inst:43} Institute of Optical Sensor Systems, German Aerospace Center (DLR), Rutherfordstrasse 2, 12489 Berlin, Germany \and
\label{inst:44} Dipartimento di Fisica e Astronomia "Galileo Galilei", Universita degli Studi di Padova, Vicolo dell'Osservatorio 3, 35122 Padova, Italy \and
\label{inst:45} Zentrum f{\"u}r Astronomie und Astrophysik, Technische Universit{\"a}t Berlin, Hardenbergstr. 36, D-10623 Berlin, Germany \and
\label{inst:46} Institut f{\"u}r Geologische Wissenschaften, Freie Universit{\"a}t Berlin, 12249 Berlin, Germany \and
\label{inst:47} Department of Physics and Kavli Institute for Astrophysics and Space Research, Massachusetts Institute of Technology, Cambridge, MA, 02139, USA \and
\label{inst:48} Department of Earth, Atmospheric and Planetary Sciences, Massachusetts Institute of Technology, Cambridge, MA, 02139, USA \and
\label{inst:49} Department of Aeronautics and Astronautics, MIT, 77 Massachusetts Avenue, Cambridge, MA, 02139, USA \and
\label{inst:50} ELTE E\"otv\"os Lor\'and University, Gothard Astrophysical Observatory, 9700 Szombathely, Szent Imre h. u. 112, Hungary \and
\label{inst:51} MTA-ELTE Exoplanet Research Group, 9700 Szombathely, Szent Imre h. u. 112, Hungary \and
\label{inst:52} SETI Institute, Mountain View, CA  94043, USA \and
\label{inst:53} Institute of Astronomy, University of Cambridge, Madingley Road, Cambridge, CB3 0HA, United Kingdom \and
\label{inst:54} Department of Astrophysical Sciences, Princeton University, 4 Ivy Ln., Princeton, NJ, 08544, USA
}

   \date{}

 
  \abstract
  {The \starname{} system was first announced with the detection of four small planets based on \tess{} data. Shortly after, the transit of an additional planet with a period of 29.54\,d was serendipitously detected by \cheops. In this way, \starname{} ($V$=9.2) became one of the brightest stars known to host five small transiting planets (R$_p$<3\,R$_{\oplus}$).}  
   {We characterize the planetary system by using all the data available from \cheops{} and \tess{} space missions. We use the flexible pointing capabilities of \cheops{} to follow up the transits of all the planets in the system, including the fifth transiting body. }
   {After updating the host star parameters by using the results from Gaia eDR3, we analyzed 16 and 43 transits observed by \cheops{} and \tess, respectively, to derive the planets' physical and orbital parameters. We carried out a timing analysis of the transits of each of the planets of \starname\   to search for the presence of transit timing variations.}
   {We derived improved values for the radius and mass of the host star ($R_{\star}=0.876\pm0.007$\,$R_{\odot}$ and
   $M_{\star}= 0.867_{-0.046}^{+0.047}$\,$M_{\odot}$). We confirm the presence of the fifth transiting planet $f$ in a 29.54\,d orbit. Thus, the \starname{} system consists of five planets of 
   R$_{b}$=1.587$\pm$0.028, R$_{c}$=2.122$\pm$0.025, R$_{d}$=2.629$\pm$0.031, R$_{e}$=3.008$\pm$0.032, and R$_{f}$=1.89$\pm$0.04 [R$_{\oplus}$]. 
   We refine the transit ephemeris for each planet and find no significant transit timing variations for planets $c$, $d,$ and $e$. For planets $b$ and $f$, instead, we measure significant deviations on their transit times (up to $22$ and $28$\,min, respectively) with a non-negligible dispersion of 9.6 and 12.6\,min in their time residuals.}
   {We confirm the presence of planet $f$ and find no significant evidence for a potential transiting planet in a 10.9\,d orbital period, as previously suggested. Further monitoring of the transits, particularly for planets $b$ and $f$, would confirm the presence of the observed transit time variations. \starname{} thus becomes a key multi-planetary system for the study of formation and evolution processes. The reported precise results on the planetary radii -- together with a profuse RV monitoring -- will allow for an accurate characterization of the internal structure of these planets.}


   \maketitle
%

\section{Introduction}
%
Multi-planet systems are key to improving our understanding and studying the formation and evolution of planetary systems, with the ultimate goal of finding resemblances with the history of our own Solar System. The dynamics and architecture of multi-planet systems are crucial for imposing strong constraints on migration processes \citep[e.g.,][]{Mills2016, Delisle2017, Leleu2021}.  In addition, depending on the system, gravitational interaction between planets can produce detectable transit timing variations (TTVs), especially when planets are located close to mean motion resonances \citep{miralda2002,Holman2005,Agol2005}. In specific cases, TTVs can be used to estimate the masses of the planets without the requirement of complementary radial velocity observations \citep[e.g.,][]{Barros2015,Gillon2017trappist,Freudenthal2018,Agol2021}.  

The amenability of the systems for in-depth characterizations, including  comparative atmospheric studies through transmission spectroscopy, strongly depends on the brightness of the host star due to signal-to-noise considerations (see, e.g., the TSM metric in \cite{Kempton2018}). Thanks to the long baseline and high signal-to-noise ratios (S/N) of their observations, to date, most of the transiting multi-planet systems with four or more planets are part of the yield of the {\it Kepler+K2} missions \citep{Borucki2010, Howell2014}, for which the magnitude distribution of the stellar hosts peaks towards $V\sim$15. 

\begin{table*}[t]
    \caption{Exoplanetary systems orbiting a bright stellar host (V$<10.5$), with more than three planets and at least one transiting planet. }
    \label{tab:table_systems}
    \centering
    \small
    \begin{tabular}{l c c c c l}
    \hline
    \hline
    System & V-mag & total number & number of  & P$_{inner}$-P$_{outer}$ &Reference  \\
    name   &    &    of planets &  transiting planets &  [d] &\\
    \hline
     HD\,219134   &  5.5 &  5(6)\tablefootmark{(a)} & 2  & 3.1--46.8 &\cite{Motalebi2015, Gillon2017HD219134}\\
     55\,Cnc      &  5.95 & 5 & 1 &  0.73--4800 &\cite{McArthur2004} \\
     Kepler-444   &  8.9  & 5 & 5 &  3.6--9.7    &\cite{Campante2015} \\
     HIP\,41378   & 8.9  & 5  & 5 &  15.6--324   &\cite{vanderburg19} \\
     \starname{}  &  9.2  & 5 & 5  & 3.79--29.5  &\cite{Daylan2021, Bonfanti2021}\\
     Kepler-37    & 9.8  &  4 & 4 &  13.3--39.8 &\cite{Barclay2013} \\
     V1298\,Tau   & 10.1 & 4 & 4  &  8.24--60   &\cite{David2019} \\
     TOI-561      &  10.3 & 4(5)\tablefootmark{(b)} & 4 &  0.45--77  &\cite{Lacedelli2022} \\
     \hline
    \end{tabular}
    \tablefoot{ The orbital periods range of the planets in each system is given in the P$_{inner}$-P$_{outer}$ column. \tablefoottext{a}{The potential presence of a sixth planet is still controversial.} \tablefoottext{b}{Hints of an additional planet in a long-period orbit were observed in radial velocities.}}

\end{table*}

In Table\,\ref{tab:table_systems}, we show the seven multi-planet systems known with more than three planets, among which there is at least one transiting a stellar host brighter than $V$=10.5. Kepler-444 is the brightest detected star ($V$=8.9) to host five transiting planets. It is an ultra-compact system formed by very small planets (radii < 0.8\,R$_{\oplus}$) in orbital periods between 3.6 and 9.7\,d. HIP\,41378 \citep[$V$=8.9,][]{vanderburg19}; it hosts two Sub-Neptune planets in orbits of 15.6 and 31.7\,d and three additional large planets in long orbital periods ($\sim$130-324\,d). Therefore, \starname{} is  the second-brightest stellar host of a multiple transiting-planet system in a compact architecture (periods < 30\,d).  The \starname{} system, also consisting of five transiting planets, is suitable for a complete and precise orbital and physical characterization thanks to the brightness of its host star.

The \starname{} system was discovered by the \tess{} survey
\citep[\tess{} Object of Interest ID TOI\,1233;][]{Guerrero2021}, and it was originally reported to consist of four transiting planets with orbital periods between 3.7-19.6\,d and radii between 1.6-3.12\,R$_{\oplus}$ \citep{Daylan2021}. Its magnitude and location in the sky, makes it a perfect target for {\it \cheops} \citep[][]{benz2021}. As a follow-up mission, its goal is to provide ultra-high precision photometric observations of known exoplanet host stars for obtaining precise determinations of transiting exoplanet sizes. \cite{Bonfanti2021} used the first three \cheops{} light curves of \starname{} system to refine the planetary transit ephemeris one year after the discovery observations. In addition, the analysis of these data revealed the presence of a fifth planet with 29.5\,d orbital period. Here, we further characterize the \starname{} system based on the analysis of the full \cheops{} observation dataset consisting of 15 light curves; together with all the available \tess{} data (i.e., with the addition of Sector 37). In this work, we confirm the presence of the fifth planet, $f,$ and we search for the 10.9\,d transit signal of a putative sixth planet, which was suggested in the discovery paper.

In Sect. \ref{sec:star}, we present the updated stellar host properties based on the updated input from Gaia eDR3. In Sect. \ref{sec:observations}, we describe the photometric observations used in this work, reporting 12 new \cheops{} light curves and all the available \tess{} data. The analysis of the data is described in Sect. \ref{sec:analysis}. We describe the timing analysis of the transits in Sect. \ref{ssec:timing_analysis} and compare different approaches to estimate the planetary masses in Sect. \ref{sec:mass_estimations}. The discussion of the results and the conclusions of the work are presented in Sects. \ref{sec:discussion} and \ref{sec:conclusions},  respectively.

\section{Host star properties}\label{sec:star}
\citet{Bonfanti2021} already provided a thorough characterization of \starname. In this work, we assume the same spectroscopic parameters and elemental abundances, and we recall that they were derived from 13 high-resolution spectra (program 1102.C-0923, PI: Gandolfi) acquired by the High Accuracy Radial velocity Planet Searcher \citep[HARPS,][]{Mayor2003}. In particular, $T_{\mathrm{eff}}$, $\log{g}$, and [Fe/H] were inferred from the ARES+MOOG tools \citep{Sousa-14}, while the elemental abundances were computed following the procedure detailed in \citet{Adibekyan-12,Adibekyan-15}. These values are reported in Table\,\ref{tab:stellarParam}.

In taking advantage of the new stellar parallax from the third Gaia early data release \citep[eDR3,][]{GaiaEDR3-2021}, we were able to update the stellar radius always through the infrared flux method \citep[IRFM][]{Blackwell1977}, which has already been summarized in \citet{Bonfanti2021}. The slight refinement brought us to $R_{\star}$=$0.876\pm0.007\,R_{\odot}$, which is fully compatible with what reported in \citet{Bonfanti2021}. Using $T_{\mathrm{eff}}$, [Fe/H], and the new $R_{\star}$ estimate as input parameters, we then updated also the isochronal mass $M_{\star}$ and age $t_{\star}$ estimates. A first pair $(M_{\star,1}, t_{\star,1})$ was computed employing the isochrone placement technique \citep{bonfanti15,bonfanti16} within PARSEC\footnote{\textit{P}adova \textit{A}nd T\textit{R}ieste \textit{S}tellar \textit{E}volutionary \textit{C}ode: \url{http://stev.oapd.inaf.it/cgi-bin/cmd}} v1.2S evolutionary models \citep{marigo17}, while a second pair $(M_{\star,2}, t_{\star,2})$ was outputted by the CLES code \citep[Code Liègeois d´Évolution Stellaire,][]{scuflaire08}, following the fitting procedure described in \citet{salmon21}. We finally combined the two respective estimates after carefully checking their mutual consistency through the $\chi^2$-based criterion outlined in \citet{Bonfanti2021}, ending up with the following results: $M_{\star}$=$0.867_{-0.046}^{+0.047}\,M_{\odot}$ and $t_{\star}$=$6.7_{-3.4}^{+3.3}$\,Gyr. All the relevant stellar parameters are listed in Table~\ref{tab:stellarParam} and we note that all the updated parameters are consistent with what computed by \citet{Bonfanti2021} well within 1$\sigma$.

\begin{table}
\caption{Stellar properties of \starname.}             
\label{tab:stellarParam}      
\centering                          
\begin{tabular}{lll}        
\hline\hline                 
\multicolumn{3}{c}{HD\,108236} \\    
\hline                        
\multirow{4}{2 cm}{Alternative names} & \multicolumn{2}{l}{TOI-1233} \\ 
 & \multicolumn{2}{l}{HIP 60689} \\ 
 & \multicolumn{2}{l}{TIC\,260647166} \\
 & \multicolumn{2}{l}{Gaia DR2 6125644402384918784} \\
\hline
Parameter & Value & Source \\
\hline
   R.A. [J2000] & 12:26:17.89 &  Gaia DR2 \\
   DEC. [J2000] & -51:21:46.21 & Gaia DR2 \\
   $V$ [mag] & 9.24 & Simbad \\
   $G$ [mag] & 9.0875 & Simbad \\
   $J$ [mag] & 8.046 & Simbad \\
   $T_{\mathrm{eff}}$ [K] & $5660\pm61$ & Spectroscopy \\
   $\log{g}$ [cgs]      & $4.49\pm0.11$ & Spectroscopy \\\relax
   [Fe/H] [dex] & $-0.28\pm0.04$ & Spectroscopy \\\relax
   [Mg/H] [dex] & $-0.27\pm0.03$ & Spectroscopy \\\relax
   [Si/H] [dex] & $-0.29\pm0.02$ & Spectroscopy \\
   \hline
   Updated values \\
   \hline
   $d$ [pc] & $64.66\pm0.05$ & Gaia parallax\tablefootmark{(a)}\\
   $\theta$\tablefootmark{(b)} [mas] & $0.1261\pm0.0010$ & IRFM \\
   $R_{\star}$ [$R_{\odot}$] & $0.876\pm0.007$ & IRFM \\
   $M_{\star}$ [$M_{\odot}$] & $0.867_{-0.046}^{+0.047}$ & Isochrones \\
   $t_{\star}$ [Gyr]        & $6.7_{-3.4}^{+3.3}$ & Isochrones \\
   $L_{\star}$ [$L_{\odot}$] & $0.707\pm0.032$ & from $R_{\star}$ and $T_{\mathrm{eff}}$\\
   $\rho_{\star}$ [g/cm$^3$] & $1.79\pm0.11$ & from $R_{\star}$ and $M_{\star}$ \\
\hline                                   
\end{tabular}
\tablefoot{\tablefoottext{a}{Parallax offset from \citet{lindegren21} applied.}\tablefoottext{b}{Angular diameter.} }
\end{table}

\section{Observations}\label{sec:observations}

Here, we describe the photometric data used for the modeling of the system from both \cheops{} and \tess{} space missions.

\subsection{CHEOPS}\label{sec:observations_cheops}

We obtained 15 \cheops{}  observation runs (visits) of \starname, accumulating $\sim$8.5\,d on target, of which $\sim$3.9\,d were not covered due to Earth occultations or South Atlantic Anomaly (SAA) crossings by the satellite.  With \cheops{}, we observed a total of 16 planetary transits of the system, as reported in the observation log in Table \ref{tab:obs_log}. 
The raw data of each visit were automatically processed by \cheops{} Data Reduction Pipeline \citep[DRP, v13][]{hoyer2020}.  In short, the DRP performs the instrumental calibration (bias, gain, linearization, and flat-fielding correction) and the environmental correction (cosmic ray hits, background, and smearing correction) before extracting the photometric signal of the target in four different aperture sizes.  Three of these apertures have a fixed radius (\texttt{RINF}=22.5\,pix, \texttt{DEFAULT}=25\,pix, \texttt{RSUP}=30\,pix) while a fourth, \texttt{ROPT}, is chosen automatically based on the contamination present in the Field of View (FoV).  The contamination is computed by DRP using simulations of the FoV based on the Gaia catalog \citep{GaiaCollaboration2018}. This information, together with other correction times series (e.g., background, smearing), are provided in the final DRP products along with the photometric time series, to be used in the analysis and detrending of the light curves. In this work, we use the light curves obtained with the \texttt{DEFAULT} aperture, as it produces the raw curves with the lowest dispersion when compared to the other photometric apertures.

\begin{table*}
\caption{Log of the \cheops{} observations of \starname. The second column shows the planetary transits observed during each \cheops{} visit. The last column shows the observation efficiency of each visit, considering the gaps produced by Earth occultations or SAA crossings. We report the unique identifiers of each \cheops{} visit: \texttt{OBSID} and \texttt{File Key}. }            
\label{tab:obs_log}      
\centering   
\small
\begin{tabular}{l c c c c c c c }
\hline\hline

id &Planets & \texttt{File Key} & \texttt{OBSID} & Start date & Duration & Exp. time & Efficiency \\ 
\# &        &           &      &   [UTC]    &  [h]     & [s] &  [\%] \\
\hline
 &        &           &      &     &      & &   \\
1\tablefootmark{(a)} & c, e, (g) & CH\_PR300046\_TG000101\_V0200 & 1015572 & 2020-03-10T18:09:15.91 & 18.33 & 42.0 & 61.1\\
2 & b & CH\_PR100031\_TG015701\_V0200 & 1084970 & 2020-04-19T08:48:11.23 & 17.04 & 49.0 & 59.6 \\
3\tablefootmark{(a)}& d & CH\_PR100031\_TG015401\_V0200 & 1088211 & 2020-04-28T07:06:11.05 & 18.63 & 49.0 & 54.8 \\
4& c & CH\_PR100031\_TG015601\_V0200 & 1088443 & 2020-04-29T16:02:11.28 & 11.24 & 49.0 & 53.7 \\
5\tablefootmark{(a)}& b, f & CH\_PR100031\_TG015702\_V0200 & 1087957 & 2020-04-30T17:06:11.04 & 16.40 & 49.0 & 54.5 \\
6& c & CH\_PR100031\_TG015602\_V0200 & 1096738 & 2020-05-05T21:36:10.33 & 10.57 & 49.0 & 57.1 \\
7& c & CH\_PR100031\_TG022401\_V0200 & 1107339 & 2020-05-12T02:10:11.27 & 07.08 & 49.0 & 57.5 \\
8& b & CH\_PR100031\_TG023001\_V0200 & 1111865 & 2020-05-19T23:34:51.37 & 08.91 & 49.0 & 56.6 \\
9& d, (g) & CH\_PR100031\_TG024301\_V0200 & 1117140 & 2020-05-26T09:18:11.28 & 26.37 & 49.0 & 51.0 \\
10& b & CH\_PR100031\_TG023002\_V0200 & 1117572 & 2020-05-27T14:31:10.32 & 08.35 & 49.0 & 47.6 \\
11& e & CH\_PR100031\_TG022701\_V0200 & 1117376 & 2020-05-28T05:21:10.39 & 13.14 & 49.0 & 50.4 \\
12& b & CH\_PR100031\_TG023003\_V0200 & 1124008 & 2020-05-31T09:24:11.59 & 08.89 & 49.0 & 52.8 \\
13& c & CH\_PR100031\_TG022901\_V0200 & 1124302 & 2020-06-05T22:42:30.53 & 10.38 & 49.0 & 51.6 \\
14& f & CH\_PR110045\_TG002601\_V0200 & 1458282 & 2021-04-20T06:58:13.84 & 10.95 & 49.0 & 62.2 \\
15& (g) & CH\_PR110045\_TG002701\_V0200 & 1456848 & 2021-04-29T18:51:51.55 & 17.98 & 49.0 & 53.9 \\
 &        &           &      &     &      & &   \\
\hline
\end{tabular}
\tablefoot{\tablefoottext{a}{Observations presented in \cite{Bonfanti2021} }}

\end{table*}

\subsection{TESS}\label{sec:observations_tess}

Similarly to \cite{Daylan2021} and \cite{Bonfanti2021}, our analysis includes the \tess{} data of \starname{}, consisting of the 2\,min light curves obtained during the monitoring of Sectors 10 and 11. Additionally, we incorporated the 2\,min Sector 37 data, collected by  \tess{} \texttt{camera 2,} between 2021-04-02 and 2021-04-28, during its second pass of the Southern Hemisphere. Here, we also use the Presearch Data Conditioning Simple Aperture Photometry \citep[PDC-SAP,][]{Stumpe2012,Stumpe2014, Smith2012} as provided by the Science Processing Operations Center \citep[SPOC,][]{jenkins16}. In the \tess{} data, we identified and analyzed 18 transits of planet {\it b}, 12 transits of planet {\it c}, 6 transits of planet {\it d}, 4 transits of planet {\it e,} and 3 transits of planet {\it f}. In order to reduce processing times in the analysis, in this work we used only the portions of the \tess{} light curves with detected transits (leaving enough out-of-transit data for further detrending), ending up with 34 single light curves which we treated individually; among them, 4 light curves present consecutive transits of two planets and other 5 light curves show overlapped transits (transit events of two planets occurring simultaneously). 

\section{System modeling}\label{sec:analysis}

\subsection{Detrending of CHEOPS light curves}\label{sec:LCdetrending}

\cheops{} rotates around the line of sight while observing. Together with the extended shape of the \cheops{} PSF, this can produce different effects that perturb the aperture photometry \citep[see e.g.,][]{hoyer2020, lendl2020, Bonfanti2021, Maxted2022}. Some of these effects are the varying contaminating flux from nearby rotating background stars, moving smear trails of bright stars, and the reflections imprinted in the detector produced by different astrophysical sources close to \cheops{} pointing direction. All of these effects vary with the roll angle of the satellite and as a result, the flux of the raw target's light curve is correlated with the roll angle. Therefore, in most cases, a proper detrending against this parameter is fundamental to exploit the exquisite \cheops{} photometry.  
Following \citet{Bonfanti2021}, we modeled the flux versus the roll angle pattern of each \cheops{} light curve through Gaussian processes \citep[GPs,][]{rasmussen05}. In particular, we adopted a Mat\'ern-3/2 kernel and used the GP predictor implemented in the \texttt{celerite} package \citep{foreman17}, which provides the GP model and its variance. Finally, we obtained the roll-angle detrended light curves, while also adjusting the flux error bars to account for the further variance introduced by the GP modeling. In order to reduce the number of fitted parameters, we used these roll angle detrended \cheops{} light curves in the following transits modeling. The \cheops{} raw and detrended light curves are reported in Appendix\,\ref{ssec:raw_detrended_lcs} (Fig.\,\ref{fig:raw_lcs}, \ref{fig:cheops_lcs1}, and \ref{fig:cheops_lcs2}).

\subsection{Transit modeling}
\label{ssec:transit_modeling}
\begin{figure*}
    \centering
    \includegraphics[width=0.47\hsize]{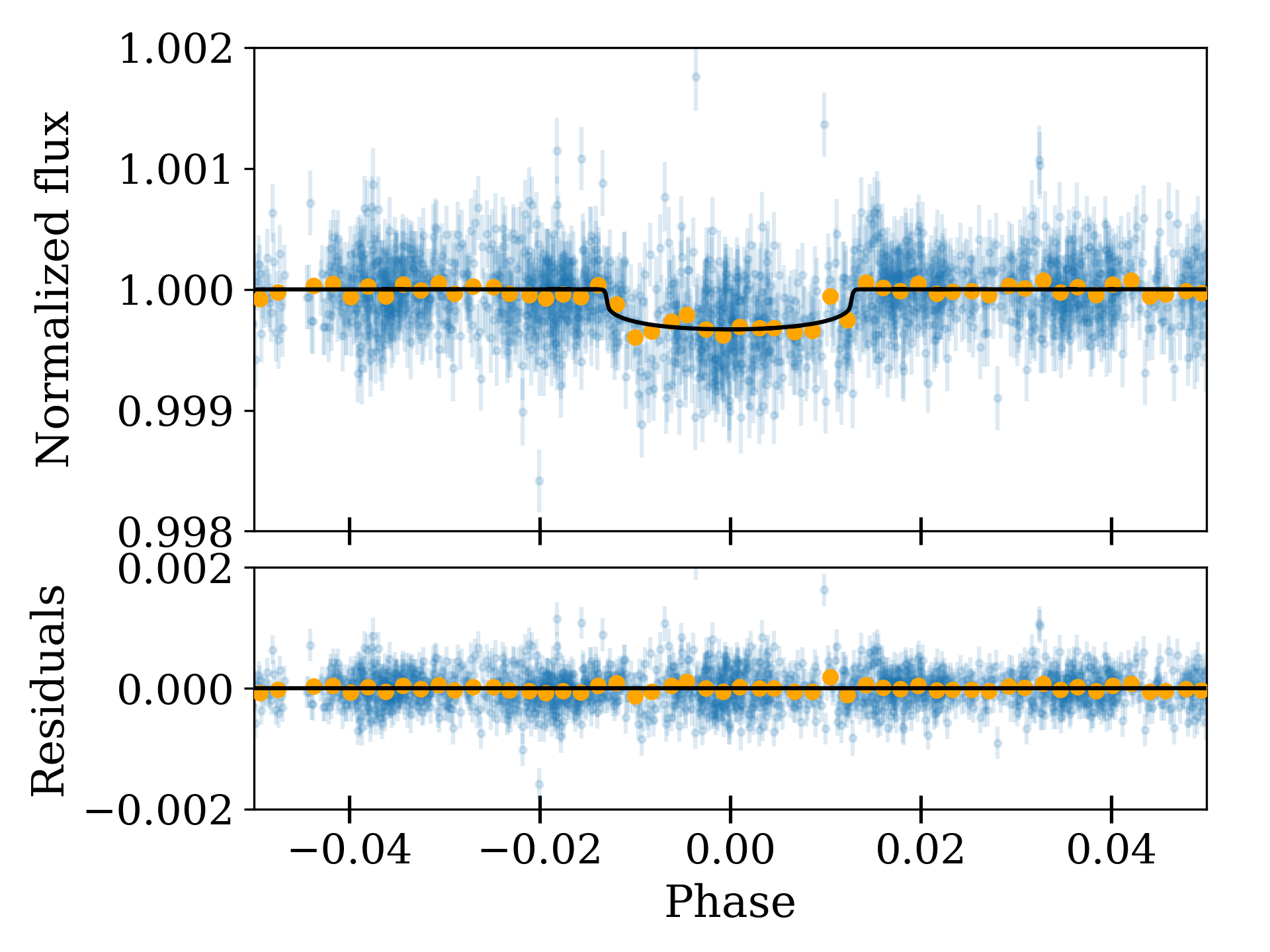} 
    \includegraphics[width=0.47\hsize]{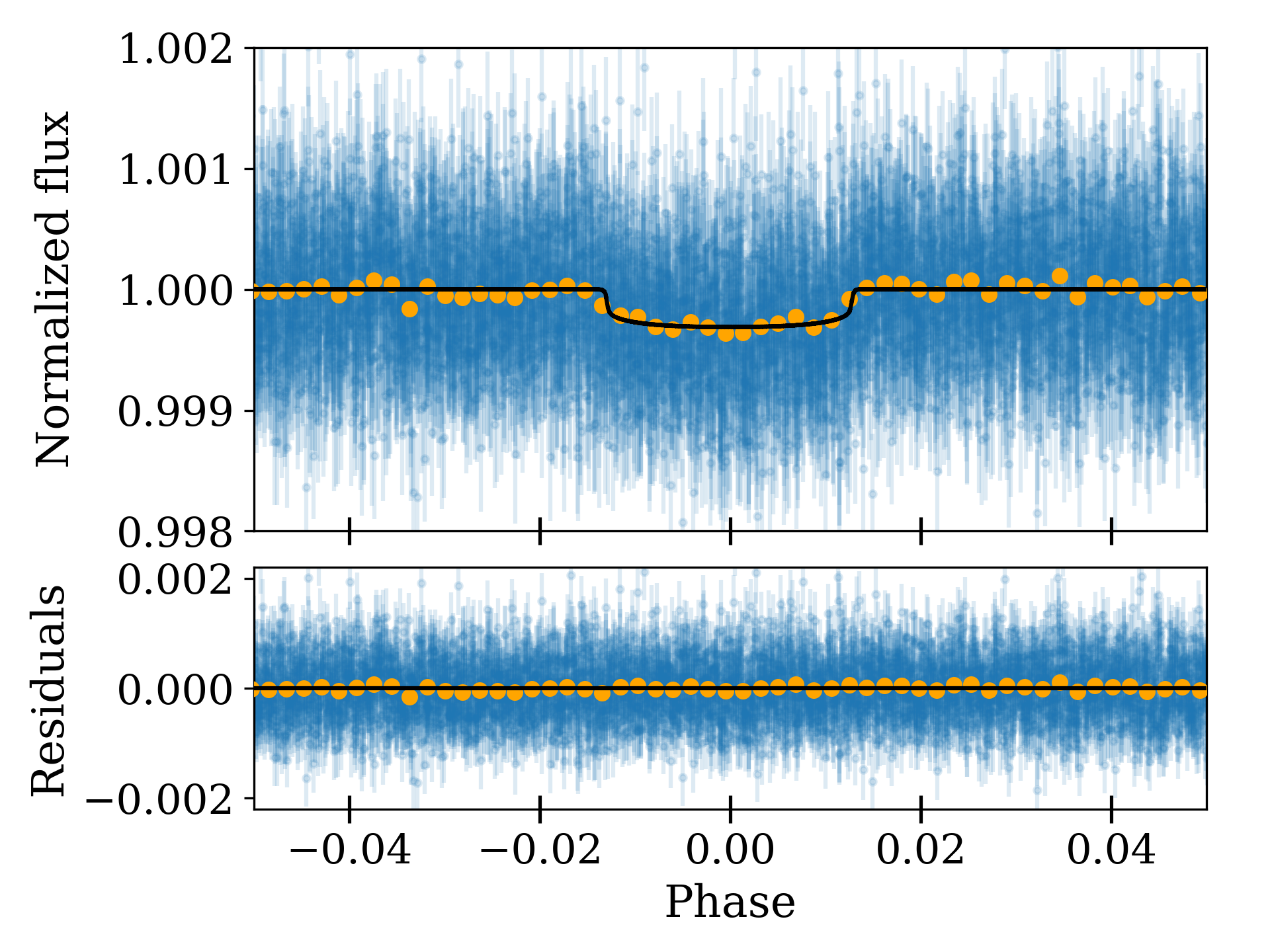} \\
    \includegraphics[width=0.47\hsize]{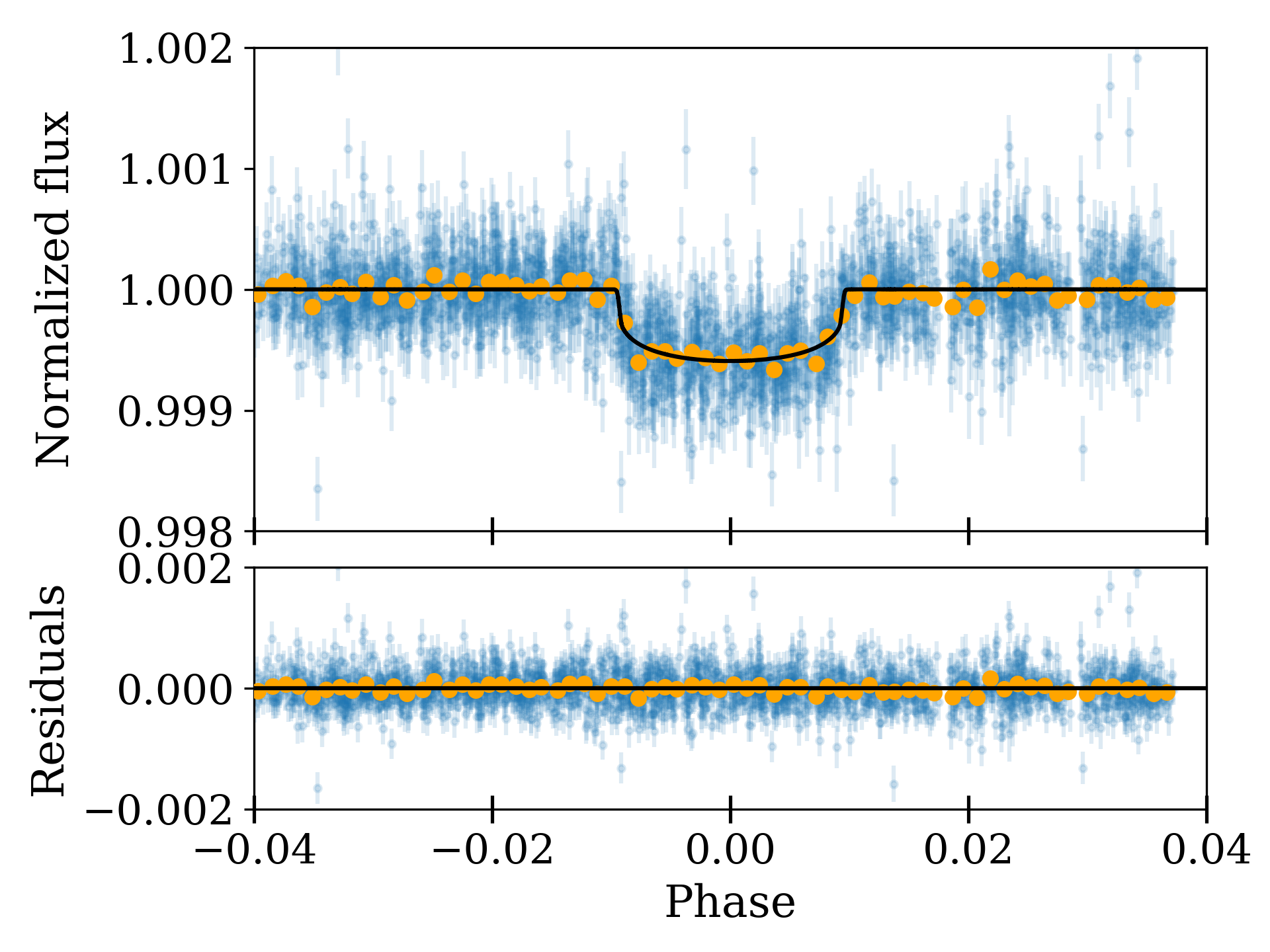}
    \includegraphics[width=0.47\hsize]{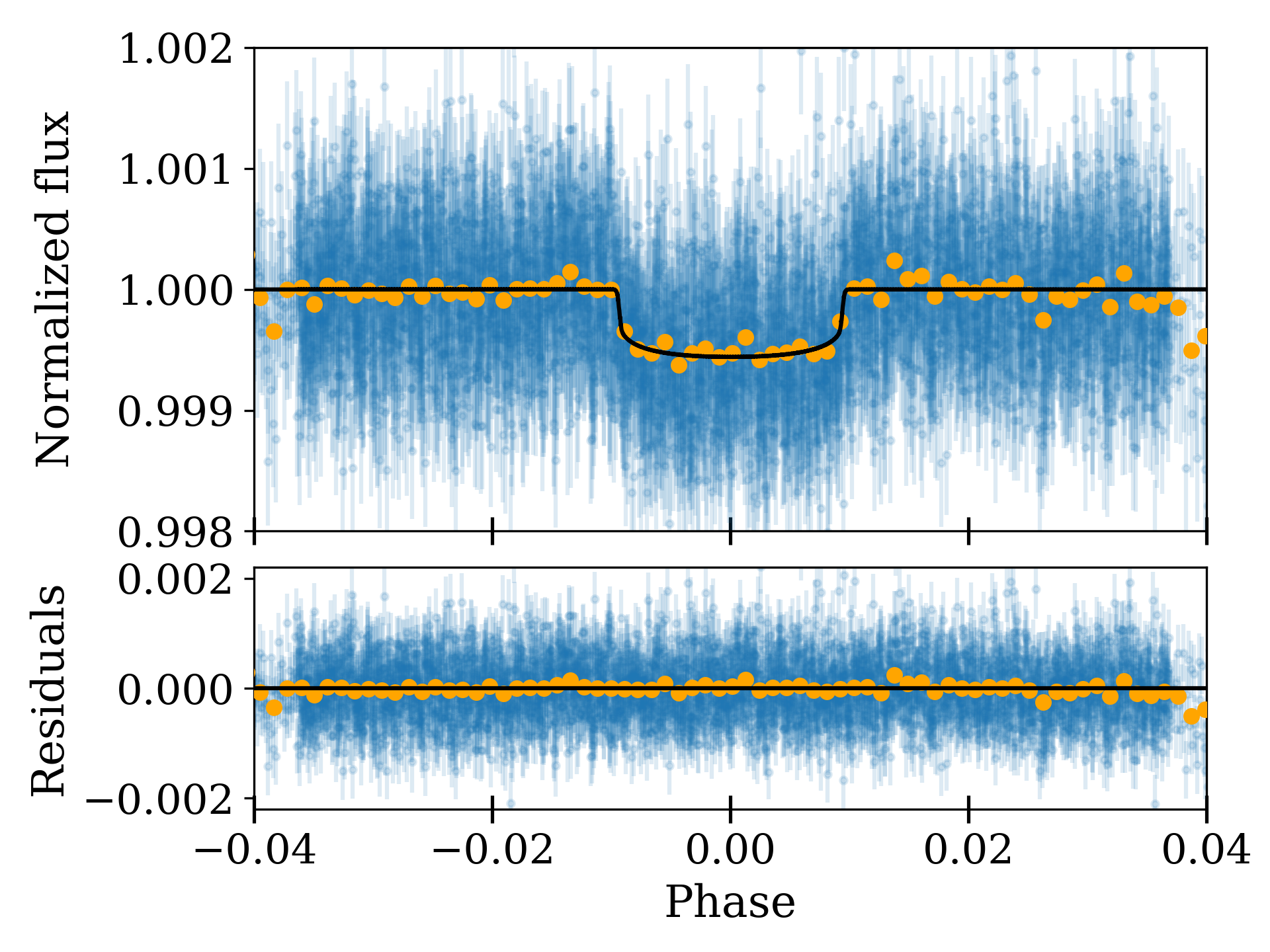} \\
    \includegraphics[width=0.47\hsize]{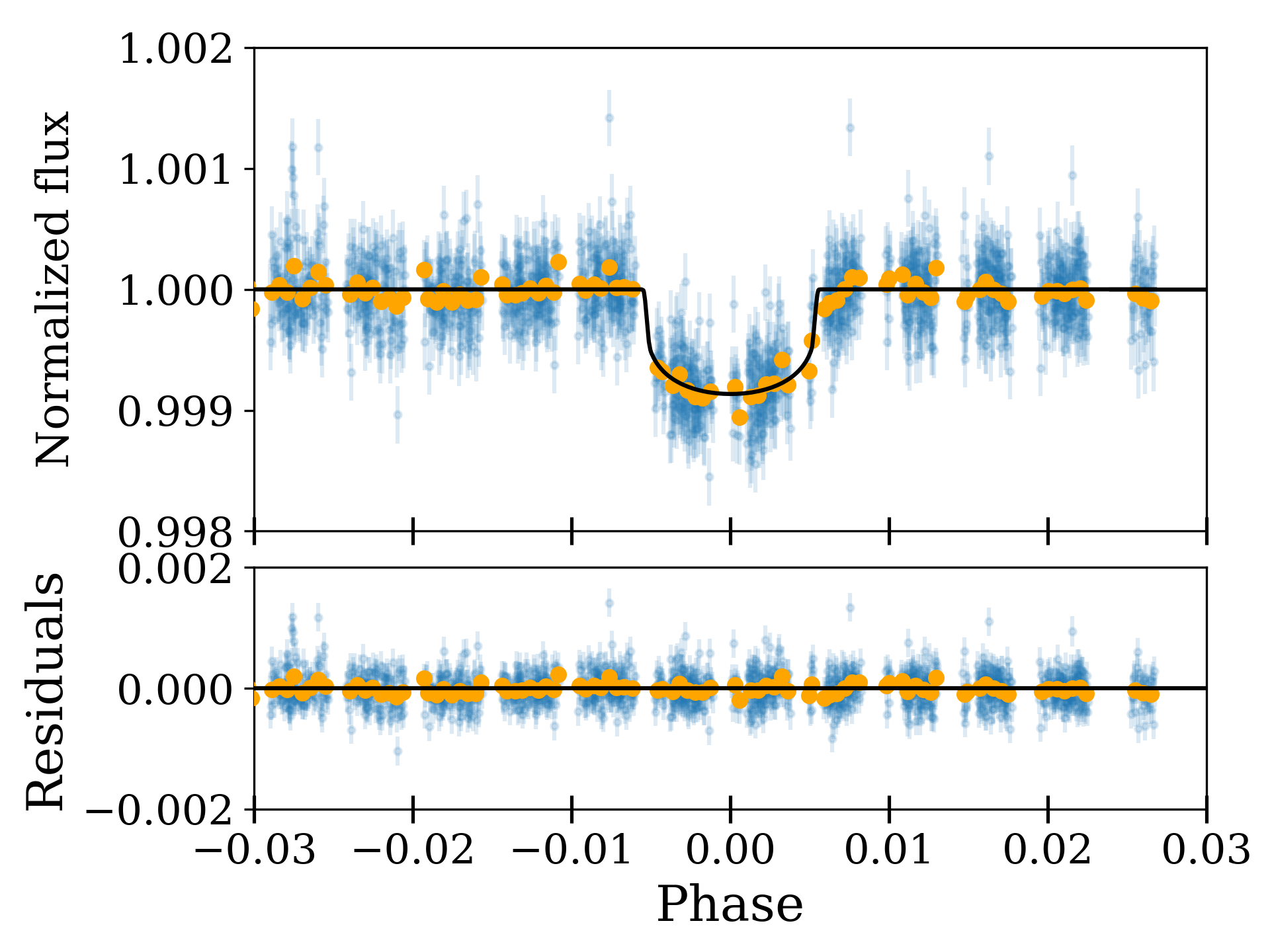}
    \includegraphics[width=0.47\hsize]{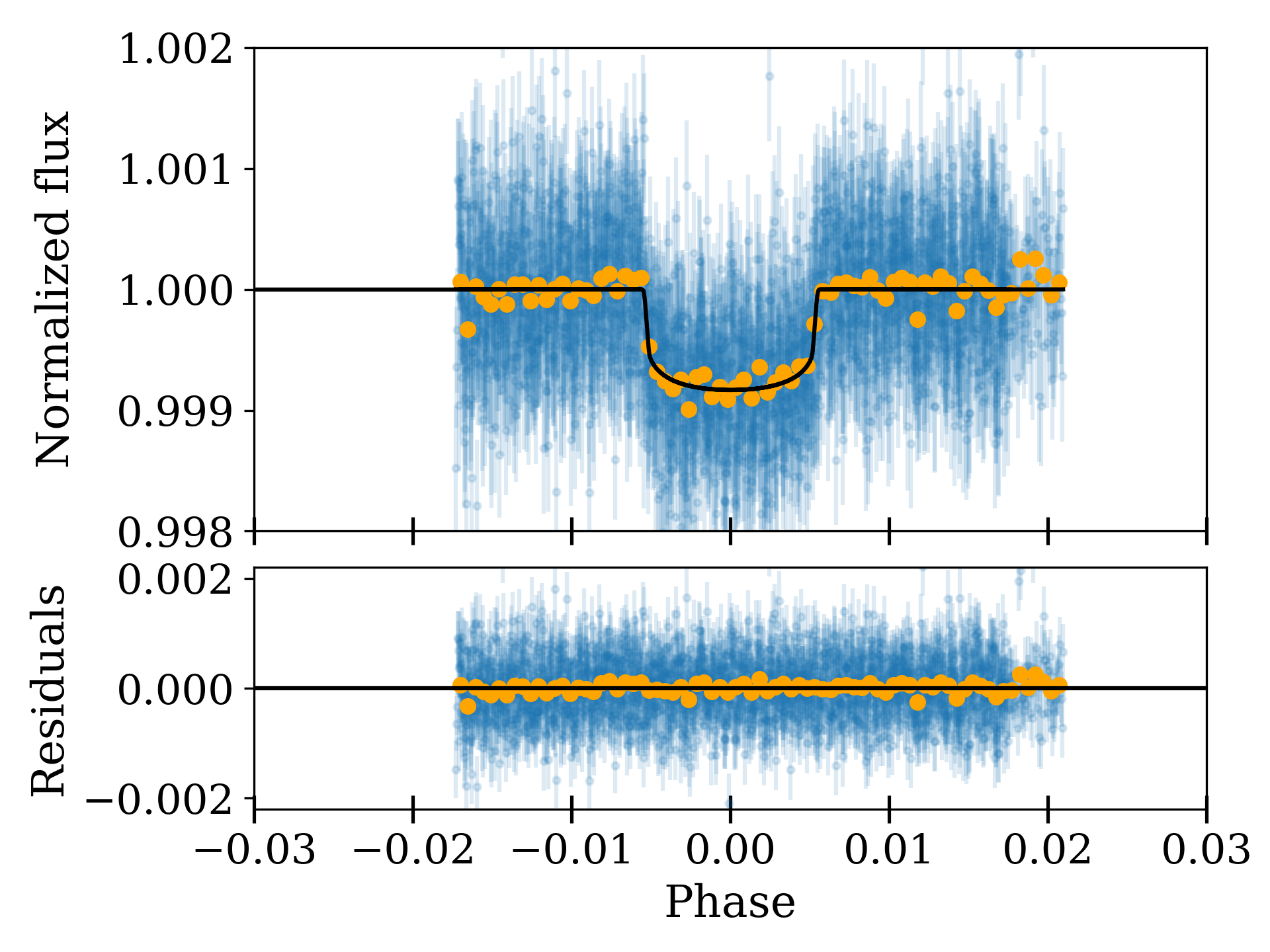} \\
    
    \caption{Phased \cheops{} (left) and \tess{} (right) light curves (blue dots) and its 10-min bins (yellow points). The best transit models (solid black curves) for planets $b$, $c$ and $d$ (from top to bottom) obtained in the global analysis of the data are shown; together with the respective residuals (thin bottom panels). Additional transit events of other planets were removed before plotting the final phased curve.}
    \label{fig:phased_curves1}
\end{figure*}

\begin{figure*}
    \centering
    \includegraphics[width=0.48\hsize]{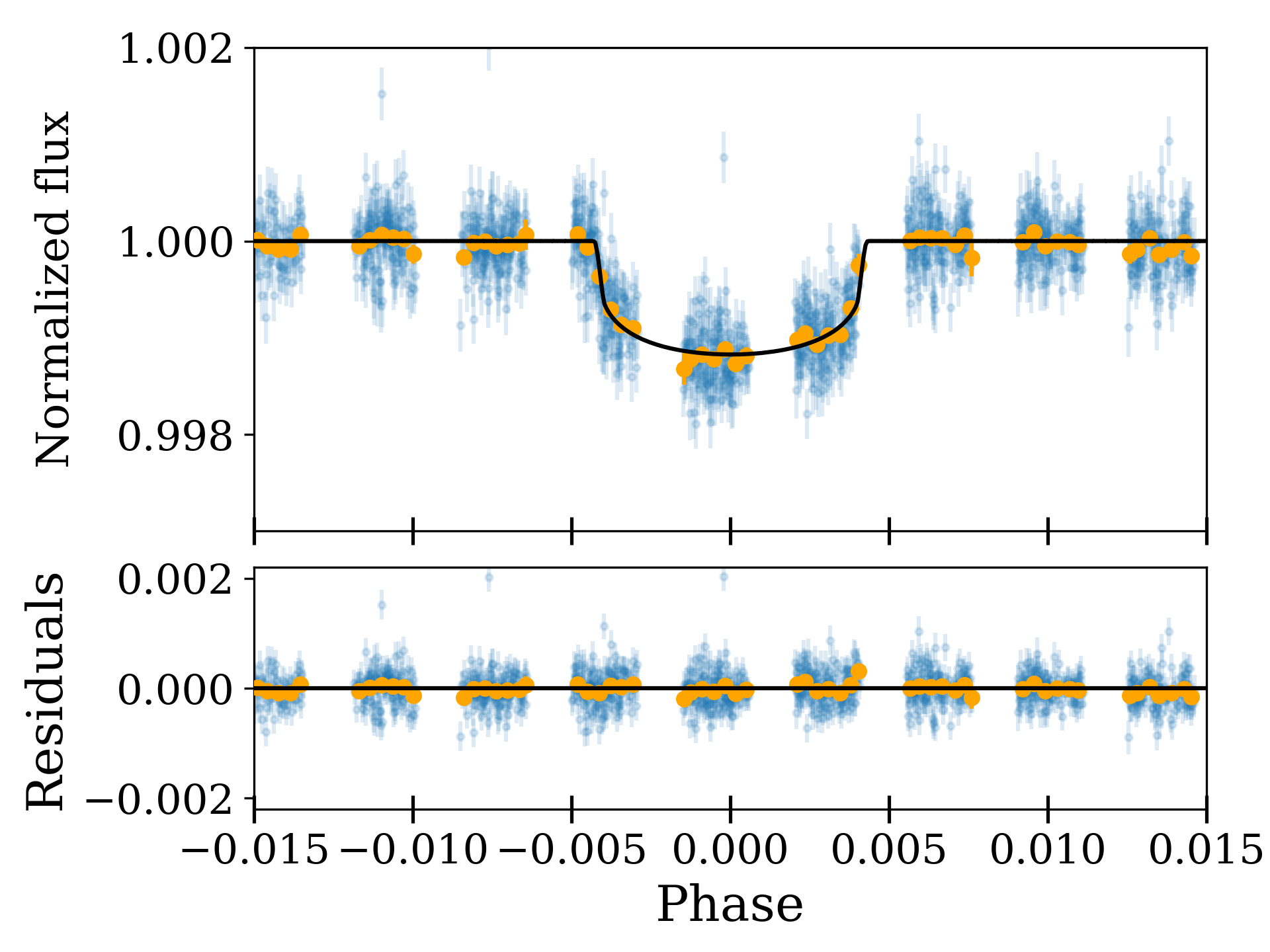} 
    \includegraphics[width=0.48\hsize]{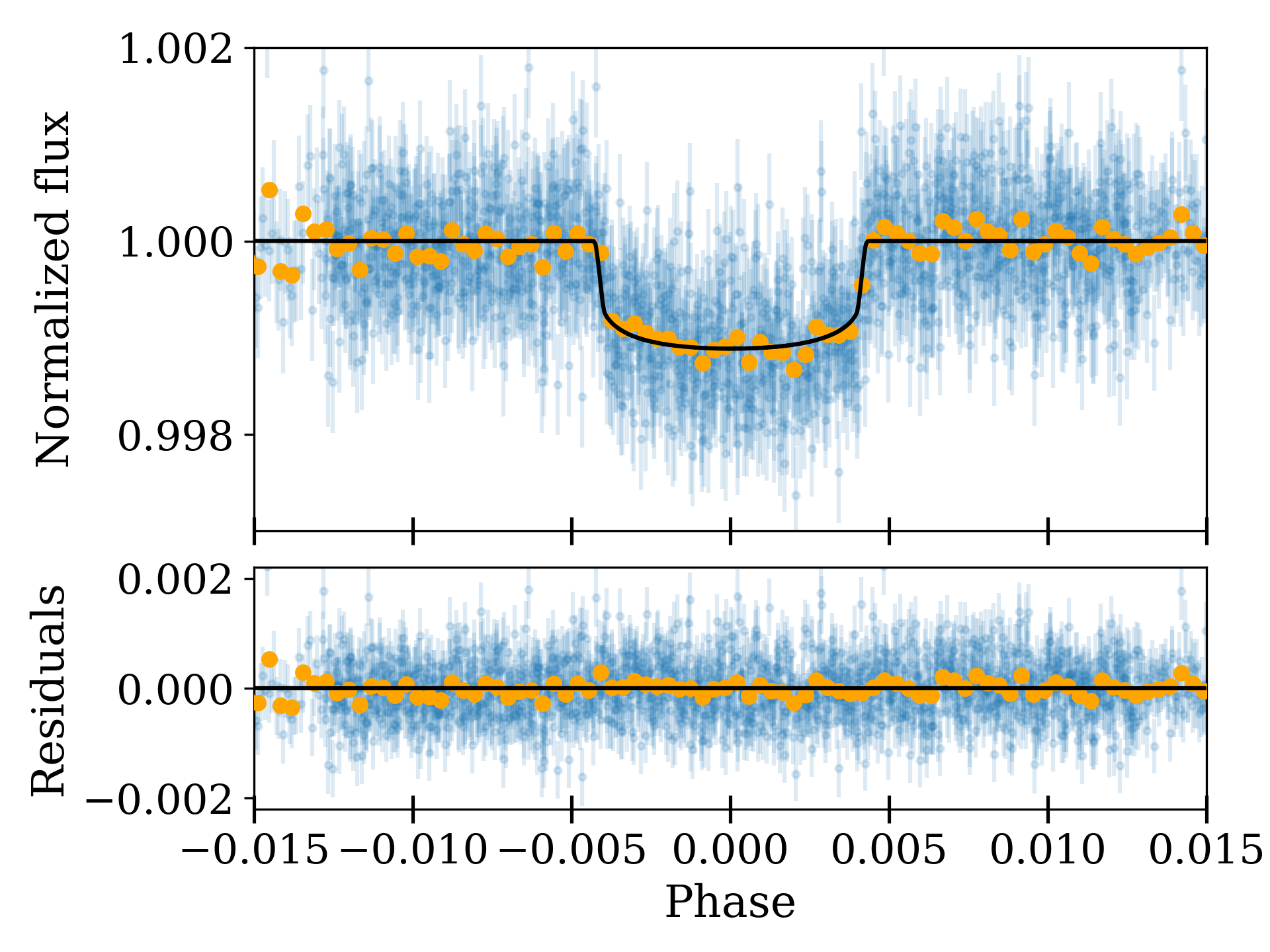} \\
    \includegraphics[width=0.48\hsize]{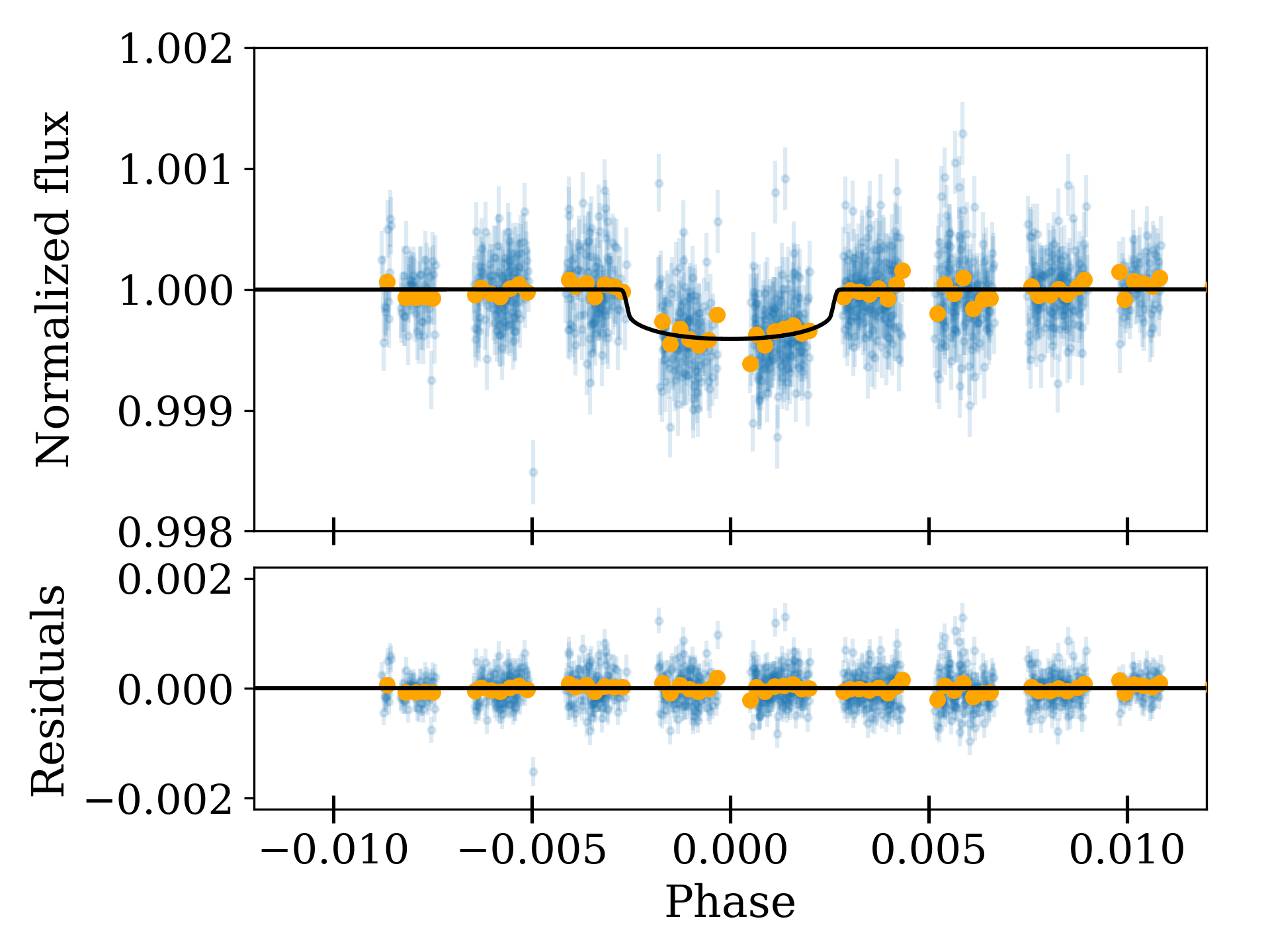} 
    \includegraphics[width=0.48\hsize]{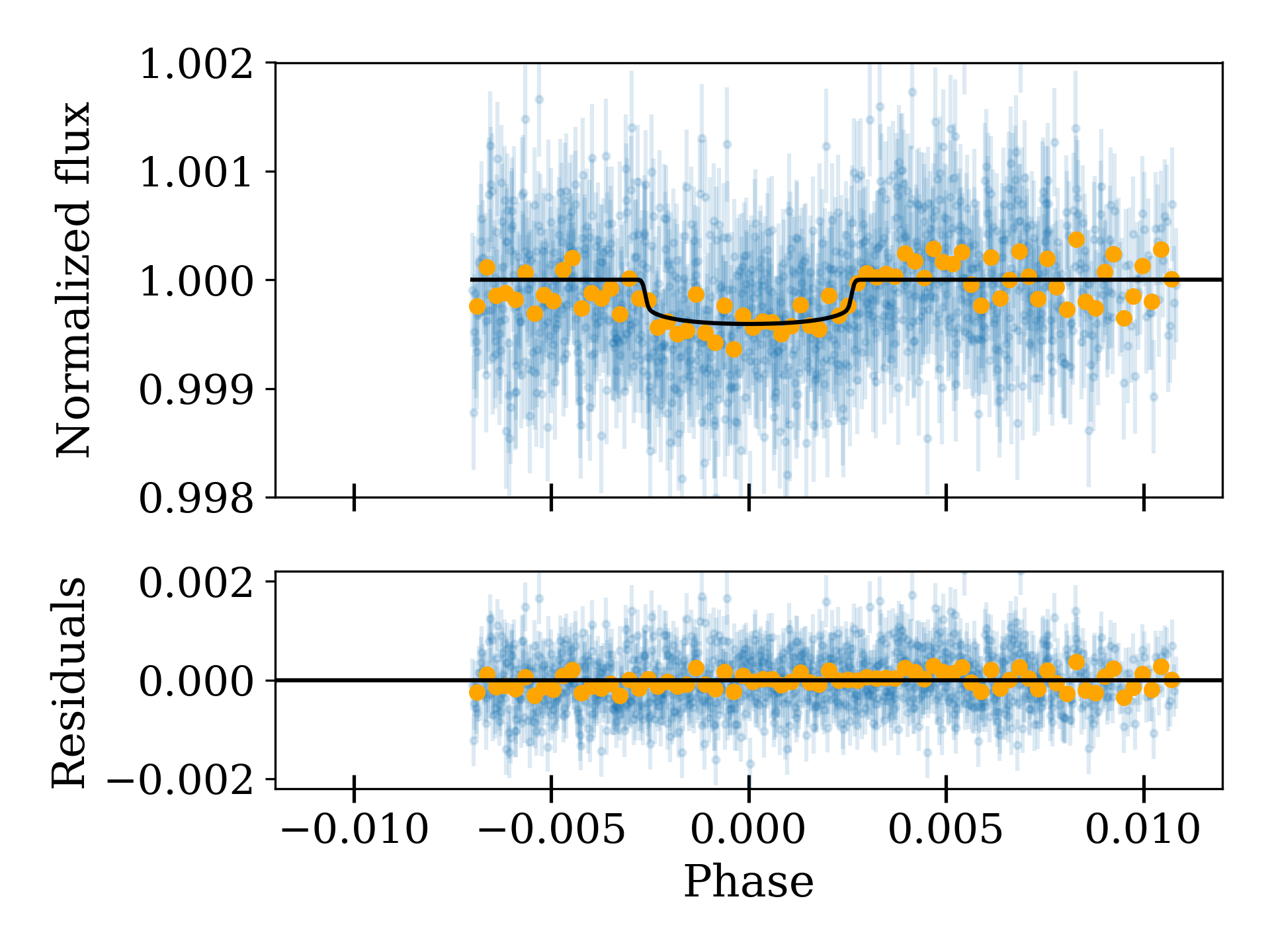} \\

    \caption{ Phased \cheops{} (left) and \tess{} (right) light curves (blue dots) and its 10-min bins (yellow points). The best transit models (solid black curves) for planet $e$ (top) and $f$ (bottom) obtained in the global analysis of the data. Same details as in Fig.~\ref{fig:phased_curves1}.}
    \label{fig:phased_curves2}
\end{figure*}

For the analysis of \cheops{} and \tess{} light curves, we used the \texttt{juliet} package \citep{Espinoza2019}, which implements the modeling of transit light curves from \texttt{batman} \citep{Kreidberg2015} in a Bayesian framework using the \texttt{dynesty} nested sampling tool \citep{Speagle2020}.
We performed the modeling of the full system (star+five planets) using all the \cheops{} (15 roll-angle detrended light curves, as described in Sect.~\ref{sec:LCdetrending}) and \tess{} (34) light curves. In this configuration, we are able to retrieve, for example, an estimation of the orbital period (P) and reference time (T$_0$) for each planet. We used as reference the values published in \cite{Bonfanti2021} to define our priors.  In general, we used normal distributions centered in their results, but wide enough to ensure they are weakly informative. For the priors of the transit reference time (T$_0$) of each planet, when possible, we used \tess{} epochs which do not correspond to overlapped transits to mitigate potential time offsets.  
    We also adopted the same constraints in the eccentricity ($e<0.1$) resulting from dynamical stability simulations of the system; and the values for the quadratic limb darkening coefficients ($q_1$, $q_2$) for \cheops{} and \tess{} passbands (see \cite{Bonfanti2021} for details). In addition, \texttt{juliet} allows us to directly fit the stellar density ($\rho_{\star}$), instead of fitting the scaled semi-major axis for each planet. Therefore, we used the value of stellar density derived from the updated stellar parameters (described in Sect. \ref{sec:star}) as a truncated-normal prior in the modeling.  We also fit for a photometric offset ($mflux$) and a "jitter"\ noise term ($\sigma_w$) for each light curve. As the contamination by external sources is small and already corrected in \cheops{} light curves by the roll angle detrending (described in Sect. \ref{sec:LCdetrending}), we fixed the dilution factor of each light curve, $mdilution$=1. In addition, we fit for a linear and quadratic temporal term ($c_1(t)$ and $c_2(t^{2})$) to take care of any residual trend remaining in the \cheops{} transits.  
Any additional free parameters of our joint fit were the orbital period (P), reference time (T$_0$), planet-to-star radius ratio (R$_p$/R$_{\star}$), and the impact parameter ($b$) of each planet. The eccentricity ($e$) and the argument of periastron passage ($\omega$) of the orbit were estimated using the parametrization ($\sqrt{e}sin(\omega)$,$\sqrt{e}cos(\omega)$). For these two parameters and the impact parameter, $b$, we used Uniform prior distributions. The value of the scaled semi-major axis of the orbit ($a$/R$_{\star}$) for each planet was derived directly from the fitted $\rho_{\star}$, P, and Kepler's third law. All the priors used in the modeling are listed in Table~\ref{tab:priors}. The fitted and derived planetary parameters from the posterior distributions of the system are shown in Table~\ref{tab:results}, while the phased light curves together with the best fit model are presented in Fig.~\ref{fig:phased_curves1} and~\ref{fig:phased_curves2} for \cheops{} and \tess{} data, respectively. The individual detrended light curves with the best fit model overplotted are shown in Figs.~\ref{fig:cheops_lcs1} and~\ref{fig:cheops_lcs2}.

To confirm our results, we also performed an independent analysis of all the data with the \texttt{Allesfitter} package \citep{allesfitter-paper, allesfitter-code}, following a similar procedure as described in \cite{Bonfanti2021}. We imposed Normal priors on the mean stellar density $\rho_{\star}$ and on the LD coefficients. For all the other system parameters, we adopted wide uniform priors. An upper limit of 0.1 on the orbital eccentricity was also imposed. 
The temporal detrending benefited of the GP treatment (Matérn-3/2 kernel) for the three \tess{} sectors, while we used splines for the 15 \cheops{} observations (already pre-detrended against roll-angle following Sect.~\ref{sec:LCdetrending}) to avoid a dramatic increase of the number of fitted parameters.
The resulting parameters are fully consistent, well within $\pm1\sigma$ level in most of the cases, with the results obtained from the \texttt{juliet} analysis.  We notice that orbital eccentricities listed in Table~\ref{tab:results} are high when compared to what is expected for multiplanetary systems. We performed a series of tests to explore the possible cause, and finally we attribute the results to the parameterizations used by \texttt{juliet}. In any case, a more detailed RV analysis will provide a final estimate of the eccentricities, something that the photometry presented in this work cannot solve on its own. We describe the performed tests in Appendix~\ref{sec:appendix_eccs}. We also confirm that the rest of the other modeled parameters are robust, no matter the approach used to fit $e$ and $\omega$.

\begin{table*}
\caption{Fitted and derived parameters of the \starname{} system from \cheops{} and \tess{} transits.} 
\label{tab:results}
\centering
\small
\begin{tabular}{lccccc}
\hline
\hline
&\multicolumn{3}{c}{ \starname }  && \\
\hline

Quadratic L. D. Coeff.& $q_{1}$& $q_{2}$& &$q_{1}$ &$q_{2}$\\
\hline
&&&&&\\
\cheops   & $0.446^{+0.016}_{-0.018}$ & $0.341^{+0.017}_{-0.014}$ &
\tess     & $0.277 ^{+0.018}_{-0.023}$  &  $0.230^{+0.016}_{-0.021}$ 
\\
&&&&&\\
\hline
 {\it Fitted parameter}  & Planet $b$  & Planet $c$ & Planet $d$ & Planet $e$ & Planet $f$ \\
\hline
&&&&&\\
Orbital Period, P\tablefootmark{(a)} {[}d{]}                        & $3.7958785^{+0.0000076}_{-0.0000070}$ & $6.2036717^{+0.0000092}_{-0.000010}$ & $14.175818^{+0.000024}_{-0.000023}$ & $19.5901277^{+0.000030}_{-0.000030}$ & $29.53935^{+0.00012}_{-0.00014}$ \\ 
T$_0$\tablefootmark{(a)} {[}BJD-2\,450\,000.{]}    & $8572.11017^{+0.00062}_{-0.00068}$    & $8572.39445^{-0.00067}_{-0.00065}$   & $8571.33678^{+0.00061}_{-0.00059}$  & $8586.56741^{+0.00056}_{-0.00055}$   & $8616.0403^{+0.0014}_{-0.0014}$  \\ 
R$_{p}$/R$_{\star}$ & $0.01661^{+0.00025}_{-0.00026}$ & $0.02221^{+0.00020}_{-0.00018}$      & $0.02751^{+0.00025}_{-0.00024}$     & $0.03148^{+0.00023}_{-0.00023}$      & $0.01979 \pm 0.00043 $ \\
$\sqrt{e} sin(\omega)$                                                               & $0.221^{+0.043}_{-0.055} $            & $0.154^{+0.043}_{-0.068}$            & $-0.159^{+0.051}_{-0.067}$          & $0.257^{+0.025}_{+0.035}$            & $-0.028^{+0.072}_{-0.076}$       \\
$\sqrt{e} cos(\omega)$                                                               & $0.120^{+0.066}_{-0.065} $            & $-0.256^{+0.047}_{-0.028}$           & $-0.217^{+0.084}_{-0.051}$          & $-0.048^{+0.062}_{-0.070}$           & $0.129^{+0.072}_{-0.076}$        \\
Impact parameter, $b$ {[}R$_{\star}${]}                                              & $0.322^{+0.060}_{-0.057}$             & $0.308^{+0.055}_{-0.046}$            & $0.519^{+0.029}_{-0.026}$           & $0.392^{+0.032}_{-0.029}$            & $0.701^{+0.020}_{-0.021}$        \\
&&&&&\\
\hline
 {\it Derived parameter }\\
\hline
&&&&&\\
Rel. semi-major  axis, a/R$_{\star}$                                              & $11.080^{+0.046}_{-0.043}$            & $15.373^{+0.064}_{-0.059}$           & $26.67^{+0.11}_{-0.10}$             & $33.10^{+0.14}_{-0.13}$              & $43.51^{+0.18}_{-0.17}$          \\ 
Eccentricity, $e$                                                                    & $0.067^{+0.025}_{-0.026}$             & $0.091^{+0.021}_{-0.033}$            & $0.075^{+0.021}_{-0.022}$           & $0.073^{+0.014}_{-0.017}$            & $0.024^{+0.024}_{-0.016}$        \\
Arg. of periastron, $\omega$ {[}$^{\circ}${]}                                    & $61^{+15}_{-14}$                      & $149 ^{+11}_{-7}$                    & $144^{+12}_{-22}$                   & $101^{+15}_{-14}$                    & $24^{+34}_{-17}$                 \\
Orbital inclination, $i$ {[}$^{\circ}${]}                                            & $88.23 \pm 0.32$                      & $88.8 \pm 0.2$                       & $88.8 \pm 0.1$                      & $89.3 \pm 0.1$                       & $89.07 \pm 0.03$                 \\
Planet radius {[}R$_{\oplus}${]}                                                     & $1.587 \pm 0.028$                     & $2.122 \pm 0.025$                    & $2.629 \pm 0.031$                   & $3.008 \pm 0.032$                    & $1.89 \pm 0.04$                  \\
Semi-major axis, a {[}AU{]}                                                          & $0.0451 \pm 0.0004$                   & $0.0626 \pm 0.0006$                  & $0.1087 \pm 0.0010$                 & $0.1348 \pm 0.0012$                  & $0.1773 \pm 0.0016$              \\
Equilibrium temp. T$_{\text{eq}}$\tablefootmark{(b)} {[}K{]} & $1202 \pm 13$                         & $1021 \pm 11$                        & $775 \pm 8$                         & $696 \pm 8$                          & $607 \pm 7$                      \\
Transit duration, T$_{\text{I-IV}}$ {[}h{]}                                          & $2.45 \pm 0.06$                       & $2.93 \pm 0.06$                      & $3.52 \pm 0.08$                     & $4.16 \pm 0.07$                      & $3.8 \pm 0.1$    \\
&&&&&\\

 \hline
\end{tabular}
\tablefoot{ The fitted parameters were obtained via the simultaneous modeling of \cheops{} and \tess{} light curves with \texttt{juliet} package (Sect. \ref{sec:analysis}). For the estimation of the derived parameters, we used the astronomical constants as defined in the \texttt{python} library \texttt{astropy.constants}. \tablefoottext{a}{The final and adopted values for P and T$_0$, which are based on the timing analysis of the transits are reported in Table\,\ref{tab:timing}.} \tablefoottext{b}{Mean temperature calculated using A$_B$=0.}}
\end{table*}

\subsection{Confirmation of planet $f$}

\begin{figure}
    \centering
    \includegraphics[width=\hsize]{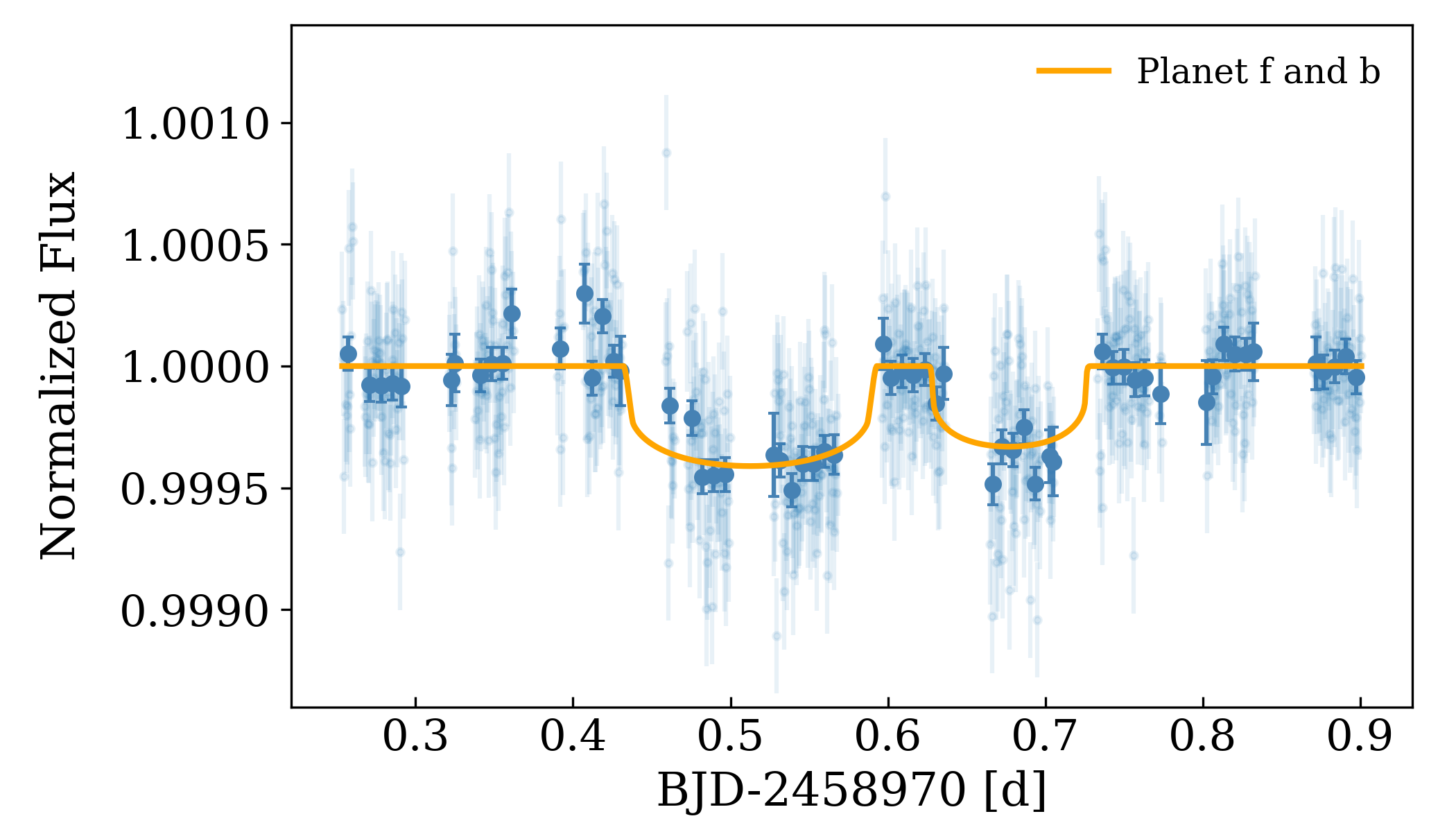}\\
    \includegraphics[width=\hsize]{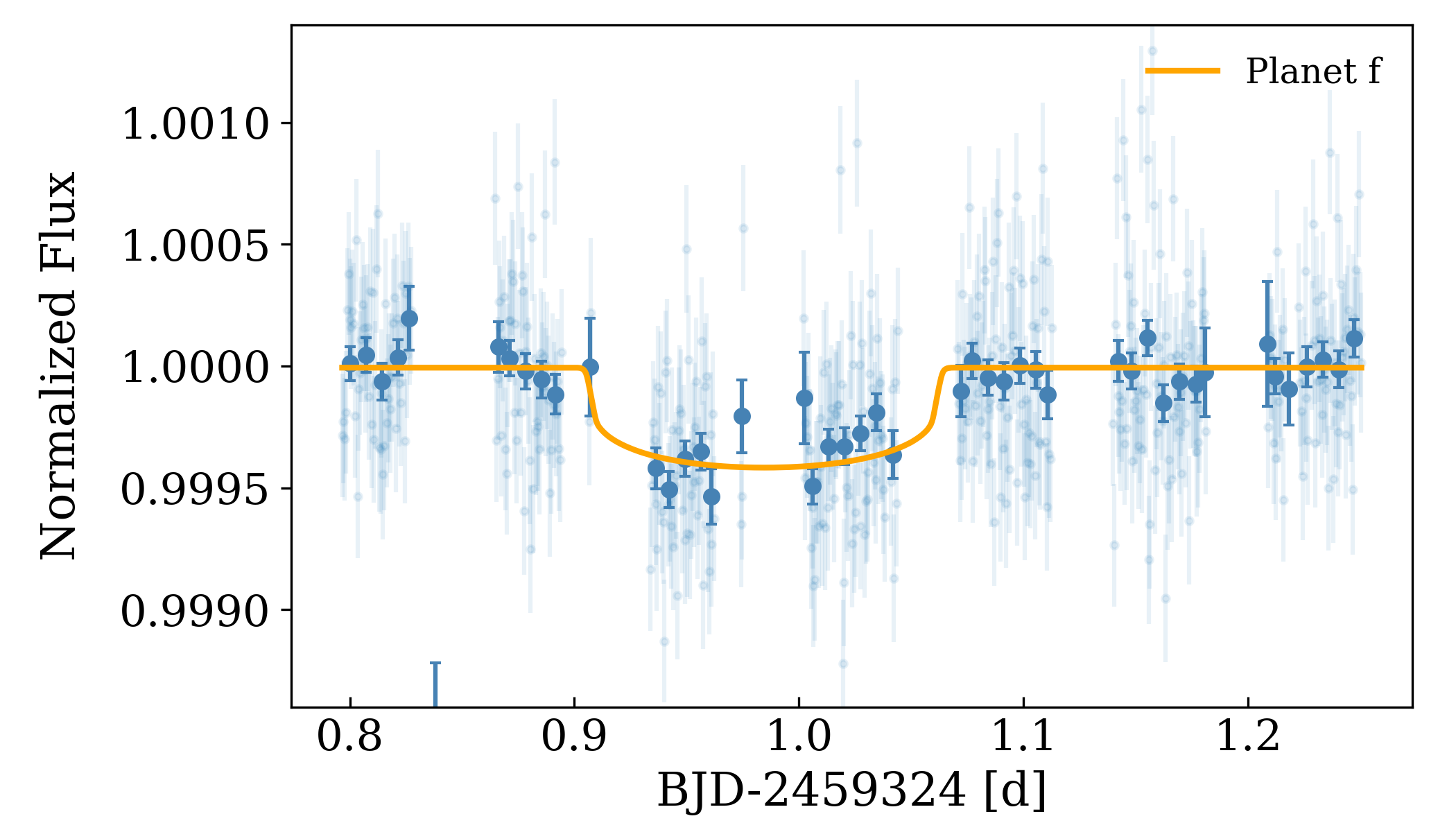}
    \caption{\cheops{} light curve of \starname{} during a transit of planet $f$ ($\sim$2\,458\,970.5\,BJD) and $b$ ($\sim$2\,458\,970.7\,BJD) (top panel), as well as during a dedicated observation of a planet $f$ transit (bottom panel). Detrended photometric points and its 10\,min bins are represented by light and dark blue dots, respectively. The best model for each planet is represented by the yellow curve.}
    \label{fig:planet_f_cheops}
\end{figure}

As described in \cite{Bonfanti2021}, an additional transit-like feature was serendipitously detected in the \cheops{} light curve when monitoring a transit of planet $b$ (see top panel in Fig.~\ref{fig:planet_f_cheops}). They also found signals in Sectors 10 and 11 of the \tess{} data that are consistent with a planetary candidate $f$ with an orbital period of 29.54\,d. We detected transits of planet $f$ in the \tess{} Sectors 10 and 11, as well as in Sector 37, which are consistent with the 29.54\,d period. The fact that two consecutive transits were identified in the first TESS Sector (as reported in Table \ref{tab:transit_times2}) and no other transit consistent with this planet was detected in the TESS light curves in that time excludes the viability of the 29.54\,d period aliases. The \tess{} phased light curve is presented at the bottom-right panel of Fig.~\ref{fig:phased_curves2}. To confirm the presence of this planet, we scheduled a dedicated \cheops{} visit (observation 14 in Table~\ref{tab:obs_log}) to observe a single transit of this planet. We present in Fig.~\ref{fig:planet_f_cheops} the light curve obtained from this observation together with the best-fit model resulting from the analysis of \cheops{} and \tess{} data. In Fig.~\ref{fig:planet_f_cheops}, we also show the light curve reported in \cite{Bonfanti2021} with a transit of both $b$ and $f$ planets. The phased version of the combined two \cheops{} light curves is shown in the bottom-left panel of Fig.~\ref{fig:phased_curves2}.

\subsection{Non-detection of a putative planet $g$}

\begin{figure}
    \centering
    \includegraphics[scale=.5]{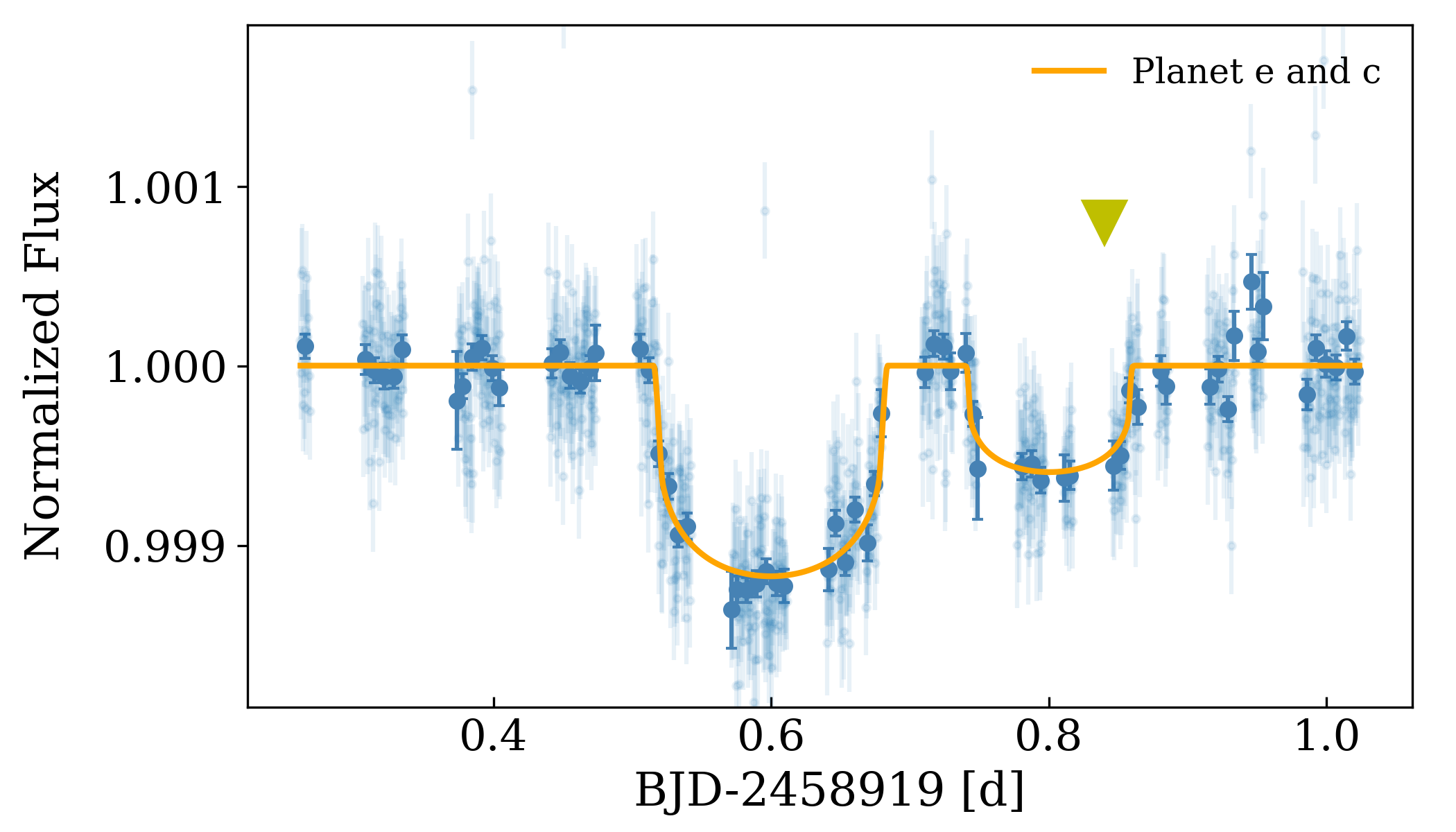}\\
    \includegraphics[scale=.5]{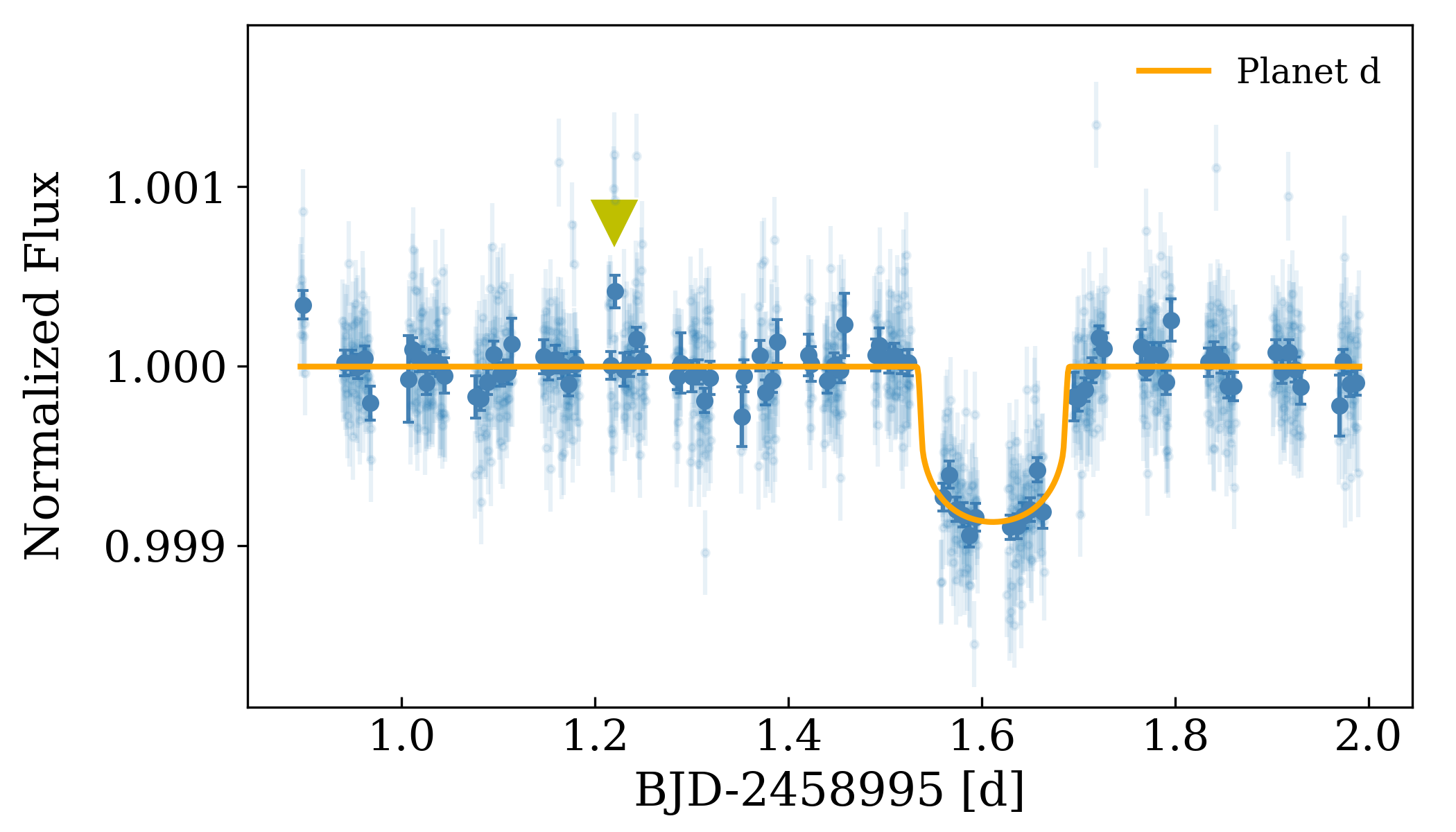}\\
    \includegraphics[scale=.5]{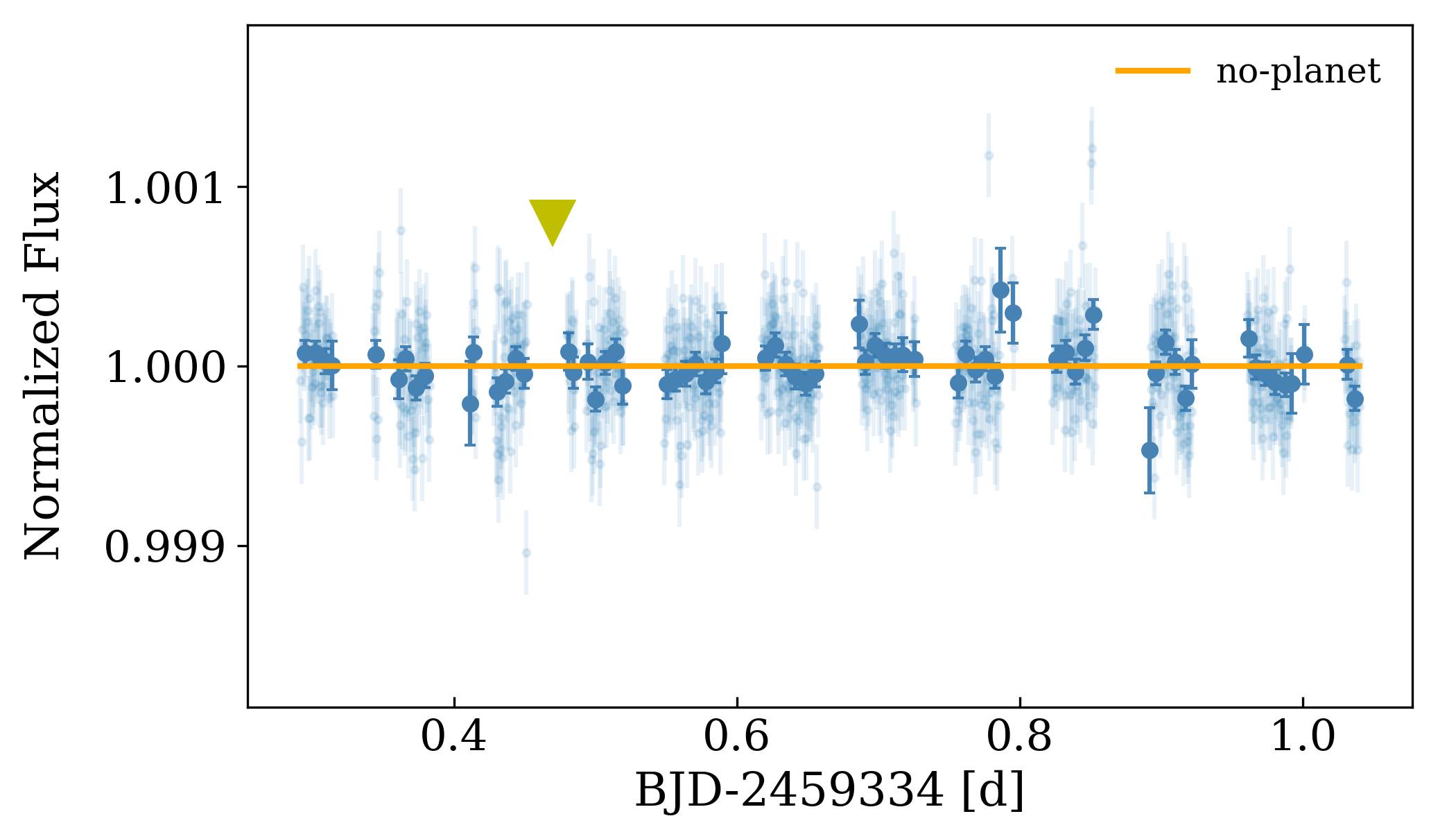}\\
    \caption{\cheops{} light curves of \starname{} during the expected transit time of a potential planet $g$ (marked with the yellow triangle). These light curves correspond to observations 1, 9, and 15 in Table~\ref{tab:obs_log} (top-to-bottom). Light and dark blue dots represent the detrended and 10\,min bins photometric points. No significant statistical evidence of a transit signal with $\sim$10.9\,d period is found in these photometric time series.}
    \label{fig:planet_g_cheops}
\end{figure}

\cite{Daylan2021} reported a hint of a $\sim$10.9\,d periodic transit signal in \tess{} data with amplitude of 230\,ppm (without excluding a possible instrumental origin). Based on the ephemeris reported in that work (T$_0$ = 2\,458\,570.6781\,BJD and P = 10.9113\,d), we can see that this planet should have transited during the \cheops{} observations 1 and 9 (see the light curves in upper and middle panels of Fig.~\ref{fig:planet_g_cheops}, respectively), but no clear evidence of this has been detected. Observation 1 has already been published by \citet{Bonfanti2021}, who noticed that the transit of this putative planet would have occurred at T$_0=2\,458\,919.84$\,BJD, being partially overlapped with the transit of planet $c$. They performed a comparative analysis of a five-planet versus six-planet scenario, finding that there was no strong evidence to reject the five-planet scenario based on the Bayes factors of the two models. According to its ephemeris, the hypothetical planet $g$ would transit at T$_0=2\,458\,996.22$\,BJD during observation 9 (Fig.~\ref{fig:planet_g_cheops}, middle panel). In the time of that \cheops{} visit, we actually noticed a short flux decrease, but it was  shifted 0.13\,d from the linear ephemeris at $\sim$2\,458\,996.35\,BJD.
The amplitude of this flux decrease is $\sim$125\,ppm, which is around half of what was reported by \cite{Daylan2021} in \tess{} data. To further probe this periodic signal, we scheduled a $\sim$18\,h \cheops{} observation, taking special care to avoid any overlap with the transits of the other five planets in the system (observation 15 in Table~\ref{tab:obs_log}). In Fig.~\ref{fig:planet_g_cheops} (bottom panel), we show the detrended light curve of this observation where we found no significant evidence of any transit-like signal. We observed a shallow flux decrease centered at $\sim$2\,459\,334.53\,BJD (linear ephemeris would predict T$_0=2\,459\,334.47$\,BJD), but also consistent with other features present in the light curve, which likely are residuals of the detrending of the well-known systematics of \cheops{} photometry as a function of its rotation angle.
Despite this, we performed additional statistical tests to confirm that the detection of this transit-like feature is not significant (see Appendix~\ref{sec:stats_planet_g} for details).  We conclude that with the data we have on hand, we cannot confirm the presence of this additional planet $g$ in the system.  In fact, a planet with a 10.9\,d period would have a transit duration between 3.7-4\,h (assuming a circular orbit and a zero impact parameter). The rms of the light curve of the \cheops{} observation 15 (bottom panel in Fig.~\ref{fig:planet_g_cheops}) at these timescales corresponds to 17--20 ppm, which translates into an upper limit in the size of an undetected planet of 0.42\,R$_{\oplus}$ (with $e=b=0$) at S/N=1.

Taking into account that a transit of this putative planet was not observed in any of the \cheops{} light curves, we can assess the period range around P=10.9113\,d. Using only \cheops{} data we can rule out a period in this range assuming that the errors in the ephemeris are given by the orbital period only, that is, fixing T$_0$ to the value reported by \cite{Daylan2021}. Thus, we can discard periods in the range P=$10.9113^{+0.052}_{-0.018}$\,d using \cheops{} data only. We note that the accumulation of these uncertainties in the orbital period will translate into offsets of more than 20\,h at \cheops{} epochs. Moreover, no transits were detected at the expected times in the \tess{} Sector 37. Additionally, we used the transit least-squares tool \citep[TLS][]{TLS2019} to search for other transits in this \tess{} light curve. After masking the transits of the known planets, no significant signals were found with periods between 0.52$-$24.5\,d. 

\section{Timing analysis}\label{ssec:timing_analysis}

To perform a detailed timing analysis for each planet, we used the results of the combined analysis in order to determine the transit times for each individual transit event. Therefore, in this case we do not fit for a global P and T$_0$ for each planet but for the  central time, T$_{\text{c}}$, of each transit event. Thus, we used as priors the results of the combined fit for all the parameters except for the single transit midtime.  With these transit times and the ephemeris (described in Sect.~\ref{sec:analysis} and displayed in Table~\ref{tab:results}) from the global fit, we generated the observed-minus-calculated (O$-$C) diagrams of the five planets in the system. A first analysis showed us that the times of the transits derived from overlapping transit events in \tess{} light curves correspond to large outliers in the O$-$C diagrams. Thus, for such light curves, we re-fit the two respective planets simultaneously. In this way, we were able to considerably reduce the temporal offsets and the uncertainties of the central times of these transits. The resulting individual transit times are reported in Tables~\ref{tab:transit_times1} and \ref{tab:transit_times2}. 

\begin{figure*}
    \centering
    \includegraphics[scale=.45]{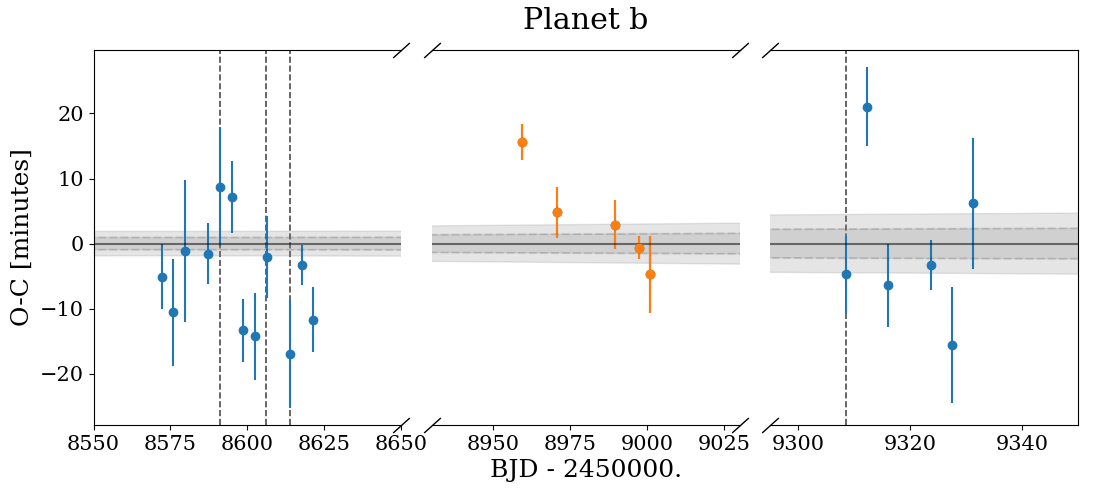} \\
    \includegraphics[scale=.45]{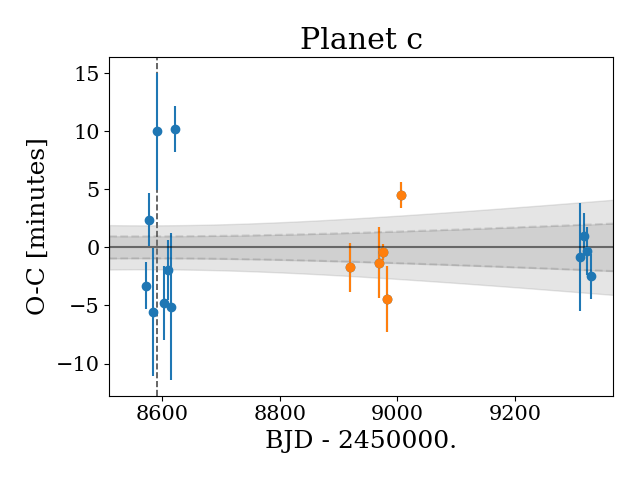}
    \includegraphics[scale=.45]{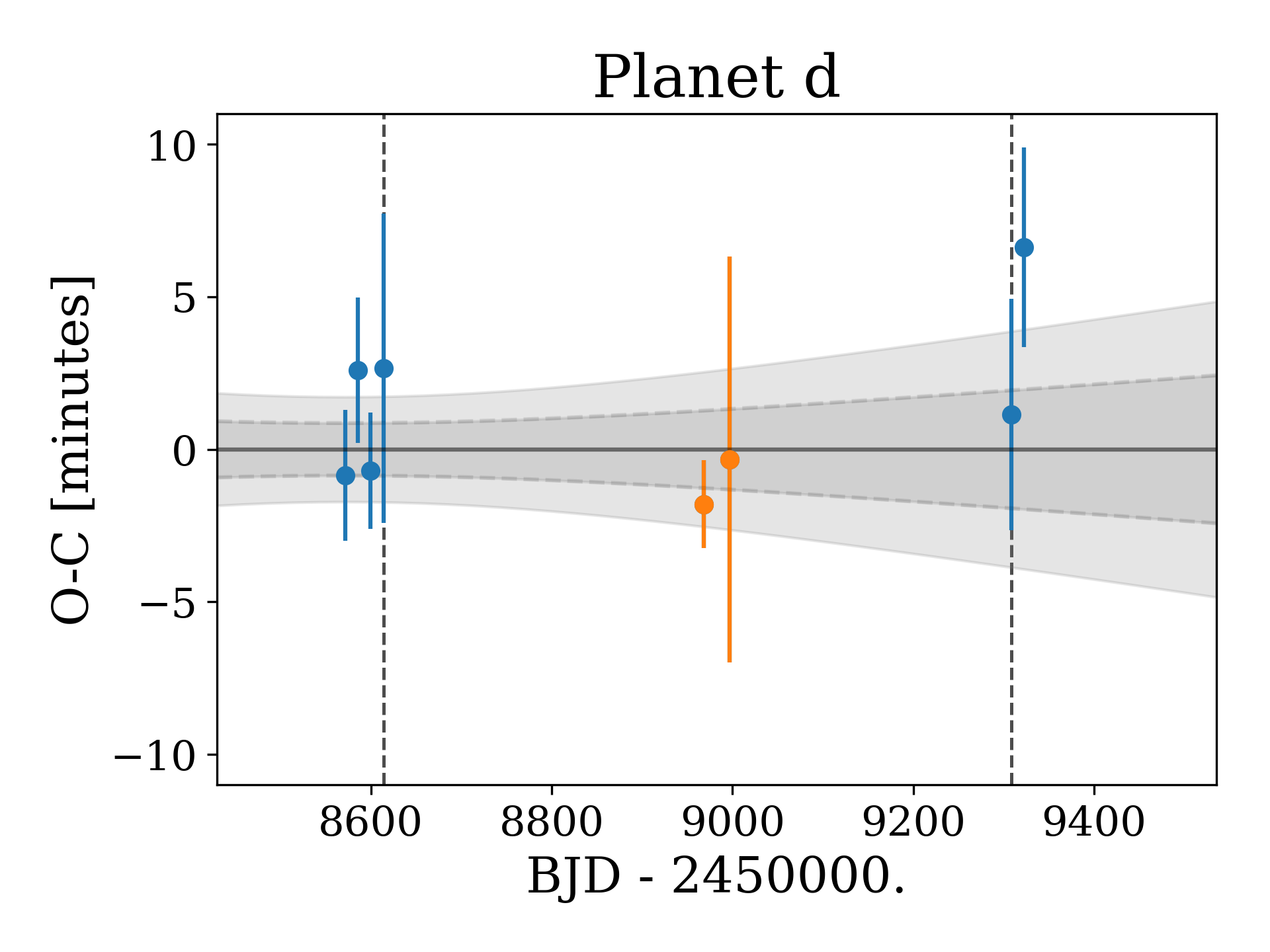} \\

    \includegraphics[scale=.45]{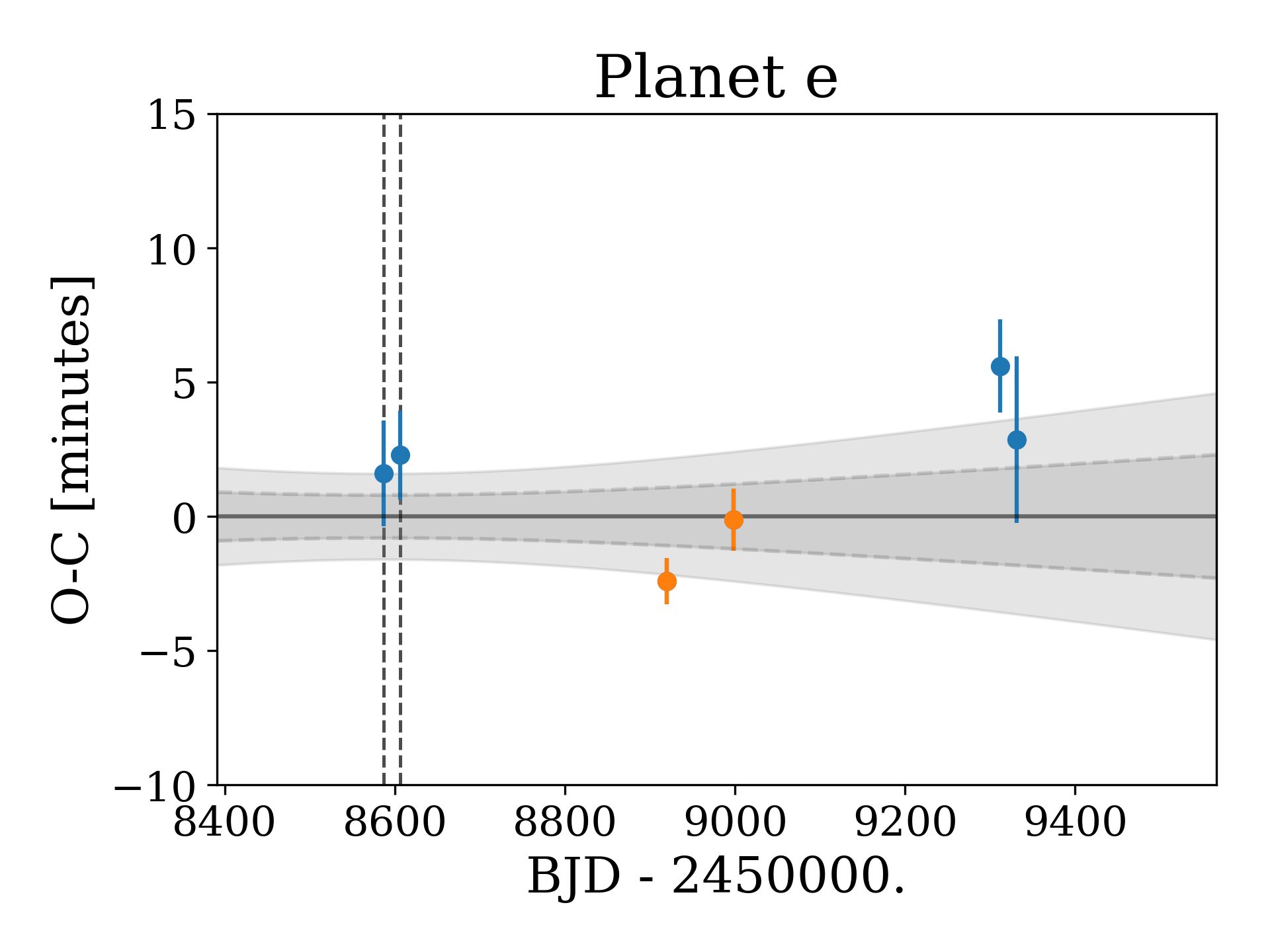}
    \includegraphics[scale=.45]{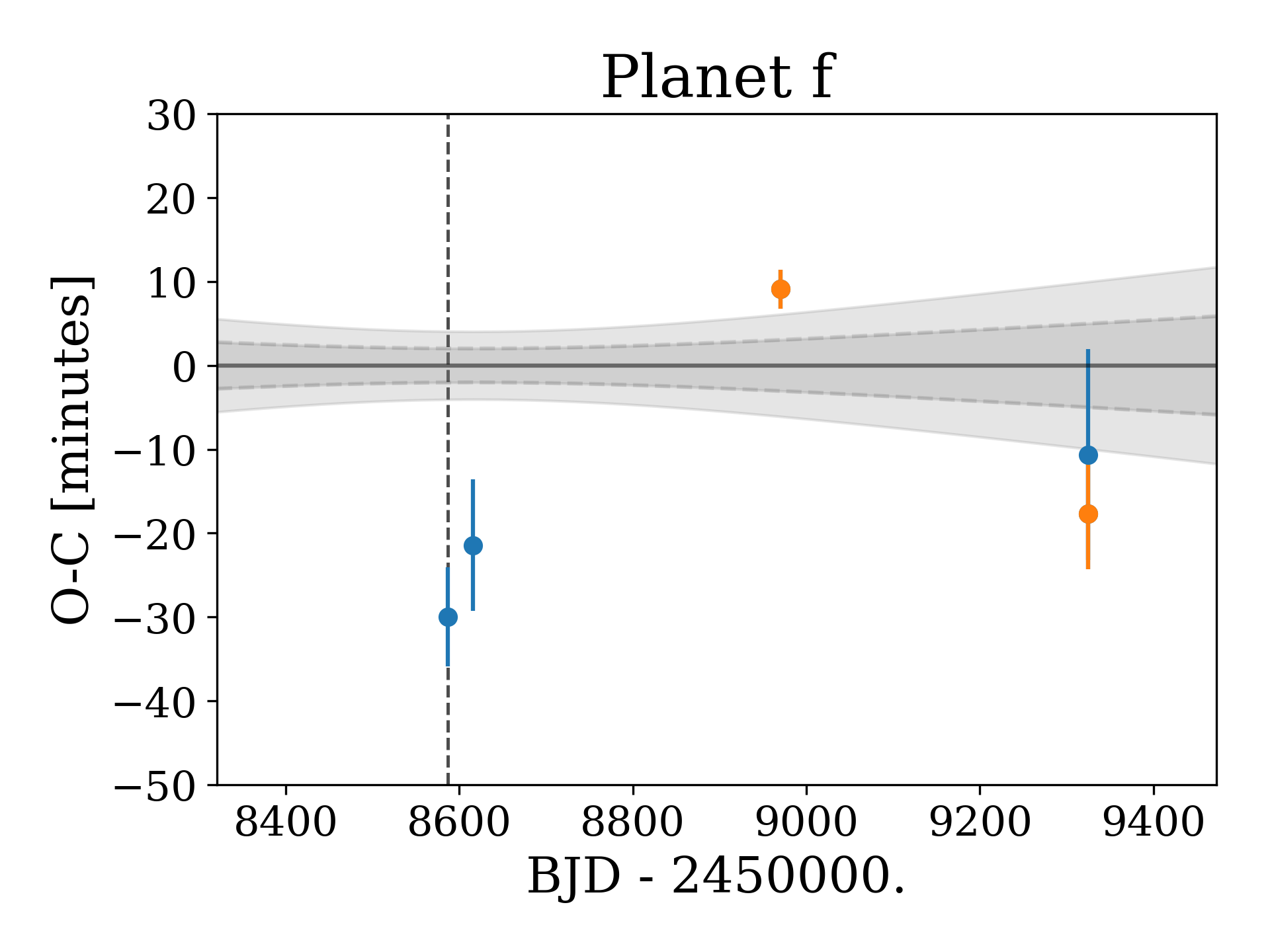}
    \caption{Observed minus calculated diagrams of the transit times of the five planets of the \starname{} system, obtained by using the ephemeris of the global fit or the updated T$_0$ values reported in Table~\ref{tab:timing}. Transit times were obtained from the individual analysis of the transits (see Sect.~\ref{sec:analysis}). Blue and orange symbols correspond, respectively, to \tess{} and \cheops{} observations. The gray regions represent the $\pm$1-2$\sigma$ uncertainties of the ephemeris equations. Vertical dashed lines mark overlapped transits in the light curves.}
    \label{fig:oc_diagrams}
\end{figure*}

We checked if we were able to fit a linear function to the timing residuals of each planet, meaning the ephemeris equation can potentially be updated. We evaluate the significance of the P or T$_0$ offsets by comparing them with their uncertainties from the global fit. No correction was needed for the orbital periods. We only found significant temporal offsets of 114.5\,s (2\,$\sigma$), -58\,s (1.1$\sigma$) and 717\,s (5.9$\sigma$) for the T$_0$ of planets $b$, $d$ and $f$, respectively.  Therefore, we  updated the T$_0$ values of the ephemeris equation for these three planets (Table\,\ref{tab:timing}). 
For comparison, in Table\,\ref{tab:timing} we also show the mean error of the transit times, $\bar{\sigma}_{\texttt{T}_{\texttt{C}}}$, for \cheops{} and \tess{} transits. The improvement on our estimations of the transit times from \cheops{} light curves in comparison to those from \tess{} transits is consistent with the results of previous works \citep[e.g.,][]{Borsato2021, Bonfanti2021}. 
The final O$-$C diagram of each planet is shown in Fig.\,\ref{fig:oc_diagrams}. There, the overlapped transits are marked with the vertical dashed lines, and the $\pm$1-2$\sigma$ uncertainties of the final ephemeris are represented by the gray regions in the diagrams.  We noticed non-negligible variations on the transit times of planets $b$ and $f$, reflected for example on the rms and mean amplitude values of their O$-$C residuals. We also noticed hints of an anticorrelation in the transit times of planets $e$ and $f$. Thus, in order to investigate further on the results in the transit times, we performed a detailed transit time variations (TTVs) analysis of the system, described in Sect.\,\ref{ssec:TTV_masses}. 

\begin{table*}
\caption{Results of the timing analysis of the transits.} 
\label{tab:timing}
\small
\centering
\begin{tabular}{c c c c c c}
\hline\hline             
 Planet & O$-$C rms & O$-$C mean & $\bar{\sigma}_{\mathrm{T}_{\mathrm{C}}}^{\mathrm{CHEOPS}}$ & $\bar{\sigma}_{\mathrm{T}_{\mathrm{C}}}^{\mathrm{TESS}}$ & Updated T$_0$\\
        & [min]  & amplitude [min] & [min] & [min] & [BJD$_{\mathrm{TDB}}$ - 2\,450\,000]\\
 \hline
 $b$ & 9.6 & 8.0 & 3.6 & 6.6 & 8572.11149 $\pm$ 0.00065 \\
 $c$ & 4.6 & 3.4 & 2.0 & 3.3 &  --\\
 $d$ & 2.4 & 2.0 & 4.0 & 3.1 & 8571.33610 $\pm$ 0.00060 \\
 $e$ & 2.4 & 2.4 & 1.0 & 2.1 &  --\\
 $f$ & 12.6 & 17.8 & 4.5 & 8.8 & 8616.0486 $\pm$ 0.0014 \\
\hline

\end{tabular}

\tablefoot{Results of the analysis described in Sect.\,\ref{ssec:timing_analysis}. The rms and the mean value of the time residuals of each planet's transit are shown (2nd and 3rd columns). For comparison, we show the mean value of the error bars for the transit times of \cheops{} and \tess{} transits (4th and 5th columns). We also report the updated values of T$_0$ for planet $b$, $d$ and $f$ based on the weighted fit of the time residuals (6th column). } 

\end{table*}

\section{Planetary masses}
\label{sec:mass_estimations}

In the absence of radial velocity (RV) measurements, model-dependent estimates of the planetary masses have been previously  delivered by \cite{Daylan2021} and \cite{Bonfanti2021}.  Later, \cite{Teske2021} presented 33 RVs measurements obtained with the Planet Finder Spectrograph (PFS) on the Magellan II telescope at Las Campanas Observatory (Chile). They performed the first Keplerian mass estimation of the \starname{} system. Among their results for this particular system, it is worth mentioning that they were only able to set an upper limit for the mass of planet $d$ and they predicted an eccentricity of $\sim$0.39 for planet $e$. 
Here, we take advantage of the refined parameters of the system to estimate the masses of the five planets by, first, performing a new fit of the RV data and; in addition, by using the M-R relations of \cite{ChenKipping2017} (CK17 hereafter). Finally, we present our  TTV analysis based on the transit timing information of  Sect.~\ref{ssec:timing_analysis}, aimed at exploring whether any constraints can be placed on the planetary masses.

\subsection{Radial velocity estimations}
\label{ssec:rv_mass}

For the fitting of the RV observations, we also used \texttt{juliet}, which implements the \texttt{RADVEL} code \citep{radvel} linked to the \texttt{dynesty} sampler.  We set the priors of all the orbital parameters (P$_{i}$, T$_{0,i}$, $b_i$, $e_i$ and $\omega_i$, $i$:1..5) as normal distributions defined by the results of the photometric fit. For the five RV semi-amplitudes, $K_i$, we set a wide uniform prior between 0-20 $m$\,$s^{-1}$. We also fit for a jitter and a systemic RV term ($\sigma\_$PFS2 and $\mu\_$PFS2, respectively). The fitted $K$'s and the derived masses for each planet are presented in Table~\ref{tab:mass_table}, where we also show the masses estimations from the literature. The RV data and the best fitted models are presented in Fig.~\ref{fig:rvs_figures1} and \ref{fig:rvs_figures2}.  Due to the paucity and poor sampling of the observations, the RV masses should be taken with caution, and therefore we treated them as only indicative values. 
It is known that the accurate measurement of masses of small planets in multiple systems requires considerable large amount of data points \citep[e.g.,][]{He2021}, which is not the case of the \starname{} system to date. In fact, the posterior distribution of the RV semi-amplitude for planet $d$, $K_{d}$, peaks around $K_{d}\sim0.41$\,$m$\,$s^{-1}$ (see Fig.~\ref{fig:corner_plot_rv}).  As noted in \cite{Teske2021}, if we allow a fitting for $K<0$ values, we then would obtain a negative-$K$ ($K_{d}\sim-0.8$\,$m$\,$s^{-1}$).  This was the reason why  \cite{Teske2021}  adopted only an upper limit for $K_{d}$ defined by the  root-mean-square (rms) of the RV residuals after removing the best Keplerian model of the other four planets.  We confirmed that the results of the other planets were almost unaffected by removing planet $d$ from the fit. Therefore, we emphasize that our reported RV result for planet $d$ is not reliable.

\subsection{Statistical estimations}
\label{ssec:ck17_mass}

To compare the RV masses with the values derived with a statistical approach, we use the CK17 M-R relations, implemented within the \texttt{Python} probabilistic tool \texttt{forecaster}\footnote{https://github.com/chenjj2/forecaster}. We used as input the derived planetary radius (and its uncertainties) of the five planets obtained from our photometric fit. The masses predicted by CK17 and their uncertainties are presented in Table~\ref{tab:mass_table}. We also checked that these values are consistent with the predictions from more recent M-R relations from \cite{Otegi2020}, although the uncertainties of the later are 2.3-2.7 times smaller.
As these M-R relations are intended to provide a broad estimation of the planetary masses, the values we derived with them should be taken with caution. 

We note that for the outer two planets, the RV derived masses are unrealistically high when compared to the models outcomes, particularly for planet $f$. This can likely be explained by the fact that the phase of planet $f$ is not well sampled by the RV measurements, and that the fit of planet $e$ seems to show a phase offset, likely resulting in the poor mass determination.
Since these two planets are located close to a mutual 3:2  mean motion resonance (MMR), it is possible to explore whether the dynamical analysis can provide additional constraints on the mass estimates, as described in Sect.~\ref{ssec:TTV_masses}.


\begin{table*}
    \caption{Mass estimations of the planets of \starname. }
    \centering
    \small
    \begin{tabular}{ccccc|cccc}
    \hline
    \hline
          & & \multicolumn{3}{c}{This work} \vline &\multicolumn{3}{c}{Literature }  \\
         \cline{2-8}
          \multirow{3}{*}{Planet} & CK17 & \multicolumn{3}{c}{RVs} \vline &T21\tablefootmark{a} (RVs) & B21\tablefootmark{b} & D21\tablefootmark{c} (CK17) \\
         \cline{3-8}
            & Mass & $K$ & Mass & Density &   \multicolumn{3}{c}{Mass}\\
            & [M$_\oplus$] & [m\,s$^{-1}$]  & [M$_\oplus$] & [gr\,cm$^{-3}$] &  \multicolumn{3}{c}{[M$_\oplus$] }\\
        \hline
        $b$ &$3.8^{+2.6}_{-1.3}$  &  $1.63^{+0.74}_{-0.72}$   & $3.6^{+1.7}_{-1.6}$   & 5.0$\pm$ 2.0& 4.33 $\pm$ 1.54 & 4.23 $\pm$ 0.40 & 5$\pm$2\\
        $c$ &$5.5^{+4.0}_{-2.2}$  & $1.94^{+0.79}_{-0.70}$    & $5.1^{+2.1}_{-1.8}$   &3.0$\pm$1.0&  3.97 $\pm$ 1.77  & 8.90 $\pm$ 0.66 & 7$\pm$2\\
        $d$ &  $7.3^{+5.6}_{-3.0}$  & $0.46^{+0.54}_{-0.31}$   & $1.6^{+1.8}_{-1.1}$   &0.5$\pm$0.5& <7.75           & 7.75 $\pm$ 0.80 & 10$\pm$2\\
        $e$ & $9.9^{+7.4}_{-4.3}$  & $3.55^{+0.78}_{-0.76}$ & $13.6^{+3.0}_{-2.9}$  & 3.0$\pm$1.0& 19.10 $\pm$ 3.91 & $8.2^{+3.8}_{-1.2}$ & 13$\pm$2 \\
        $f$ & $5.1^{+3.8}_{-2.0}$  & $4.2^{+1.8}_{-1.6}$ & $18.4^{+7.9}_{-7.1}$  &15.0$\pm$6.0& 10.85 $\pm$ 5.55 & 3.95 $\pm$ 0.4 & --\\
        \hline
    \end{tabular}
    \tablefoot{The masses from the radial velocities (RVs) and from the probabilistic method based on the M-R relations of \cite{ChenKipping2017} (CK17) (Sect.~\ref{ssec:planets_pars}) are shown. The fitted RV semi-amplitudes, $K$, and the derived planet densities are also shown. We recall in the last 3 columns the values from the literature: \tablefoottext{a}{\cite{Teske2021}}, \tablefoottext{b}{\cite{Bonfanti2021},} and    \tablefoottext{c}{\cite{Daylan2021}.} 

}

    \label{tab:mass_table}
\end{table*}

\subsection{TTV Analysis}

\label{ssec:TTV_masses}

\begin{figure}
    \centering
    \includegraphics[width=0.35\textwidth]{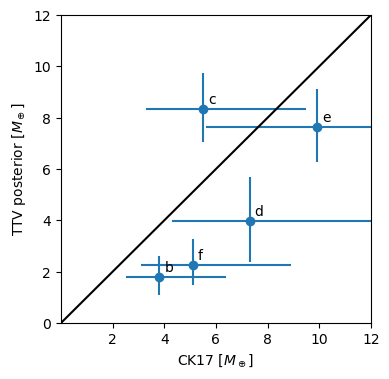}
    \caption{Mass posterior vs. mass prior of the TTV analysis. The error bars show the 1-$\sigma$ interval.}
    \label{fig:TTVs_masses}
\end{figure}

\begin{figure}
    \centering
    \includegraphics[width=0.42\textwidth]{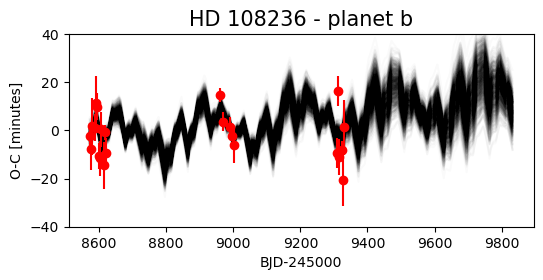}
\includegraphics[width=0.42\textwidth]{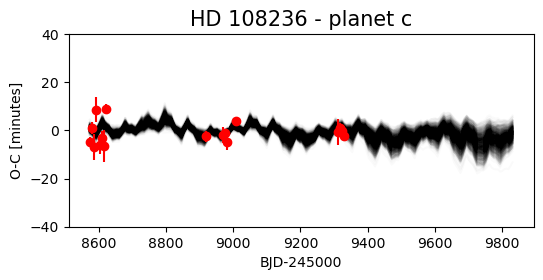}
\includegraphics[width=0.42\textwidth]{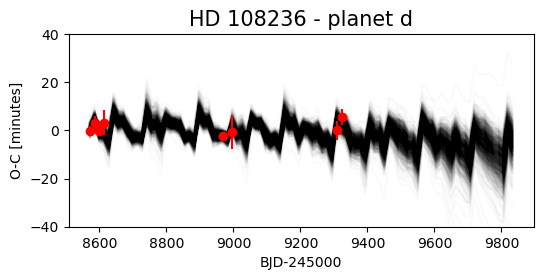}
\includegraphics[width=0.42\textwidth]{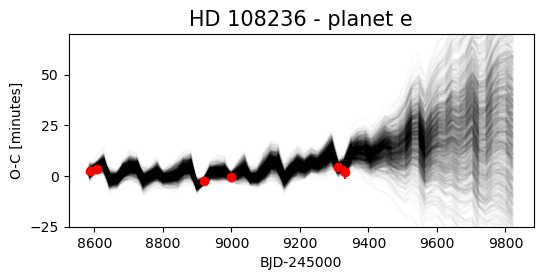}
\includegraphics[width=0.42\textwidth]{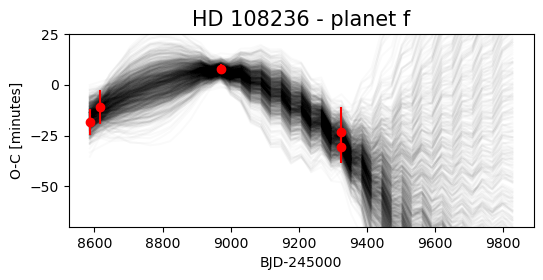}
    \caption{Posterior of the transit timings fit presented in Tables \ref{tab:transit_times1} and \ref{tab:transit_times2}.}
    \label{fig:TTVs}
\end{figure}

The period ratio between consecutive pairs of planets (P$_{i+1}$/P$_i$) in the \starname{} system is (P$_c/$P$_b$, P$_d/$P$_c$, P$_e/$P$_d$, P$_f/$P$_e$) = (1.634, 2.285, 1.382, 1.508). Thus, the outer planets, $f$ and $e$, are close to the exact 3/2 commensurability (P$_f/$P$_e \approx 1.508$). The proximity to a two-body mean motion resonances (MMRs), $P_{out}/P_{in}\sim (p+q)/p$, with p and q integers, can generate TTVs that can be observed and modeled to constrain the orbit and masses of the involved planets \citep[e.g.,][]{Agol2005,Lithwick2012,NeVo2016}. 
We do not currently have enough TTVs measurement to estimate masses. Here, we report the mass posterior of this TTV analysis with the sole purpose to compare them with the CK17 masses from Table \ref{tab:mass_table} (that we use as prior) in order to check whether these masses are consistent with the observed TTVs.

We fit the timings reported in Tables \ref{tab:transit_times1} and \ref{tab:transit_times2} using the code presented in \cite{Leleu2021b}: the transit timings are estimated using the {\ttfamily TTVfast} algorithm \citep{ttvfast2014}, and the \texttt{samsam}\footnote{\url{https://gitlab.unige.ch/Jean-Baptiste.Delisle/samsam}} MCMC algorithm \citep[see][]{Delisle2018} is used to sample the posterior. 
The mean longitudes, periods, arguments of periastron and eccentricities of the planets have flat priors. The masses and eccentricities posteriors of the TTV analysis are shown in Table \ref{tab:mass_table},  and 1000 randomly chosen samples of the fitted model are shown in Fig.~\ref{fig:TTVs}.  

The prior-posterior comparison is made in Fig. \ref{fig:TTVs_masses}. For planets $b$, $d$, $e,$ and $f$, we see that the mass posterior points toward a lower value than the prior, although the difference lies close to (or within) the 1-$\sigma$ interval. Planet $c$ is the only one for which TTVs mass is larger than the prior, which appears to be due to the tentative chopping in the TTV signal of planet $b$ (see Fig. \ref{fig:TTVs}). 

For the outer two, near-resonant planets $e$ and $f$, there are hints of anti-correlated TTVs, which are also apparently not fully consistent with the mass priors.
For the derived set of masses, significant TTVs are nonetheless expected to come in the next years for these two planets, as can be seen on the TTV projections in Fig.~\ref{fig:TTVs} at dates 9600 [BJD-2\,450\,000] onward. 
Overall, the TTVs observed for \starname{} seem to be somewhat at odds with the expected masses and eccentricities, although the current phase coverage is not sufficient to draw strong conclusions and warrant additional observations.

In addition, in order to better constrain the architecture of the system, we used the TTVs posterior to check if the planets $e$ and $f$ are inside the 3:2 MMR using the Hamiltonian formulation of the second fundamental model for Resonance of \cite{HenLe1983}. We applied the "one degree of freedom" model of first-order resonances presented in \cite{DePaHo2013}, and computed the value of the Hamiltonian parameter $\Gamma'$ over the posterior of the TTV analysis (Eq. 36 in the aforementioned paper). The resonant state appears for $\Gamma' \geq 1.5$ \citep{DePaHo2013}, and for this pair of planets ($e$ and $f$) we calculated $\Gamma'=-2.76 \pm 0.38$, implying that the planets are outside the 3:2 MMR.

\section{Discussion}\label{sec:discussion}

\subsection{Planetary parameters}
\label{ssec:planets_pars}

\begin{figure}
    \centering
    \includegraphics[scale=0.4]{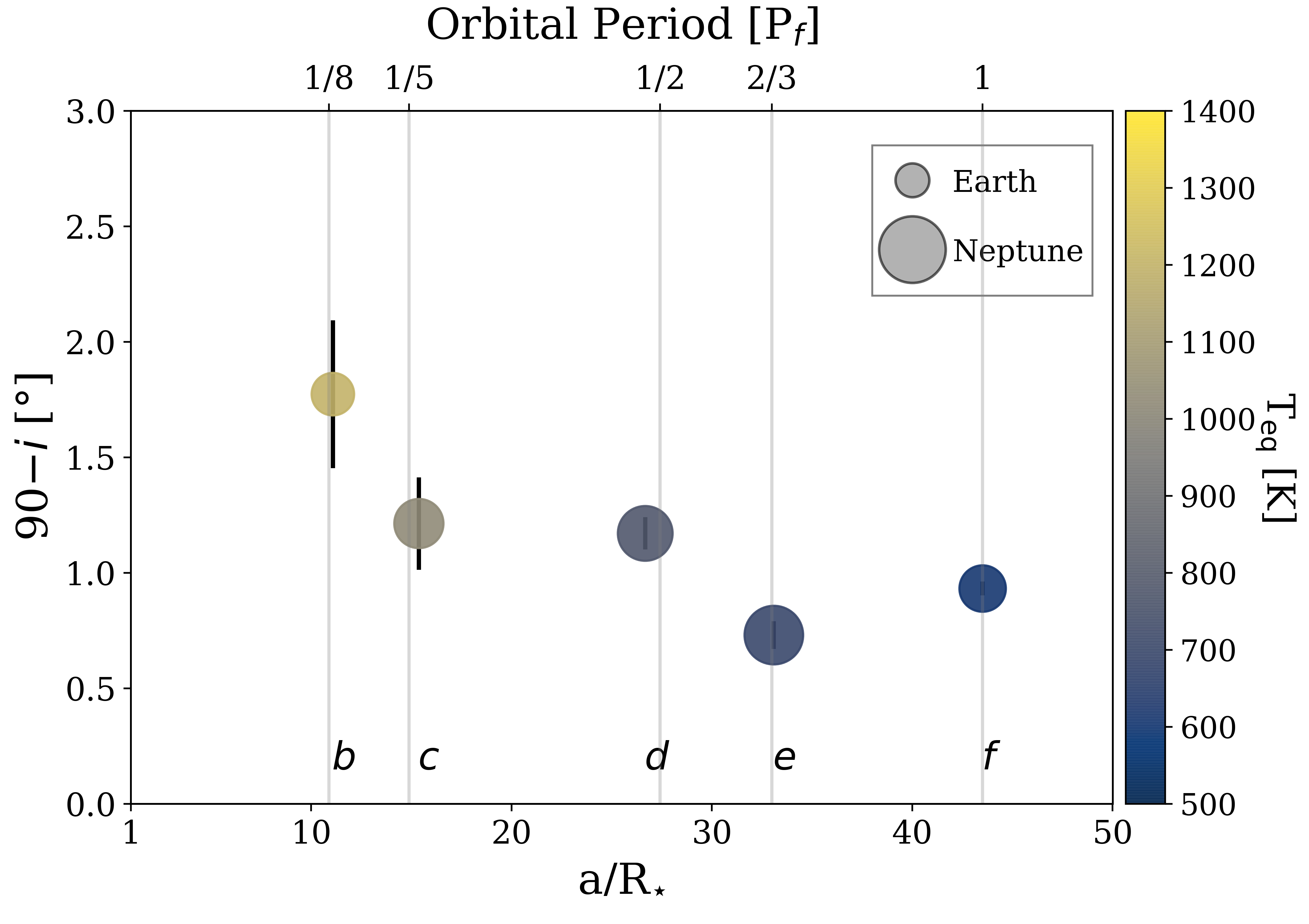}

    \caption{Diagram of the \starname{} system.  We show the scaled semi-major axis, a/$R_{\star}$ (bottom axis) and the respective orbital period of the planets (in units of planet $f$'s period, top axis). We mark the relevant main mean motion resonances with vertical lines. The relative orbital inclination (90$^{\circ}-i$) of each planet is shown in the left axis. The symbol size scales with the planetary radius in units of Earth radius; and the color with the equilibrium temperature of the planet (with $A_{B}=0$).  }
    \label{fig:system_diagram}
\end{figure}

As our final planetary parameters, we adopted those resulting from the combined analysis of \cheops{} and \tess{} data described in Sect.~\ref{sec:analysis} and reported in Table~\ref{tab:results}.  All the parameters are consistent with those presented in \cite{Bonfanti2021} within 1-2$\sigma$.  Notably, the uncertainties we report, which represent the 68$\%$ intervals of the resulting posterior distributions of each parameter, are reduced between 30-80\,\% with respect to previously published uncertainties. A diagram of the orbital configuration, together with a comparative of the sizes of the planets, is presented in Fig.\,\ref{fig:system_diagram}. 

The uncertainties on the transit ephemeris also improved when compared to \cite{Bonfanti2021} thanks to the extended baseline of the monitored transits. Moreover, we also performed a transit timing analysis (see Sect.~\ref{ssec:timing_analysis}), which provided us with the updated final T$_0$ reported in Table\,\ref{tab:results} for planets $b$, $d,$ and $f$. For the other planets, we still referred to the values in Table \ref{tab:results}. 

Despite the fact that none of the three approaches we used to estimate the planetary masses are conclusive, we noticed that (as shown in Table~\ref{tab:mass_table}) the masses of the three most internal planets are consistent among the different estimations (except for RV mass of planet $d$ as discussed in Sect.~\ref{ssec:rv_mass}). For the two outer planets, the RV masses seem unrealistically high when compared to the other methods, although they are poorly constrained by the data.
Below, we use the RV masses to probe the planetary internal structure.

\subsection{Rocky or gaseous planets}
\label{ssec:interior_analysis}

\begin{figure*}
\centering
\includegraphics[width=0.6\textwidth]{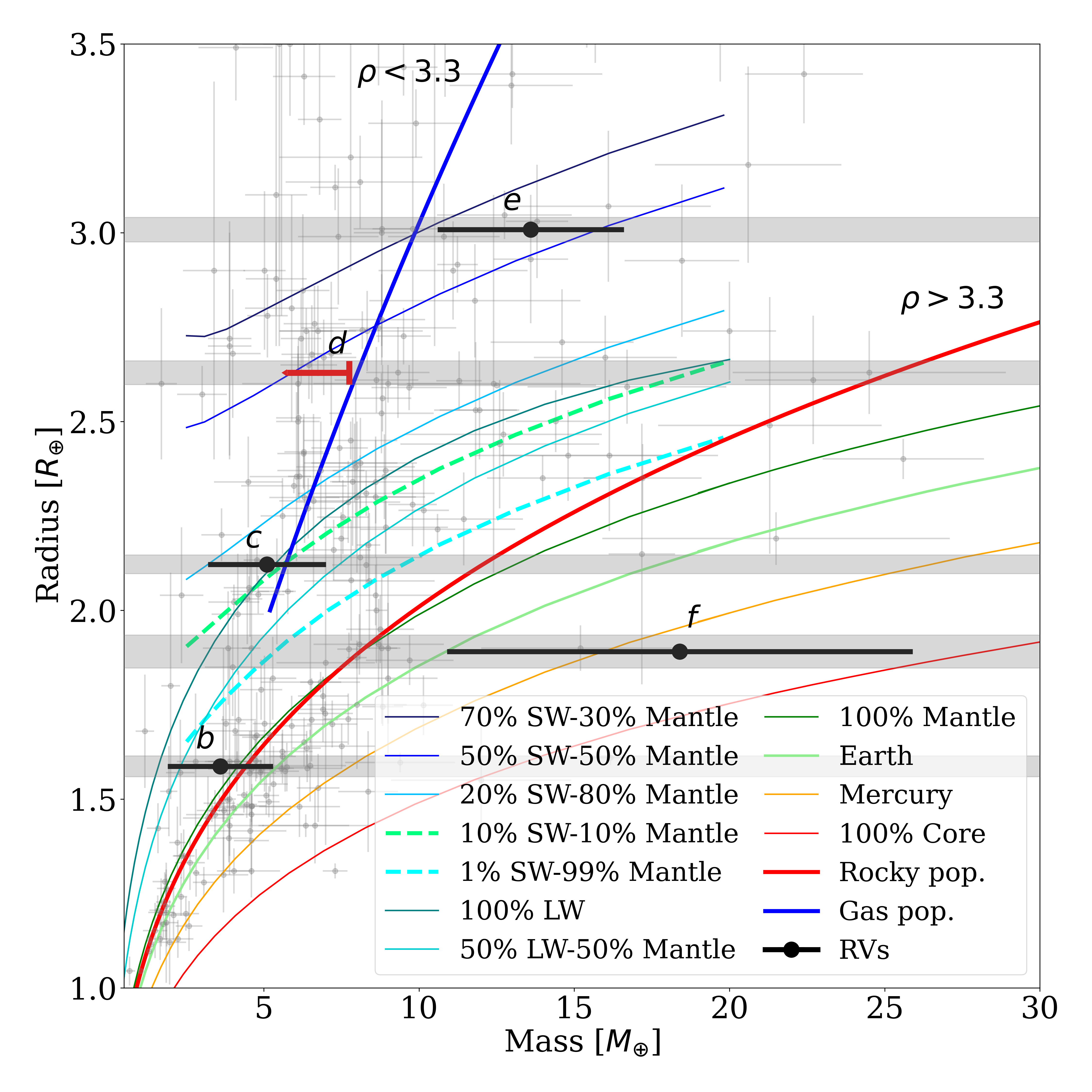}
\caption{Mass-radius diagram for planets with radius and masses below 3.5\,R$_{\oplus}$ and 30\,M$_{\oplus}$, respectively. The radius of the \starname's planets obtained from the transits modeling are represented by the gray horizontal regions, while the mass values suggested by the RV fit are represented by the black symbols. The upper limit in the mass of planet $d$ (based on estimations from \cite{Teske2021} and \cite{Bonfanti2021}) is marked with the red arrow. We compared the \starname{} planets with the current planet population (with estimates of mass and radius better than 30\%) represented by the gray symbols and the MR relations from \cite{Otegi2020} (blue and red thick lines). The planet interiors models from \cite{brugger17} and \cite{acuna2021} are represented by the thin curves. }
\label{fig:mr_diagram}
\end{figure*}

\begin{figure}
\centering
\includegraphics[width=0.5\textwidth]{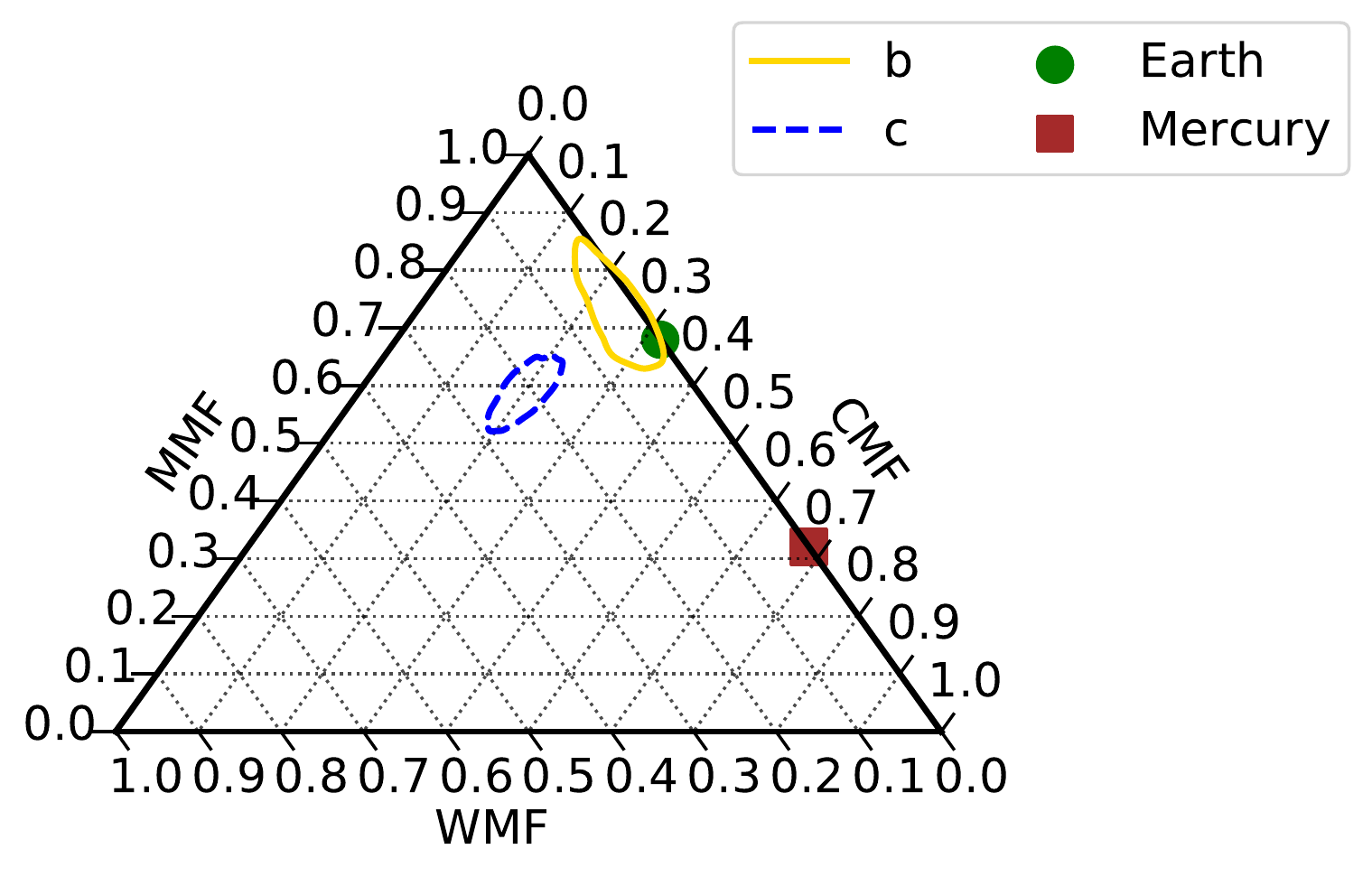}
\caption{Ternary diagram representing the internal structure of \starname{} $b$ and $c$, where we show the estimated core mass fraction (CMF), water mass fraction (WMF) and mantle mass fraction (MMF) defined as MMF = 1 $-$ CMF $-$ WMF.  The colored contours represent the 1$\sigma$ uncertainties derived from the marginal posterior distribution function for each planet. For comparison, the green dot and red square indicate the position of Earth and Mercury in the ternary diagram, respectively.}
\label{fig:ternary_comp}
\end{figure}

Taking advantage of the precision obtained in the planets' sizes and our RVs estimations of their masses, we placed the five planets of the system in an M-R diagram (Fig.\,\ref{fig:mr_diagram}). For planet $d,$ we used 7.75\,M$_{\oplus}$ value as an upper limit based on the estimations from \cite{Teske2021} and \cite{Bonfanti2021}. We also plotted the M-R relations from \cite{Otegi2020}, red and blue thick curves, representing the "rocky"\ and "gaseous"\ planet populations, respectively.  
Despite the results of our RV fit are not conclusive, and therefore, the calculated masses and their uncertainties should be treated as indicative values only, we are still able to shed light on the nature of the \starname{} planets.  
For this purpose, we overplotted in Fig.\,\ref{fig:mr_diagram} the planetary internal models from \cite{brugger17} and \cite{acuna2021}. These models consider a planetary interior structure divided in three layers: a Fe-rich core, a silicate mantle and a volatile layer \citep{brugger17}. The volatile layer is composed of water, which, depending on the surface pressure and temperature, presents different phases, including liquid, ice, steam, and supercritical. 
Thus, the models in Fig.\,\ref{fig:mr_diagram} represent different ratios of core, mantle and volatile rich envelope with water either in the supercritical (SW) or liquid (LW) form. In the case of the SW mass-radius relationships, we have considered an equilibrium temperature of 1200\,K, which is approximately the equilibrium temperature of planets $b$ and $c$.

Thus, based on its location on the M-R diagram, planets $d$ and $e$ likely have a significant volatile envelope surrounding a large mantle.  Unfortunately, with the data in hand, for these planets, it is not possible to say anything further about the relative fraction of these two layers. For planet $f$, for which the RVs suggest a rocky composition although the proportion between mantle and core components is not possible to constrain due to the large uncertainties we still have on its mass.

On the contrary, despite their mass uncertainties, it is possible to discriminate between a rocky planet with no volatiles and a volatile-rich planet for planets $b$ and $c$. For these planets, we performed a more detailed interior structure analysis within a MCMC Bayesian framework, as described in \cite{acuna2021}. For this analysis, we considered the masses estimated from the RVs (Table~\ref{tab:mass_table}) and the radii presented in Table~\ref{tab:results}. With an equilibrium temperature of about 1200\,K, these planets are highly irradiated; therefore, in our interior structure model, we assumed that the volatile layer is not water in a condensed form, but in supercritical and steam phases \citep{Mousis2020,acuna2021}. The upper region of the volatile layer consists of an atmosphere in radiative-convective equilibrium that is coupled with the interior at 300 bar. The atmospheric temperature at the bottom of the volatile layer, $T_{300}$, the Bond albedo, and the atmospheric thickness were calculated with a 1D k-correlated model \citep{Pluriel19,Marcq17}. In addition, we also considered the Fe/Si mole ratio as an observable in our MCMC analysis, which is calculated as described in \cite{sotin07} and \cite{brugger17}, using our \starname's abundance estimations of Fe, Si, and Mg (Table~\ref{tab:stellarParam}). We obtain a Fe/Si=0.79$\pm$0.08, which is below the solar value Fe/Si$_{\odot}$=0.96 \citep{sotin07}.  The core mass fraction (CMF), the water mass fraction (WMF), and the atmospheric parameters, resulting from our interior structure analysis, are shown in Table \ref{tab:interior_res}. 

We found that the densities of all planets in the system can be accounted by a core and a mantle with a Fe/Si mole ratio equal to that of the stellar host, together with a volatile layer on top. Under this assumption, in the Fe/Si mole ratio, planet $b$ could have a thin atmosphere (< 300 bar) or an envelope that could constitute up to 10\% of its mass. To explore the possibility of planet $b$ being a "dry"\ super-Earth, that is, composed of a core and a mantle only, we removed the Fe/Si mole ratio as a constraint in our MCMC Bayesian analysis, leaving only the observable constraints imposed by mass and radius. In addition, we fixed the WMF to zero. In this case, planet $b$ is compatible with a dry planet with CMF=0.04$^{+0.22}_{-0.04}$ and a Fe/Si = 0.67$\pm$0.54, having a low Fe content compared to the Earth value (CMF$_{\oplus}$ = 0.32). We observe that planet $c$ is more volatile-rich than planet $b$ (see Table~\ref{tab:interior_res} and Fig.~\ref{fig:ternary_comp}). On the other hand, it would be necessary to better constrain the masses of planets $d$ and $e$ to narrow down the mean value and uncertainties of their volatile mass fractions. If their volatile contents are confirmed to be higher than that of planet $c$, the \starname{} system could present an increasing volatile mass fraction trend with increasing semi-major axis, which has been noted in other multi-planetary systems with low-mass planets \citep[e.g.,][]{acuna2022}. In addition, if the rocky nature of planet $f$ is confirmed, for example via a more complete RV follow-up, it would be very interesting to explore the processes in the formation and/or evolution of the system, such a Jeans atmospheric escape, required to explain why the more massive and less irradiated planet does not present a gaseous envelope unlike its two precedent planets.

\begin{table*}
\caption{Compositional and atmospheric parameters for \starname{} planets $b$ and $c$.}
\label{tab:interior_res}
\centering
\begin{tabular}{cccccc}

\hline \hline
Planet & CMF\tablefootmark{a} & WMF\tablefootmark{b} & $T_{300}$\tablefootmark{c} [K] & $z_{atm}$\tablefootmark{d} [km] & $A_{B}$\tablefootmark{e} \\ \hline
$b$ & 0.24$\pm$0.06 & 0.05 $\pm$ 0.05 & 4071$\pm$135 & 1035$\pm$395 & 0.22$\pm$0.01\\
$c$ & 0.20$\pm$0.02 & 0.21$\pm$0.05 & 3804$\pm$123 & 733$\pm$179 & 0.22$\pm$0.01\\ \hline
\end{tabular}

\tablefoot{We report here the parameters retrieved by our interior-atmospheric MCMC Bayesian analysis: \tablefoottext{a}{core mass fraction}; \tablefoottext{b}{water mass fraction}; \tablefoottext{c}{atmospheric temperature at 300 bar}; \tablefoottext{d}{atmospheric thickness from 300 bar to 20 bar}; and \tablefoottext{e}{Bond albedo}.}

\end{table*}

\section{Conclusions}\label{sec:conclusions}

In this work, we present the most comprehensive characterization of the \starname{} system based on the photometric data gathered by \cheops{} and \tess{} space missions. We confirm the existence of a fifth transiting planet with an orbital period of 29.54\,d. On the contrary, we found no indication of transit signals of a putative additional planet in a 10.9\,d orbit in \cheops{} dedicated observations we performed. A planet larger than 0.42\,R$_{\oplus}$ (with $e$=$b$=0) would have been detectable in our dataset at S/N=1. The general picture of the \starname{} system is summarized in Fig.\,\ref{fig:system_diagram}. 

With our analysis, we greatly improved the characterization of the planetary properties, in particular, we refined the size estimations of the five planets in the system with uncertainties between 1.5--3\%. Also, the extensive and detailed timing analysis facilitated an update for the ephemeris of all the planets and to report that TTVs were observed for planets $b$ and $f$. 
Moreover, in the absence of a profuse RV monitoring to date, we performed mass estimations of the planets by different approaches. As illustrated in Table\,\ref{tab:mass_table}, the best agreement between the different applied methods was obtained for the inner planets. On the other hand, a large spread in the mass estimations is seen for the two outer planets. In this case, the constraints set by the timing analysis for that pair of planets in a 3:2 MMR, suggest a low density configuration for these bodies. A TTV analysis that would include additional transits for these two bodies would allow stronger constraints to be placed on their gravitational interaction and, thus, their density as well as the architecture of the system.

Thanks to its brightness, \starname{} system is fitting for an intense and precise radial velocity follow-up, which together with our precise estimations of the planetary radii will provide definitive constraints on the density of the transiting bodies.

Based on the refined planetary properties, we computed the transmission spectroscopy metric (TSM) of \cite{Kempton2018} to estimate its suitability for a potential atmospheric follow-up with {\it James Webb Space Telescope} \citep[\textit{JWST,}][]{Gardner2006}. By using the RV masses (Table\,\ref{tab:mass_table}) and the upper limit of planet $d$ (M$_{d}=7.75\pm1.0$) from the estimations of \cite{Teske2021} and \cite{Bonfanti2021}, we obtained the following TSM values for each planet ($b$ to $f$): 54 $\pm$ 26, 77 $\pm$ 30, 73 $\pm$ 10, 56 $\pm$ 13, and 9 $\pm$ 4, where the errors are dominated by the uncertainties on the planetary mass, with all of them below the suggested threshold for this range of planet sizes (TSM>90). The potential for an \textit{JWST} atmospheric follow-up will be finally defined when the densities of these planets are better determined. So far, our analysis suggests the presence of significant atmospheres in planets $c$ to $e$. On the contrary, planet $b$ is compatible with the presence of a thin atmosphere, or a Fe-poor dry planet, while planet $f$ seems to have a rocky structure.

\begin{acknowledgements}
CHEOPS is an ESA mission in partnership with Switzerland with important contributions to the payload and the ground segment from Austria, Belgium, France, Germany, Hungary, Italy, Portugal, Spain, Sweden, and the United Kingdom. The CHEOPS Consortium would like to gratefully acknowledge the support received by all the agencies, offices, universities, and industries involved. Their flexibility and willingness to explore new approaches were essential to the success of this mission. 
Funding for the TESS mission is provided by NASA's Science Mission Directorate. We acknowledge the use of public TESS data from pipelines at the TESS Science Office and at the TESS Science Processing Operations Center. Resources supporting this work were provided by the NASA High-End Computing (HEC) Program through the NASA Advanced Supercomputing (NAS) Division at Ames Research Center for the production of the SPOC data products. This paper includes data collected by the TESS mission that are publicly available from the Mikulski Archive for Space Telescopes (MAST). 
SH gratefully acknowledges CNES funding through the grant 837319. 
The authors acknowledge support from the Swiss NCCR PlanetS and the Swiss National Science Foundation. 
LMS gratefully acknowledges financial support from the CRT foundation under Grant No. 2018.2323 ‘Gaseous or rocky? Unveiling the nature of small worlds’. 
This project was supported by the CNES. 
This work was also partially supported by a grant from the Simons Foundation (PI Queloz, grant number 327127). 
ACC and TW acknowledge support from STFC consolidated grant numbers ST/R000824/1 and ST/V000861/1, and UKSA grant number ST/R003203/1. 
S.G.S. acknowledge support from FCT through FCT contract nr. CEECIND/00826/2018 and POPH/FSE (EC). 
YA and MJH acknowledge the support of the Swiss National Fund under grant 200020\_172746. 
We acknowledge support from the Spanish Ministry of Science and Innovation and the European Regional Development Fund through grants ESP2016-80435-C2-1-R, ESP2016-80435-C2-2-R, PGC2018-098153-B-C33, PGC2018-098153-B-C31, ESP2017-87676-C5-1-R, MDM-2017-0737 Unidad de Excelencia Maria de Maeztu-Centro de Astrobiologí­a (INTA-CSIC), as well as the support of the Generalitat de Catalunya/CERCA programme. The MOC activities have been supported by the ESA contract No. 4000124370. 
S.C.C.B. acknowledges support from FCT through FCT contracts nr. IF/01312/2014/CP1215/CT0004. 
XB, SC, DG, MF and JL acknowledge their role as ESA-appointed CHEOPS science team members. 
ABr was supported by the SNSA. 
ACC acknowledges support from STFC consolidated grant numbers ST/R000824/1 and ST/V000861/1, and UKSA grant number ST/R003203/1. 
The Belgian participation to CHEOPS has been supported by the Belgian Federal Science Policy Office (BELSPO) in the framework of the PRODEX Program, and by the University of Liège through an ARC grant for Concerted Research Actions financed by the Wallonia-Brussels Federation. 
L.D. is an F.R.S.-FNRS Postdoctoral Researcher. 
This work was supported by FCT - Fundação para a Ciência e a Tecnologia through national funds and by FEDER through COMPETE2020 - Programa Operacional Competitividade e Internacionalizacão by these grants: UID/FIS/04434/2019, UIDB/04434/2020, UIDP/04434/2020, PTDC/FIS-AST/32113/2017 \& POCI-01-0145-FEDER- 032113, PTDC/FIS-AST/28953/2017 \& POCI-01-0145-FEDER-028953, PTDC/FIS-AST/28987/2017 \& POCI-01-0145-FEDER-028987, O.D.S.D. is supported in the form of work contract (DL 57/2016/CP1364/CT0004) funded by national funds through FCT. 
B.-O.D. acknowledges support from the Swiss National Science Foundation (PP00P2-190080). 
This project has received funding from the European Research Council (ERC) under the European Union’s Horizon 2020 research and innovation programme (project {\sc Four Aces}. 
grant agreement No 724427). It has also been carried out in the frame of the National Centre for Competence in Research PlanetS supported by the Swiss National Science Foundation (SNSF). DE acknowledges financial support from the Swiss National Science Foundation for project 200021\_200726. 
MF and CMP gratefully acknowledge the support of the Swedish National Space Agency (DNR 65/19, 174/18). 
DG gratefully acknowledges financial support from the CRT foundation under Grant No. 2018.2323 ``Gaseousor rocky? Unveiling the nature of small worlds''. 
M.G. is an F.R.S.-FNRS Senior Research Associate. 
KGI is the ESA CHEOPS Project Scientist and is responsible for the ESA CHEOPS Guest Observers Programme. She does not participate in, or contribute to, the definition of the Guaranteed Time Programme of the CHEOPS mission through which observations described in this paper have been taken, nor to any aspect of target selection for the programme. 
This work was granted access to the HPC resources of MesoPSL financed by the Region Ile de France and the project Equip@Meso (reference ANR-10-EQPX-29-01) of the programme Investissements d'Avenir supervised by the Agence Nationale pour la Recherche. 
ML acknowledges support of the Swiss National Science Foundation under grant number PCEFP2\_194576. 
PM acknowledges support from STFC research grant number ST/M001040/1. 
GSc, GPi, IPa, LBo, VNa and RRa acknowledge the funding support from Italian Space Agency (ASI) regulated by “Accordo ASI-INAF n. 2013-016-R.0 del 9 luglio 2013 e integrazione del 9 luglio 2015 CHEOPS Fasi A/B/C”. 
IRI acknowledges support from the Spanish Ministry of Science and Innovation and the European Regional Development Fund through grant PGC2018-098153-B- C33, as well as the support of the Generalitat de Catalunya/CERCA programme. 
S.S. has received funding from the EuropeanResearch Council (ERC) under the European Union’s Horizon 2020 researchand innovation program (grant agreement No 833925, project STAREX). 
GyMSz acknowledges the support of the Hungarian National Research, Development and Innovation Office (NKFIH) grant K-125015, a PRODEX Institute Agreement between the ELTE E\"otv\"os Lor\'and University and the European Space Agency (ESA-D/SCI-LE-2021-0025), the Lend\"ulet LP2018-7/2021 grant of the Hungarian Academy of Science and the support of the city of Szombathely. 
V.V.G. is an F.R.S-FNRS Research Associate. 
NAW acknowledges UKSA grant ST/R004838/1. 
This research has made use of computing facilities operated by CeSAM data center at LAM, Marseille, France. We also thank J.C.~Meunier and J.C.~Lamber for the support provided in the use of these facilities.

\textit{Software}: We gratefully acknowledge the open-source software which made this work possible: \texttt{astropy} \citep{astropy2013, astropy2018}, \texttt{ipython} \citep{ipython}, \texttt{numpy} \citep{harris2020array}, \texttt{scipy} \citep{scipy}, \texttt{matplotlib} \citep{matplotlib}, \texttt{jupyter} \citep{jupyter2016}. \\ 

\end{acknowledgements}

\bibliography{biblio}
\bibliographystyle{aa}

\begin{appendix}
\section{Raw and detrended CHEOPS light curves }
\label{ssec:raw_detrended_lcs}

\begin{figure*}[th!]
    \centering
    \includegraphics[angle=90,origin=c,scale=.95]{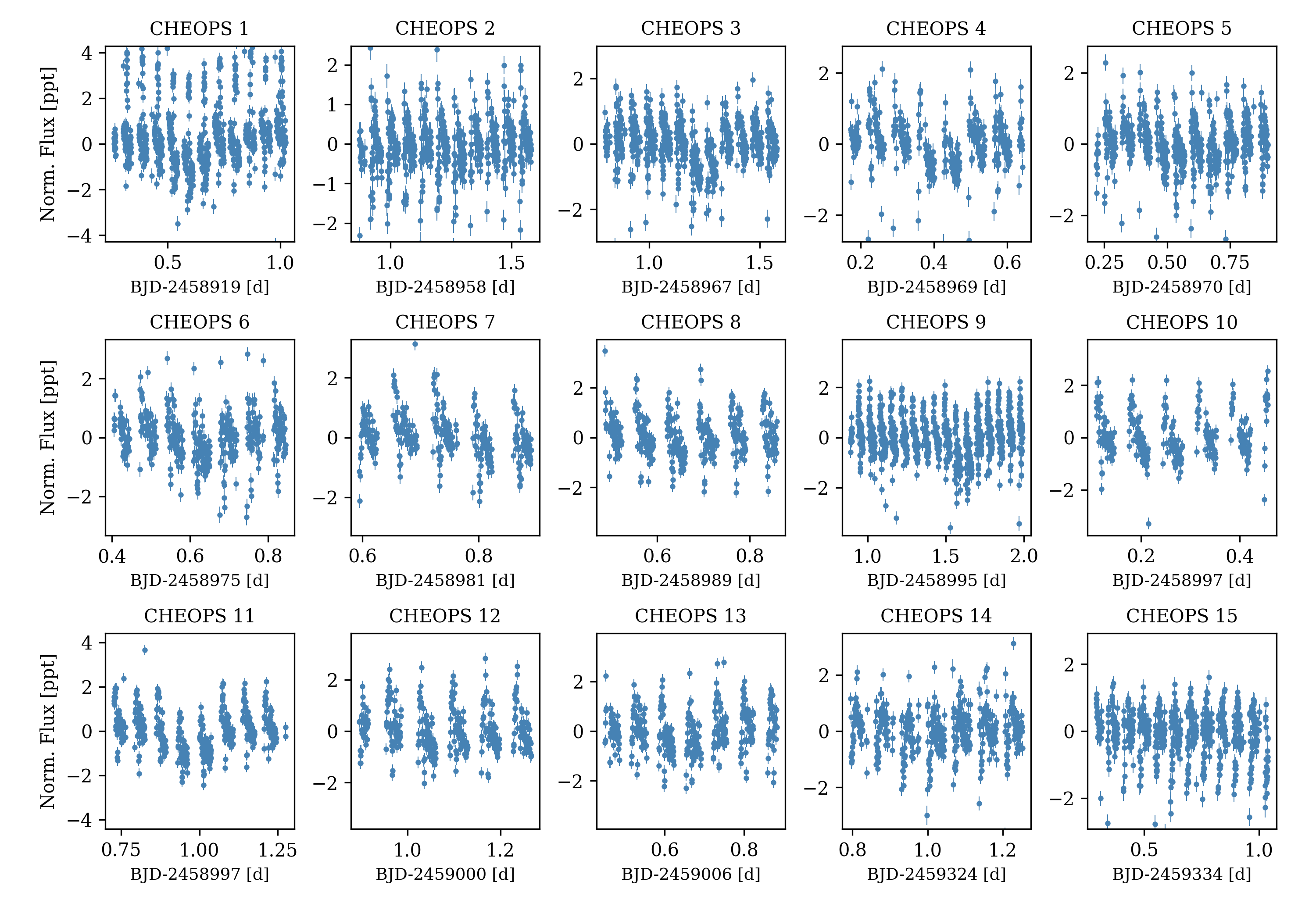}
     \caption{Normalized \cheops{} raw light curves of \starname{} used in this work. Large outliers have been clipped out to improve visualization. The \cheops{} label number on top of each panel correspond to the ID given in Table~\ref{tab:obs_log}. }
    \label{fig:raw_lcs}
\end{figure*}

\begin{figure*}[h!]
    \centering
    \includegraphics[width=.43\textwidth]{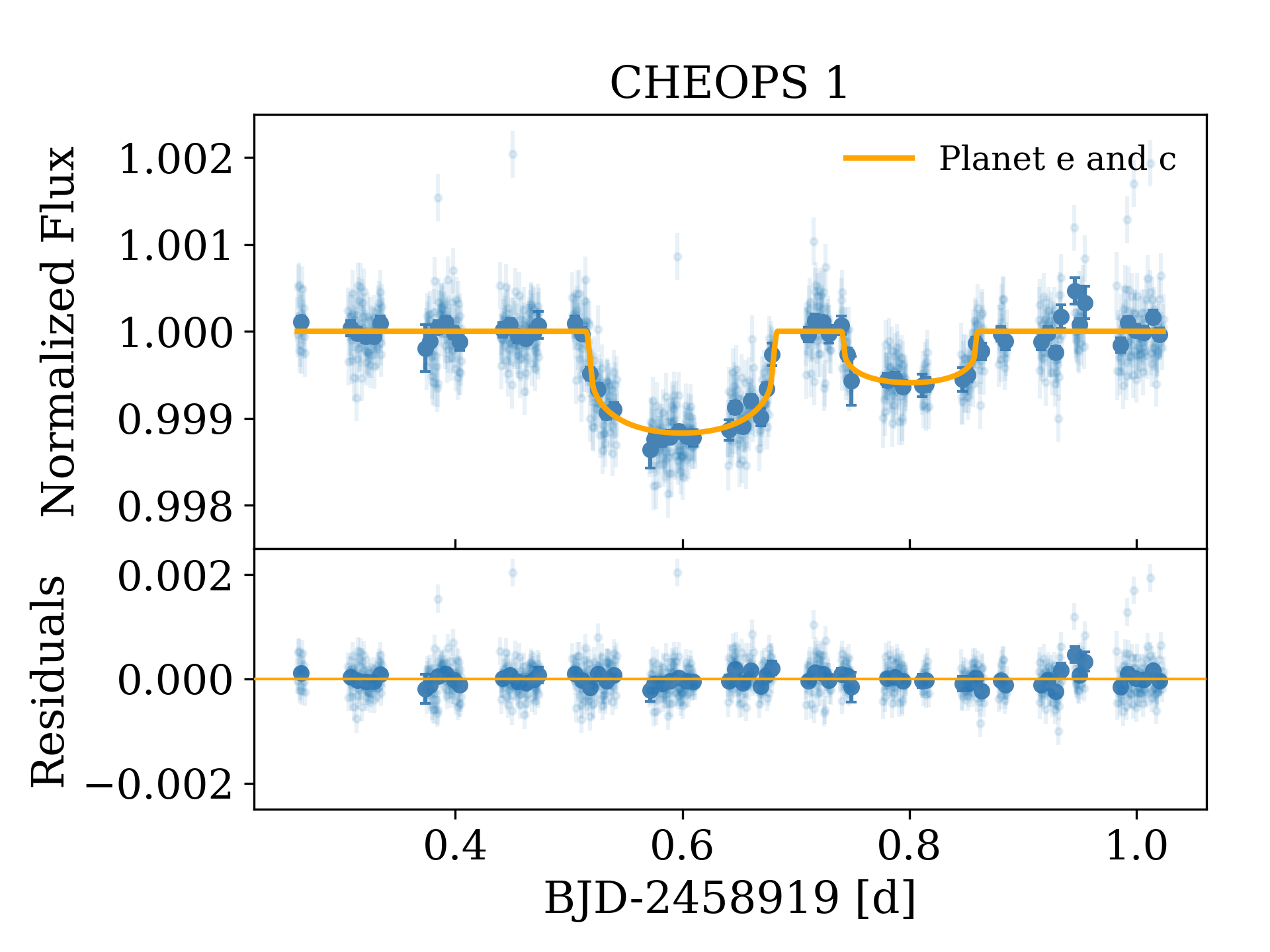}
    \includegraphics[width=.43\textwidth]{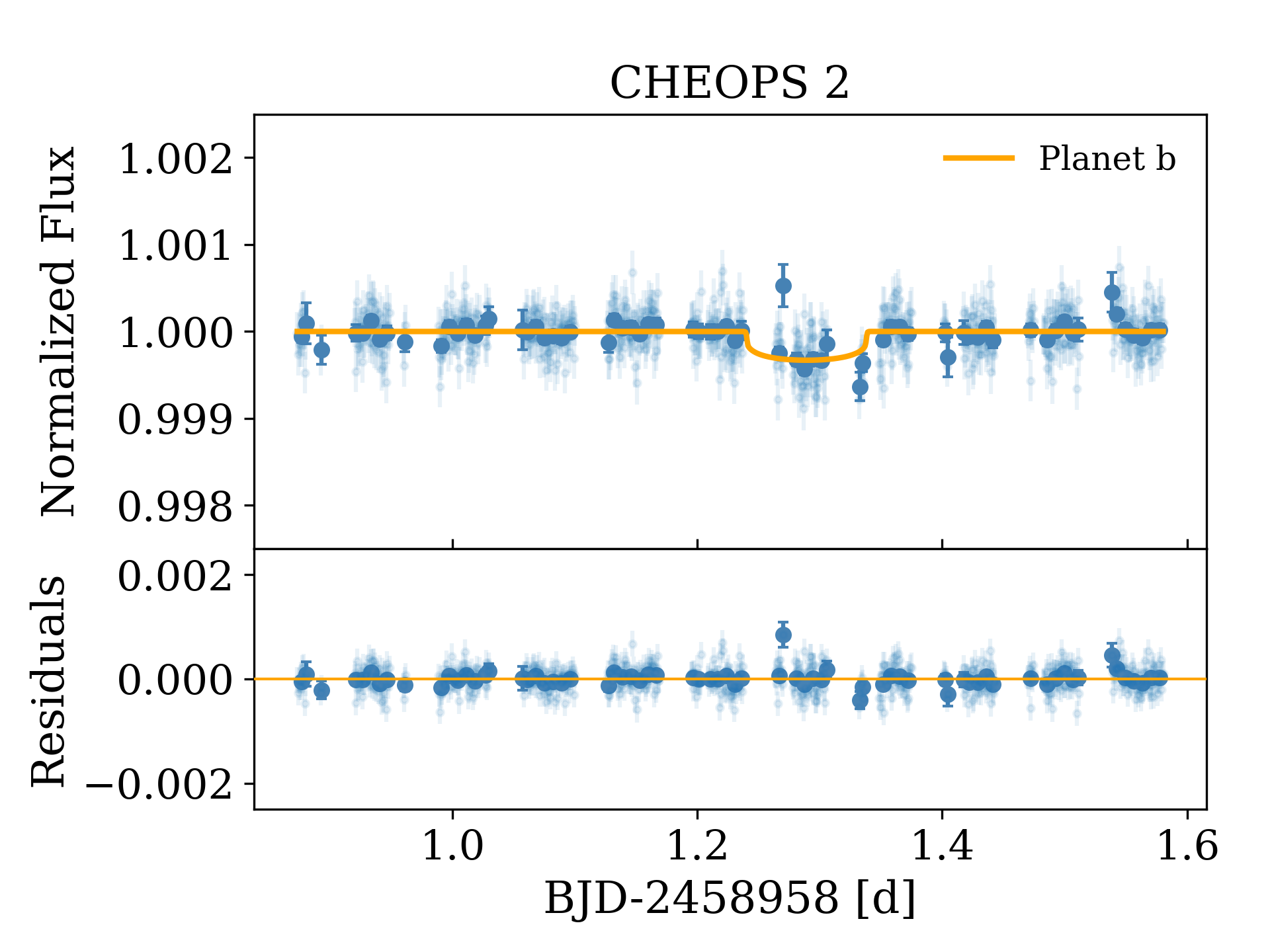} \\
    \includegraphics[width=.43\textwidth]{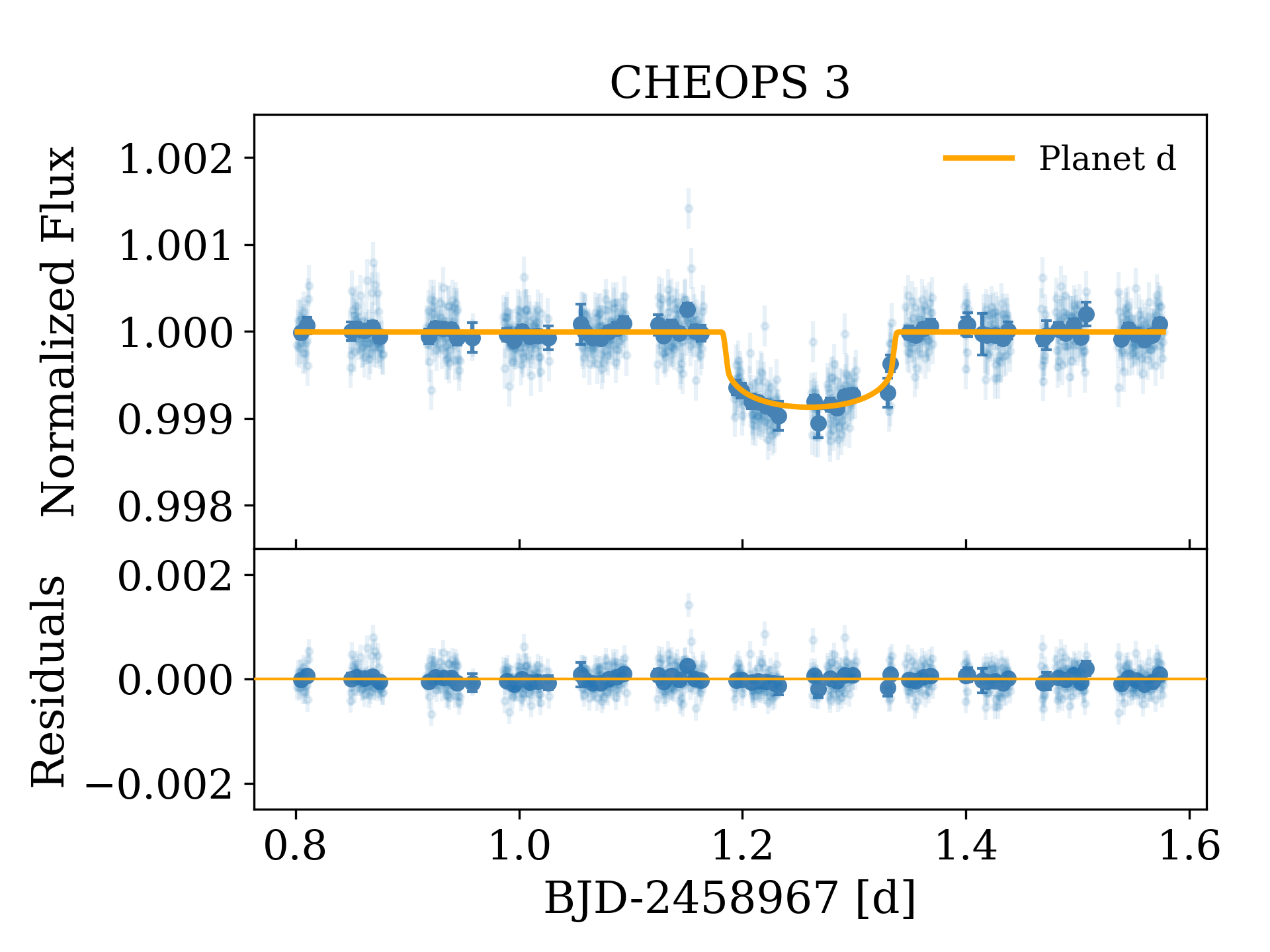}
    \includegraphics[width=.43\textwidth]{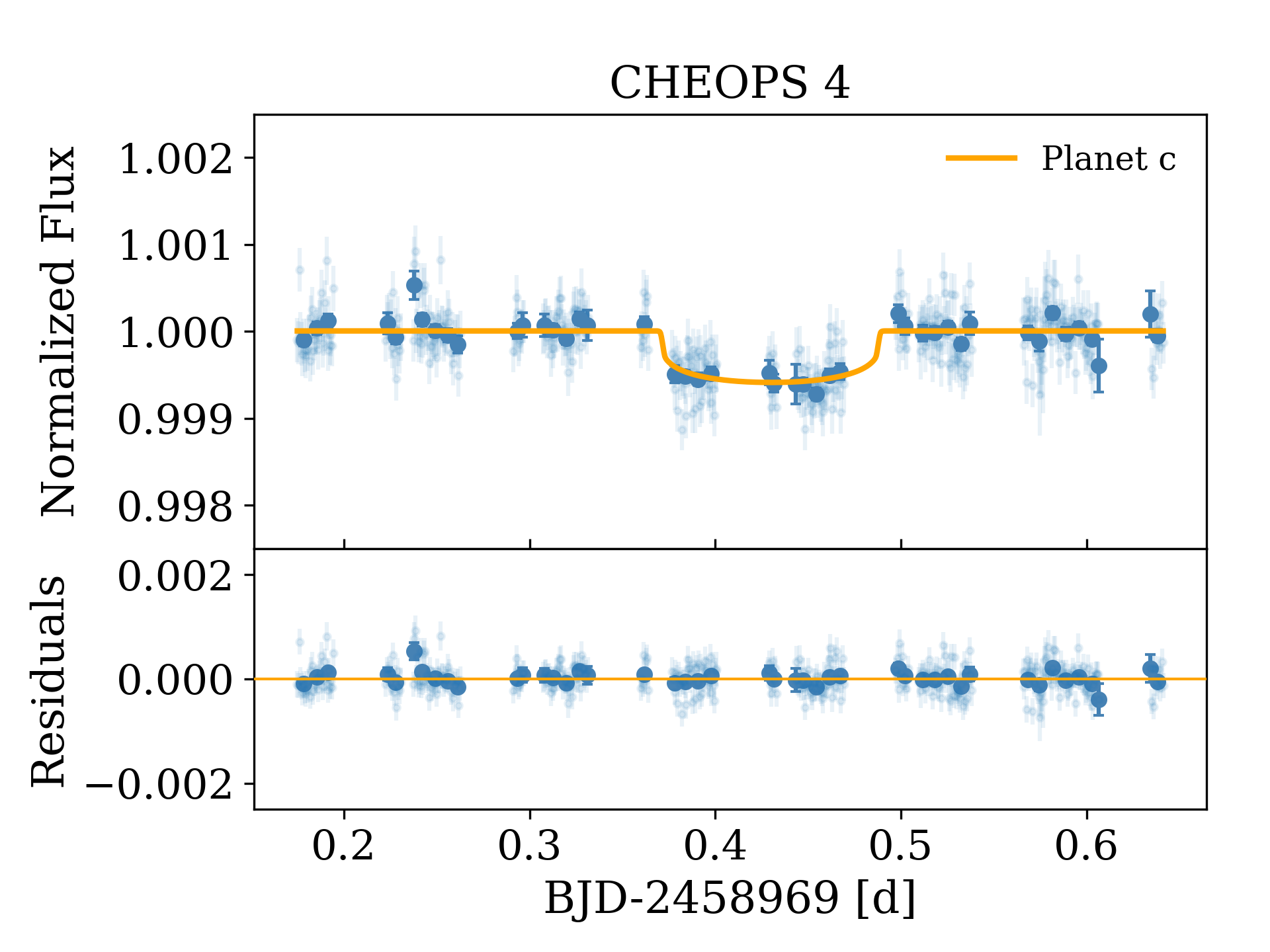} \\
    \includegraphics[width=.43\textwidth]{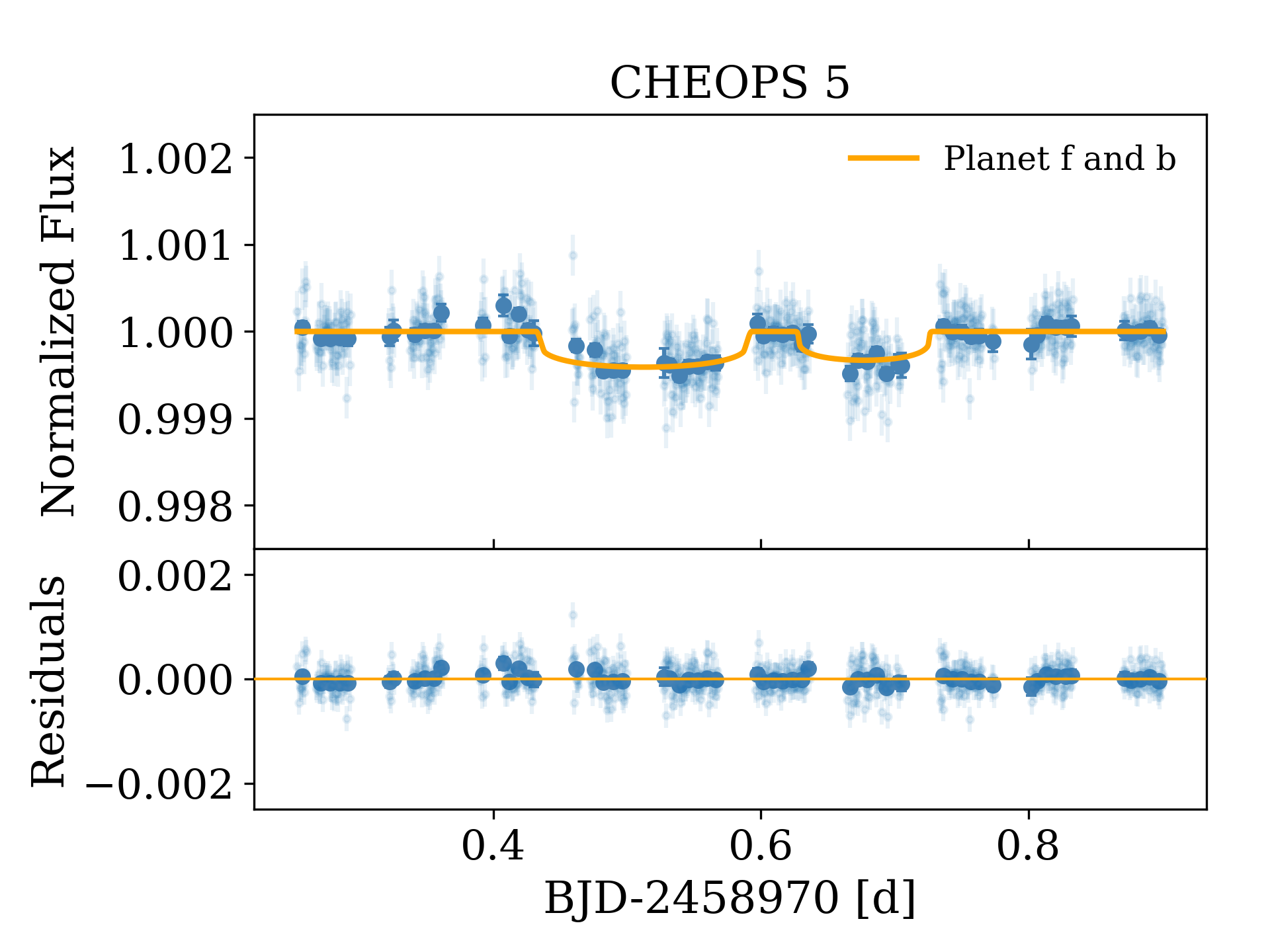}
    \includegraphics[width=.43\textwidth]{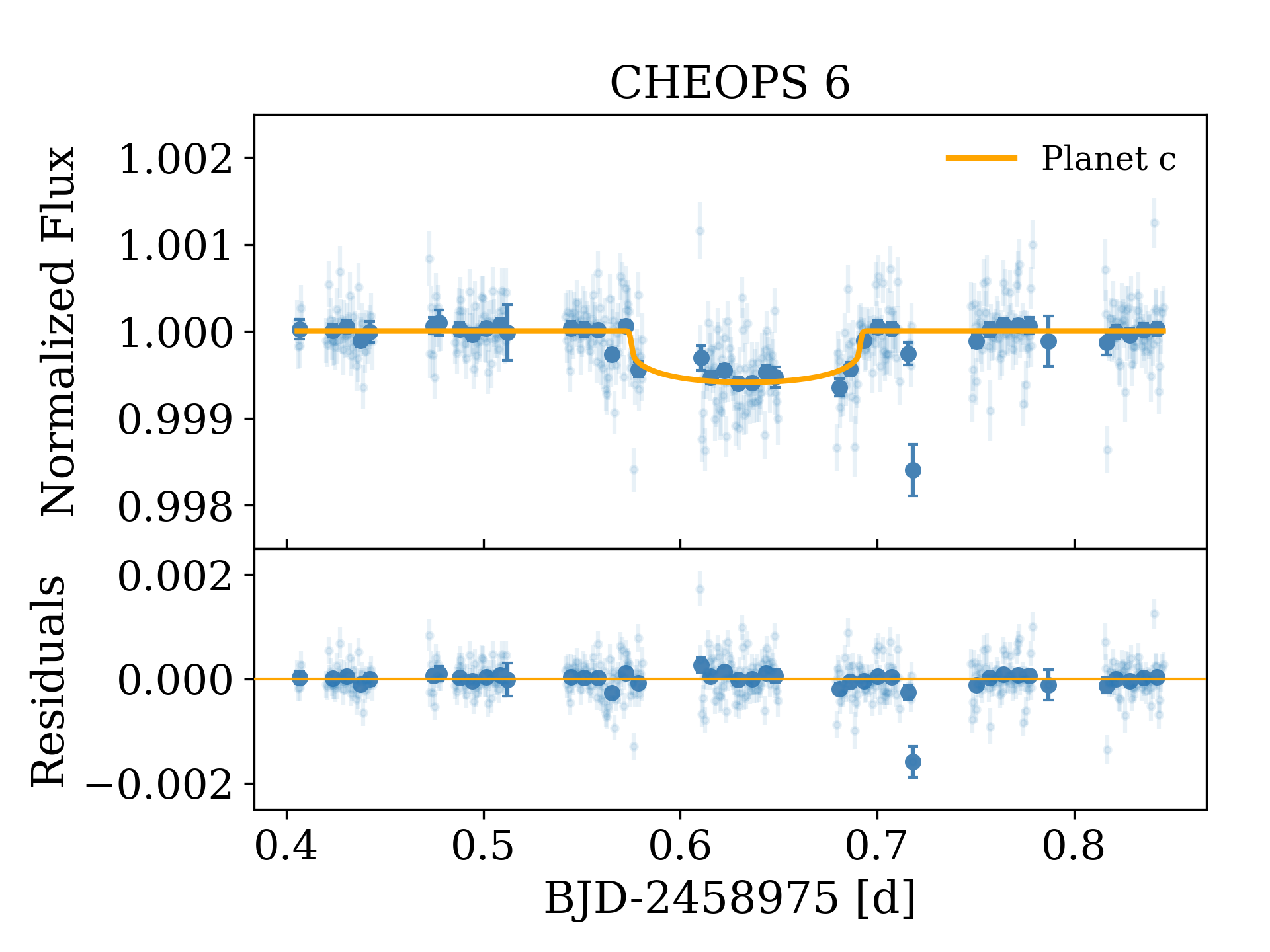} \\
    \includegraphics[width=.43\textwidth]{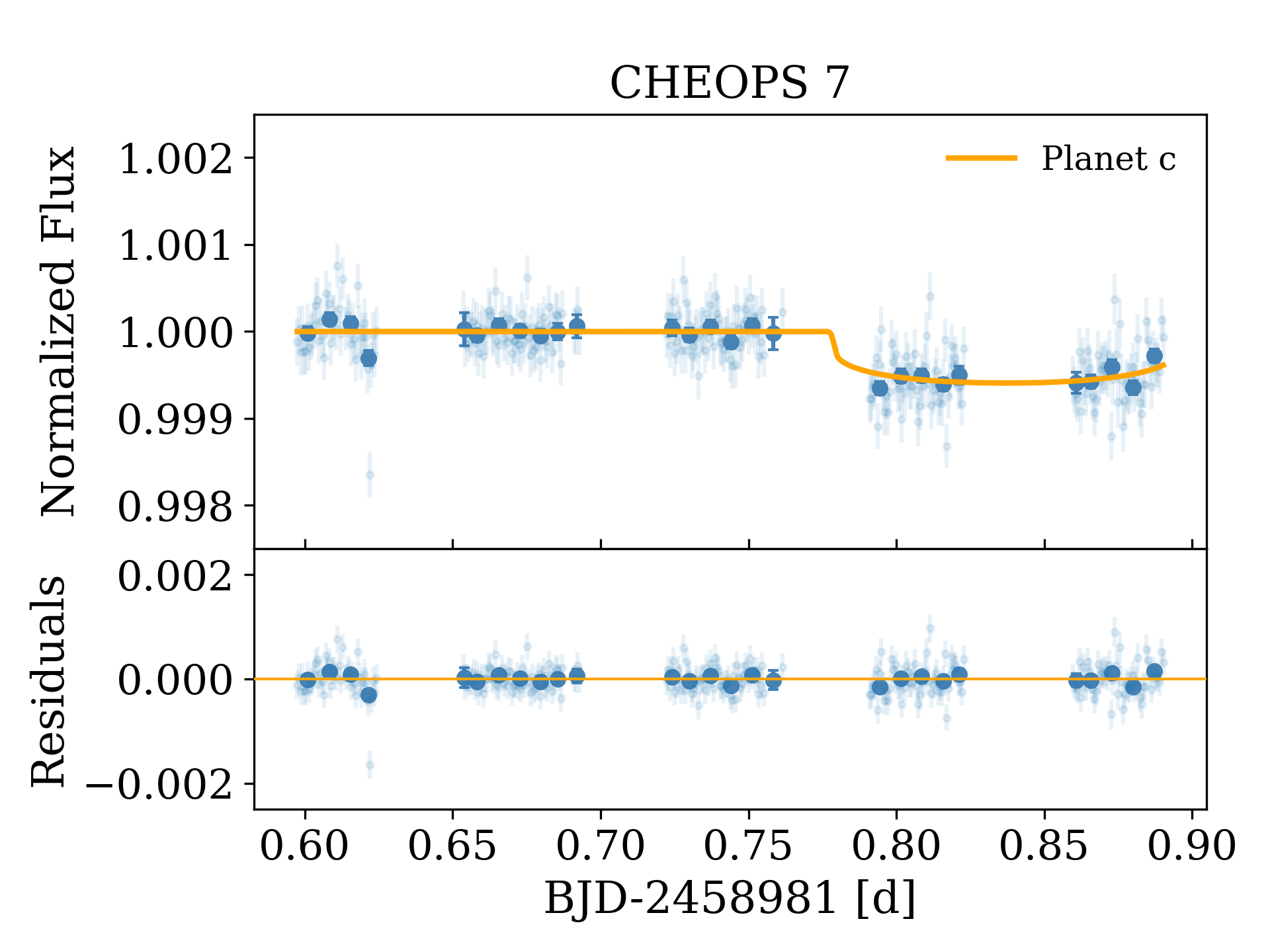}
    \includegraphics[width=.43\textwidth]{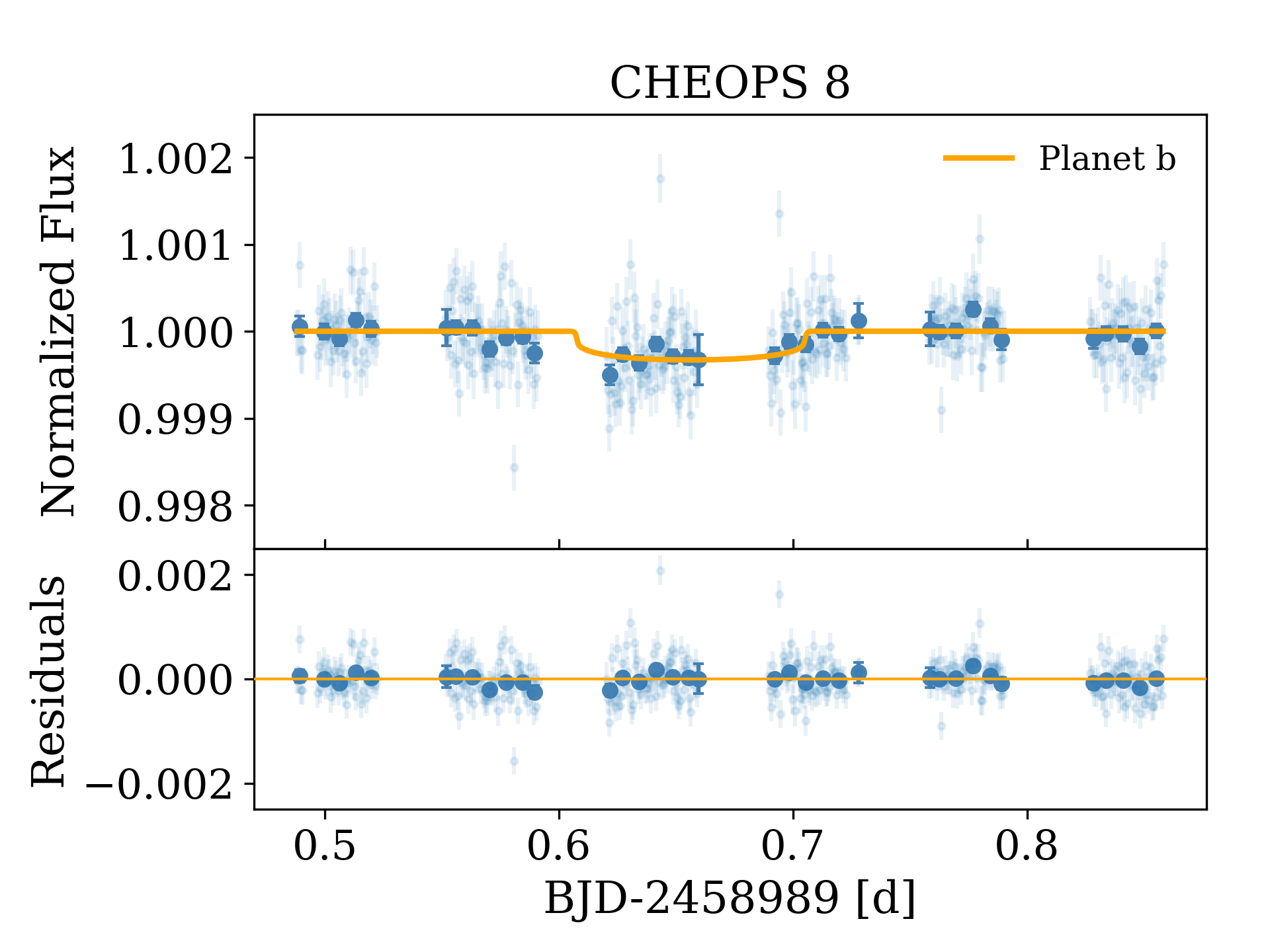} \\
   
   \caption{Normalized \cheops{} detrended  light curves of \starname{} used in this work and its 10\,min bins are shown, respectively, with light and dark blue symbols. The best transit models obtained from the global analysis of the system are represented by the yellow curve. The residuals are shown in the bottom panels.}
    \label{fig:cheops_lcs1}
\end{figure*}

\begin{figure*}[h!] 
    \centering
    \includegraphics[scale=.47]{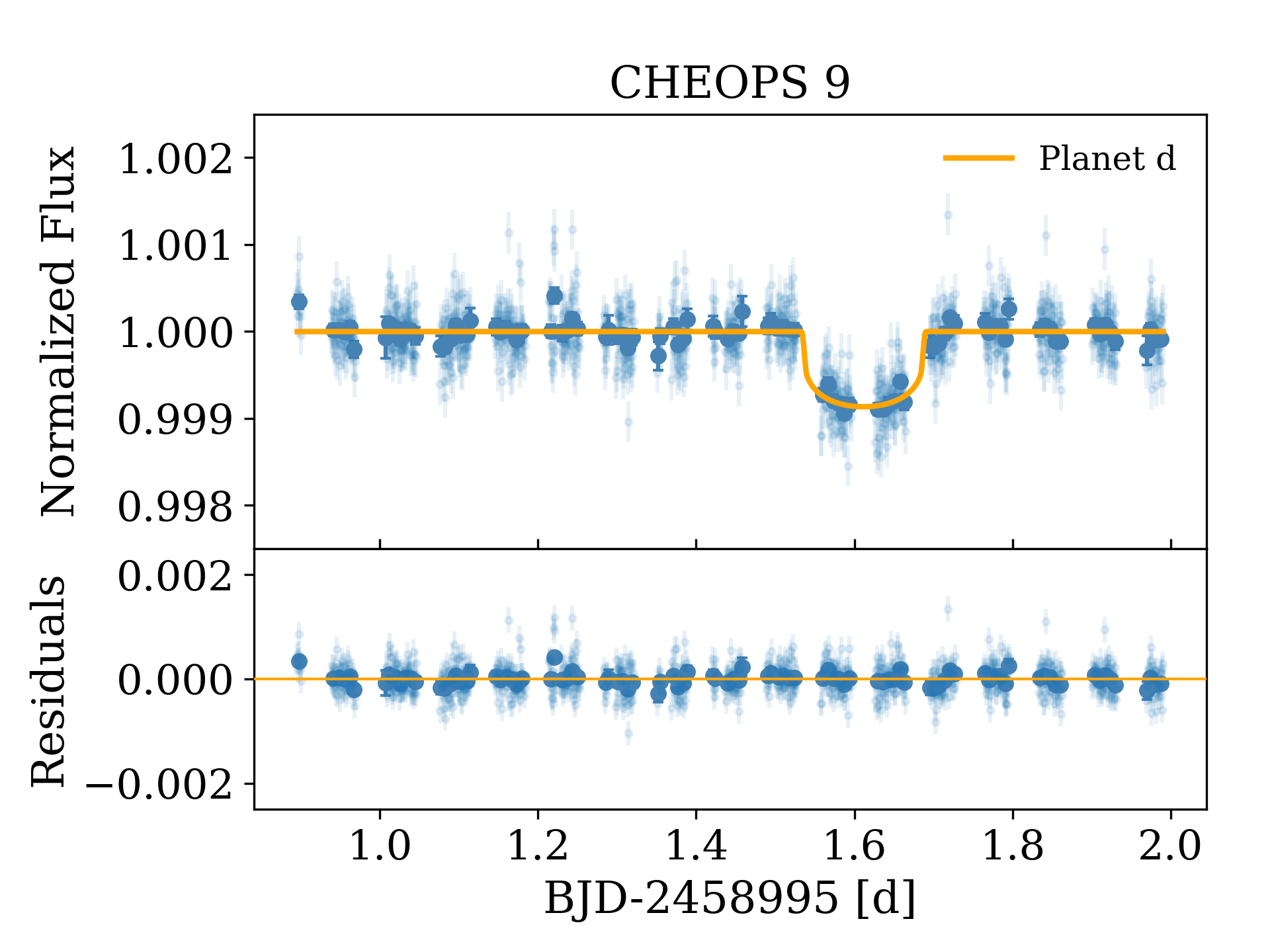}
    \includegraphics[scale=.47]{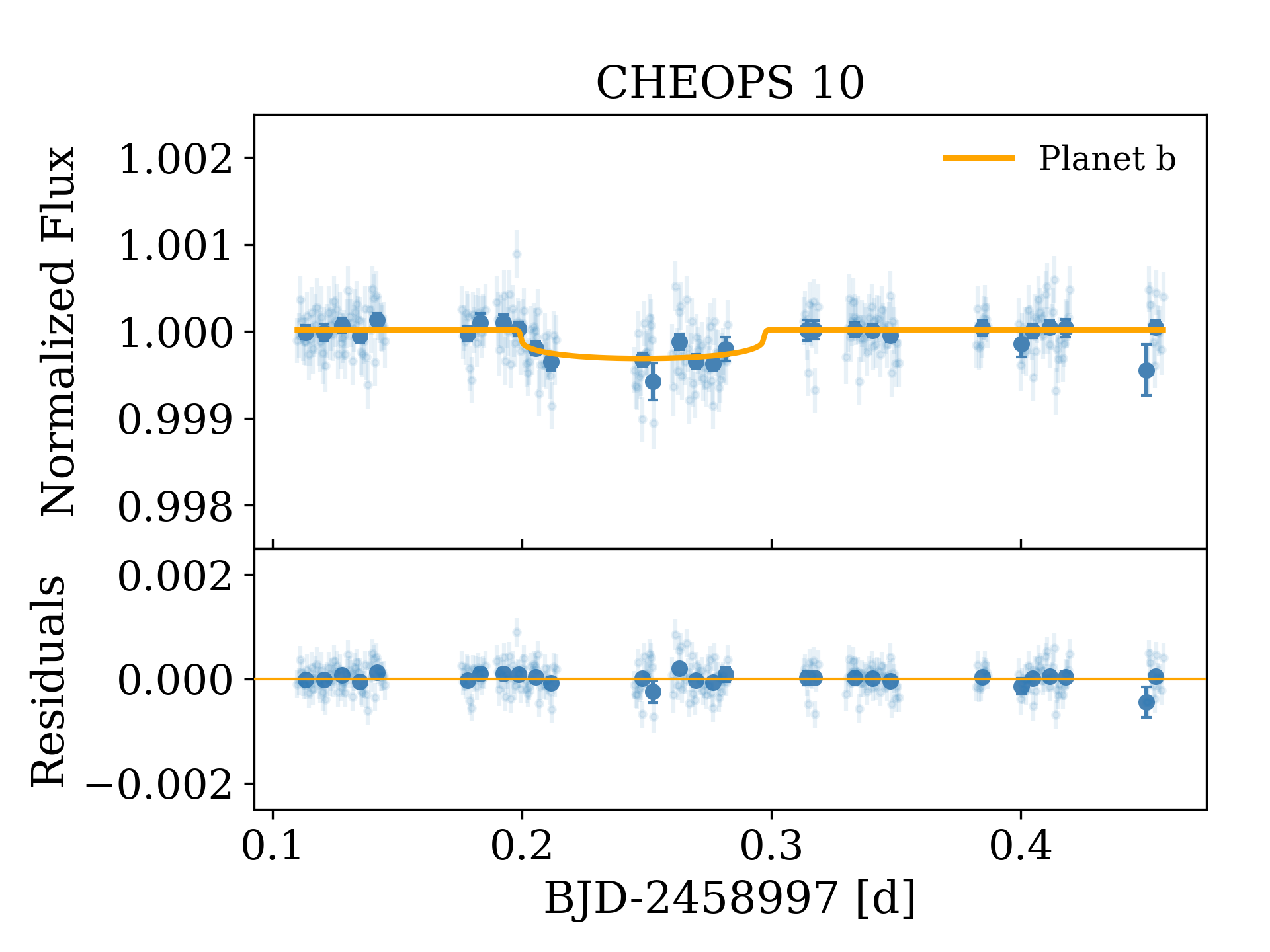} \\
    \includegraphics[scale=.47]{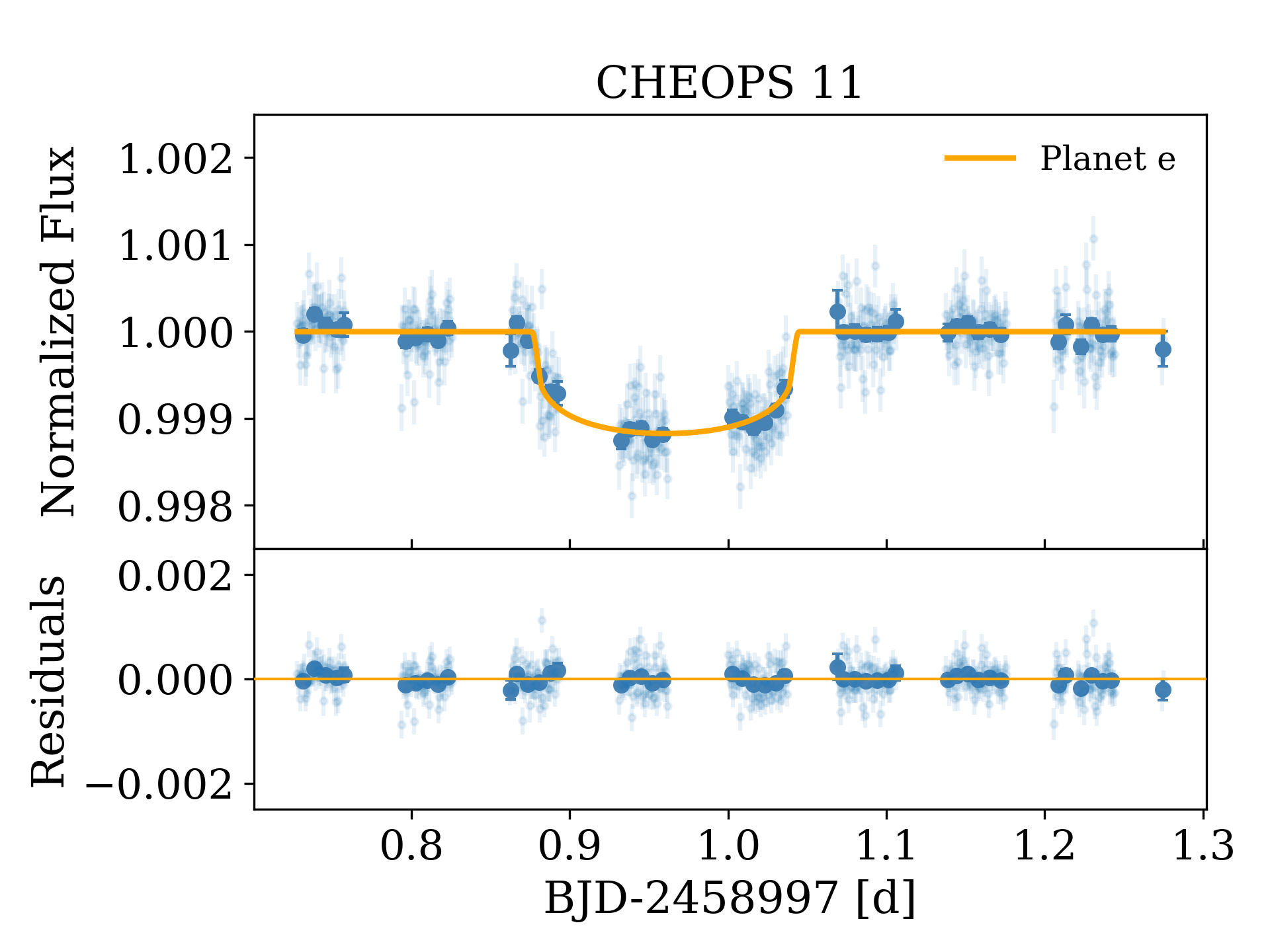} 
    \includegraphics[scale=.47]{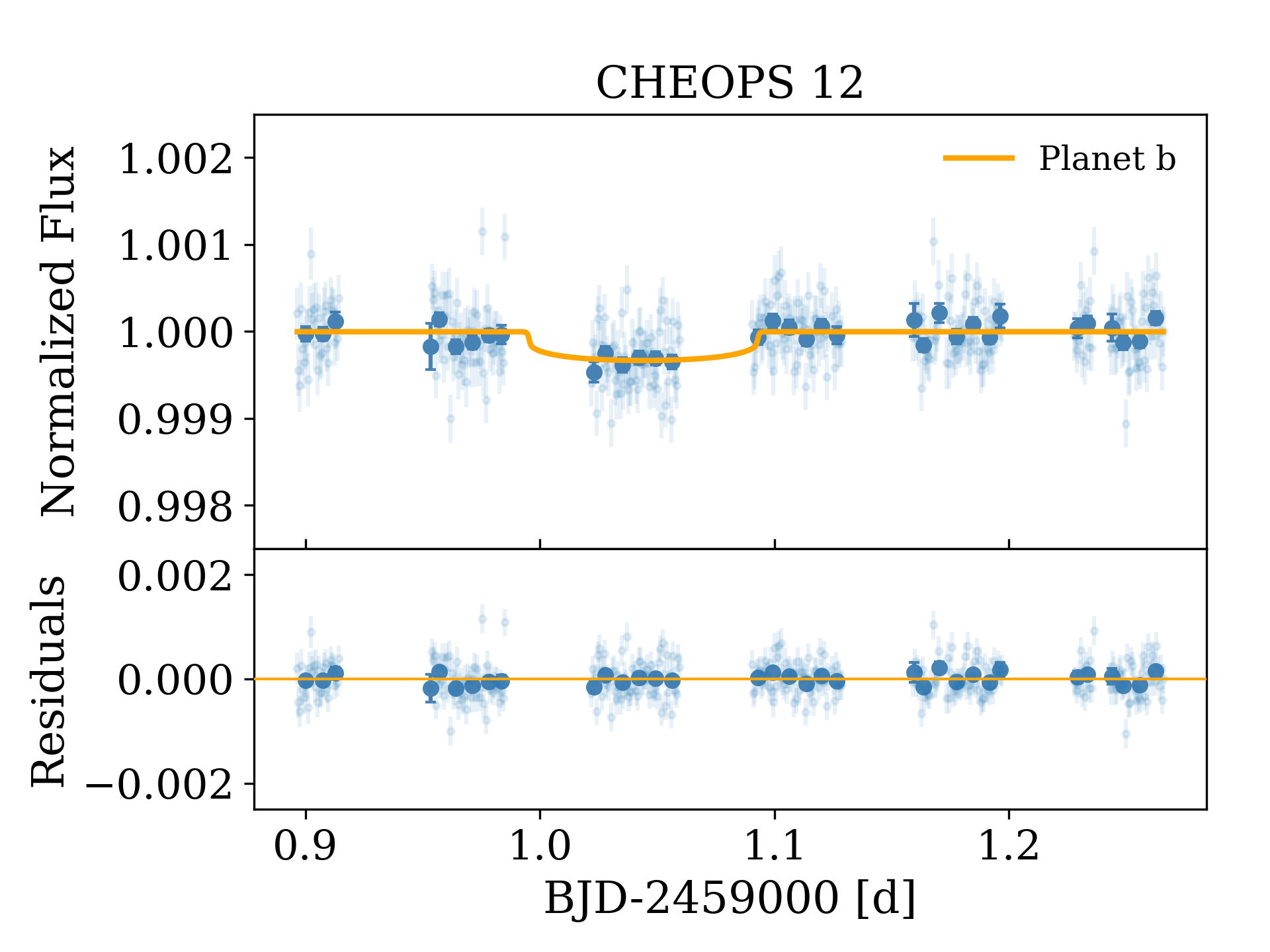} \\
   \includegraphics[scale=.47]{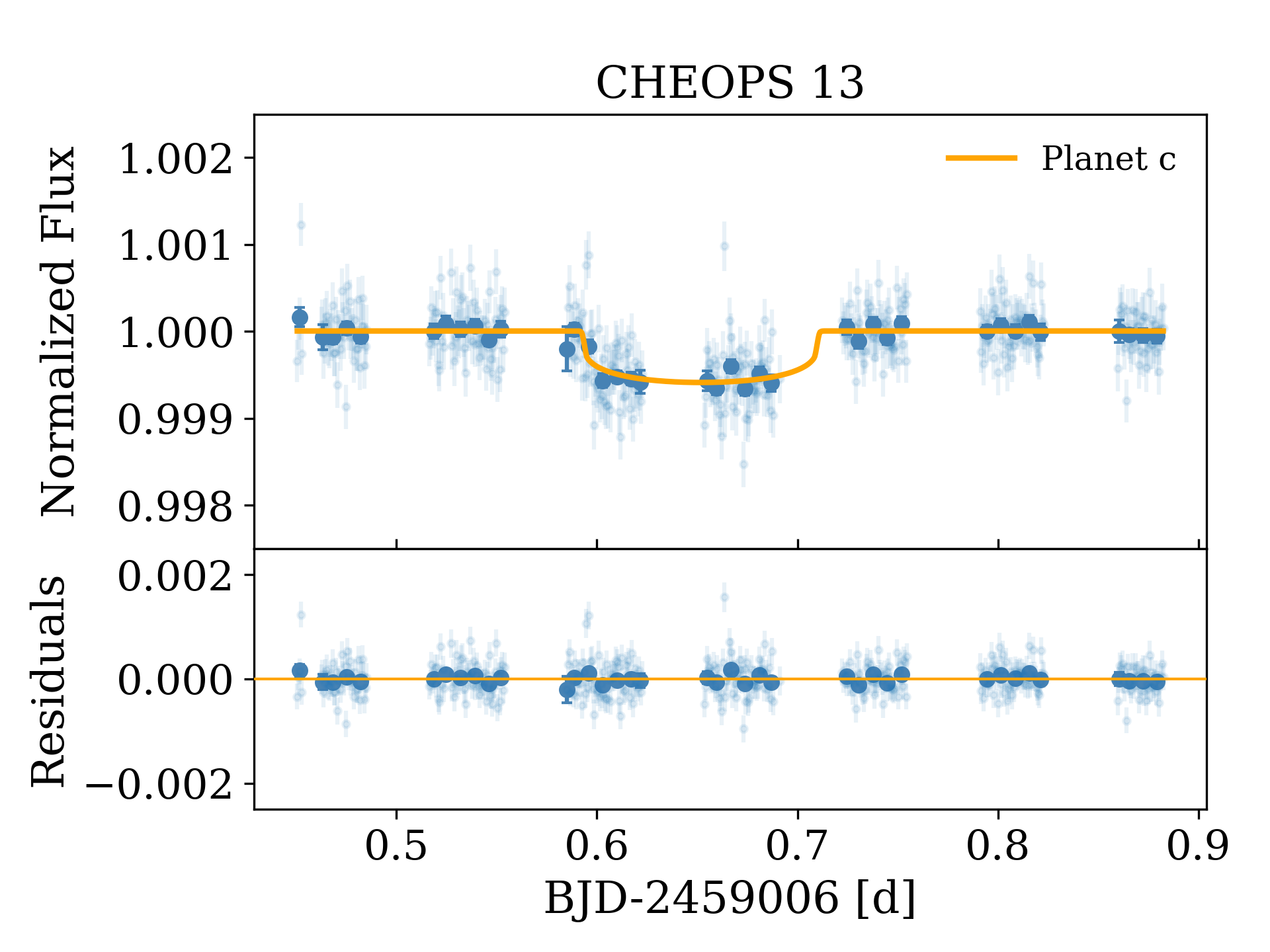} 
    \includegraphics[scale=.47]{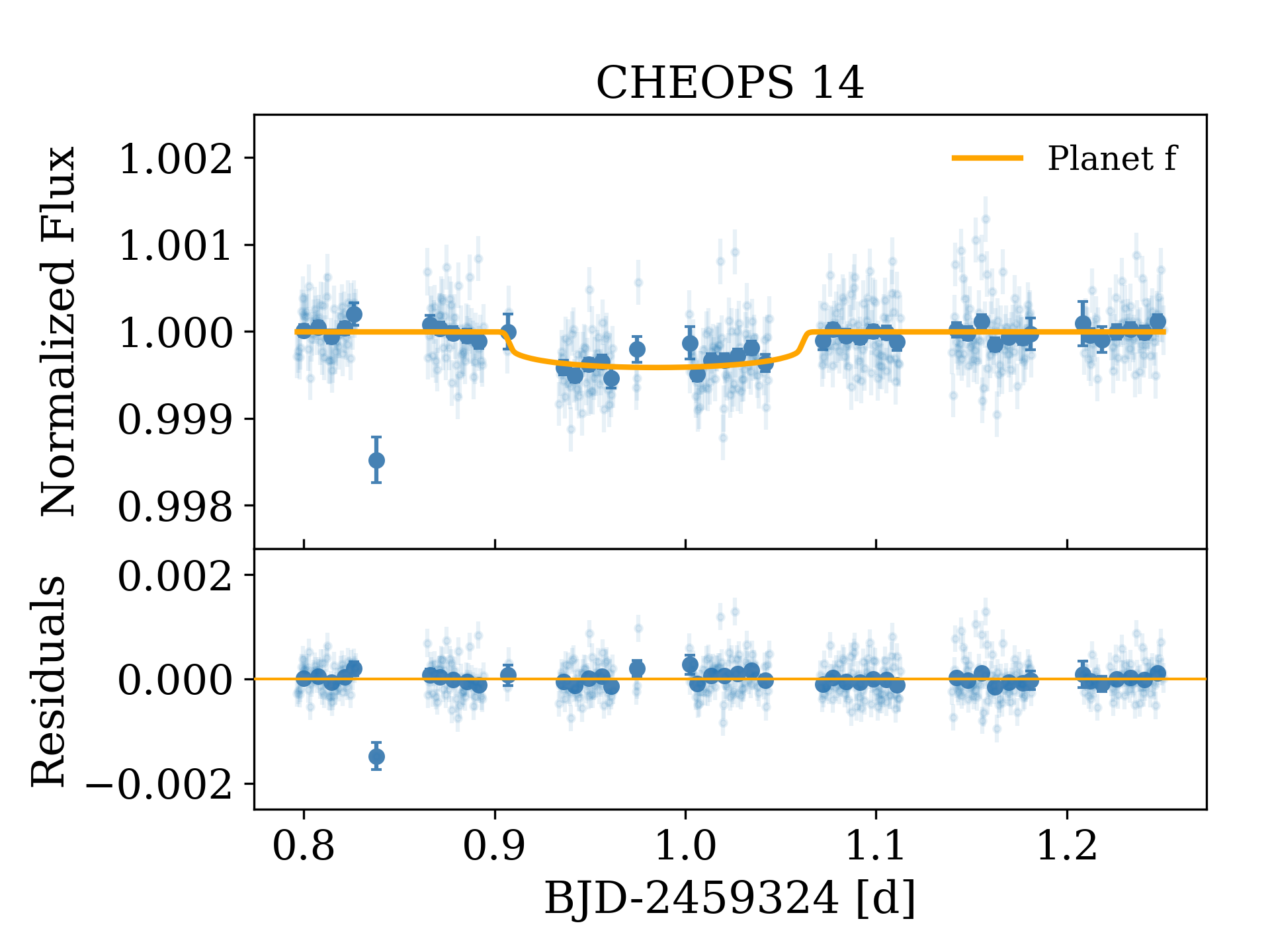} \\

    \caption{Normalized \cheops{} detrended  light curves of \starname{} used in this work and its 10\,min bins are shown, respectively, with light and dark blue symbols. The best transit models obtained from the global analysis of the system are represented by the yellow curve. The residuals are shown in the bottom panels.   }
    \label{fig:cheops_lcs2}
\end{figure*}

In Fig.~\ref{fig:raw_lcs} we show the raw light curves as output by the \cheops{} Data Reduction Pipeline \citep[DRP-v13,][]{hoyer2020}.  These light curves are usually affected by systematics as a function of the rotation angle of the Field of View along the pointing direction. In Figs.~\ref{fig:cheops_lcs1} and~\ref{fig:cheops_lcs2} we show the detrended version of the \cheops{} light curves (see Sect.\ref{sec:LCdetrending} for details), together with the best models obtained from the global modeling of the system (Sect.~\ref{sec:analysis}).

\FloatBarrier

\section{Initial priors of the LC analysis}
In Table\,\ref{tab:priors}, we list the priors distributions used for the global fit of all the photometric data described in Sect.\,\ref{sec:analysis}, consisting of 34 \tess{} and 15 \cheops{} light curves.

\begin{table}[h!]
\caption{Priors for each of the parameters fitted in the global photometric analysis of \cheops{} and \tess{} data.  }

\label{tab:priors}
\centering
\begin{tabular}{l c}
\hline\hline            
 Parameter & Prior \\
  \hline
 \multicolumn{2}{c}{Star} \\
 $\rho_{\star}$ [$kg$\,$m^{-3}$] & $\mathcal{TN}(1790,110,1680,1900)$ \\
 $q_{1,{\mathrm{TESS}}}$ & $\mathcal{N}(0.34,0.05)$ \\
 $q_{2,{\mathrm{TESS}}}$ & $\mathcal{N}(0.27,0.05)$ \\
 $q_{1,{\mathrm{CHEOPS}}}$ & $\mathcal{N}(0.46,0.05)$ \\
 $q_{2,{\mathrm{CHEOPS}}}$ & $\mathcal{N}(0.32,0.05)$ \\
\hline
\multicolumn{2}{c}{Planet $b$} \\
$R_p/R_{\star}$ & $\mathcal{N}(0.01688,0.01)$ \\
T$_0$-2\,450\,000. [BJD] & $\mathcal{N}(8572.1128,0.01)$ \\
P [$d$] & $\mathcal{N}(3.796,0.001)$ \\
\hline
\multicolumn{2}{c}{Planet c} \\
$R_p/R_{\star}$ & $\mathcal{N}(0.0217,0.01)$ \\
T$_0$-2\,450\,000. [BJD] & $\mathcal{N}(8572.3949,0.01)$ \\
P [$d$] & $\mathcal{N}(6.203,0.001)$ \\
\hline
\multicolumn{2}{c}{Planet d} \\
$R_p/R_{\star}$ & $\mathcal{N}(0.0266,0.01)$ \\
T$_0$-2\,450\,000. [BJD] & $\mathcal{N}(8571.3368,0.01)$ \\
P [$d$] & $\mathcal{N}(14.1758,0.001)$ \\
\hline
\multicolumn{2}{c}{Planet e} \\
$R_p/R_{\star}$ & $\mathcal{N}(0.0322,0.01)$ \\
T$_0$-2\,450\,000. [BJD] & $\mathcal{N}(8586.5677,0.01)$ \\
P [$d$] & $\mathcal{N}(19.59,0.001)$ \\
\hline
\multicolumn{2}{c}{Planet f} \\
$R_p/R_{\star}$ & $\mathcal{N}(0.0211,0.01)$ \\
T$_0$-2\,450\,000. [BJD] & $\mathcal{N}(8616.039,0.01)$ \\
P [$d$] & $\mathcal{N}(29.5398,0.001)$ \\
\hline   
\multicolumn{2}{c}{All planets} \\
$b$ [$R_{\star}$] & $\mathcal{U}(0.001,0.99)$ \\
$\sqrt{e}sin(\omega)$ & $\mathcal{U}(-0.3,0.3)$ \\
$\sqrt{e}cos(\omega)$ & $\mathcal{U}(-0.3,0.3)$ \\
\hline
\multicolumn{2}{c}{Instrumental} \\
$mflux\_\small{\texttt{TESS}_{i}}$ &  $\mathcal{N}(0.0,0.1)$ \\
$\sigma_{w}\_\small{\texttt{TESS}_{i}}$ [ppm] &  $log\mathcal{U}(0.1,1000)$ \\
$mflux\_\small{\texttt{CHEOPS}_{i}}$ &  $\mathcal{N}(0.0,0.1)$ \\
$\sigma_{w}\_\small{\texttt{CHEOPS}_{i}}$  [ppm] &   $log\mathcal{U}(0.1,1000)$ \\
$c_{1}(t)_\_\small{\texttt{CHEOPS}_{i}}$ & $\mathcal{U}(-1.0,1.0)$ \\
$c_{2}(t^{2})_\_\small{\texttt{CHEOPS}_{i}}$ &  $\mathcal{U}(-1.0,1.0)$\\

\hline
\end{tabular}
\tablefoot{The following distributions were used: {\it Normal} ($\mathcal{N}(\mu_0,\sigma_0)$), {\it Truncated Normal} ($\mathcal{TN}$($\mu_0$,$\sigma_0$,limit\_low, limit\_upper)), {\it Uniform} ($\mathcal{U}$(limit\_low, limit\_upper)), {\it log-Uniform} ($log\mathcal{U}$(limit\_low, limit\_upper)). The instrumental parameters are defined for each fitted light curve.}

\end{table}

\FloatBarrier

\section{Role of the parametrization in the fitted parameters: $e$, $\omega$ and $b$.}
\label{sec:appendix_eccs}

\begin{figure*}
    \centering
    \includegraphics[width=\textwidth]{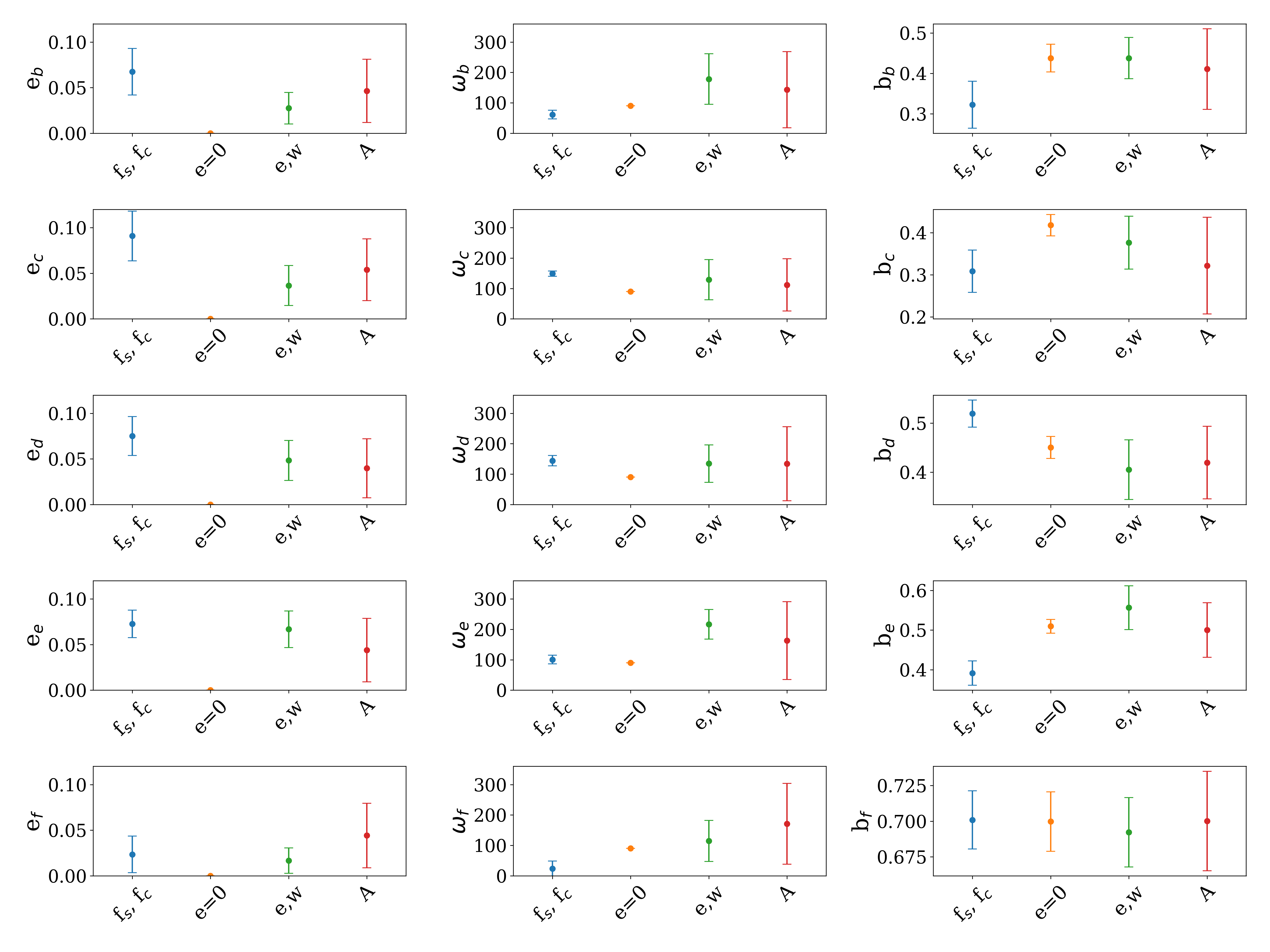}
    \caption{Comparison of fitted values using different parameterizations/priors.  We show the fitted eccentricity, $e$ (left panels), argument of periastron $\omega$ (middle panels), and impact parameter $b$ (right panels) of each planet.  In the x-axis we denote the used method: the (f$_s$,f$_c$) parametrization, fixing the eccentricity to zero ($e=0$) and by jumping directly on $e$ and $\omega$ ($e$,$\omega$). We also show the resulting values of the modeling with \texttt{Allesfiter} code ($A$).}
    \label{fig:ecc_comparison}
\end{figure*}

We explored different configurations of the system modeling with \texttt{juliet} to compare the resulting values of the orbits' eccentricities and any possible effect it might have in other planetary parameters.  As described in Sect.~\ref{ssec:transit_modeling}, we used the parametrization f$_s$ = $\sqrt{e} sin(\omega)$ and f$_c$ = $\sqrt{e} cos(\omega)$ to fit for the eccentricity ($e$) and the argument of the periastron ($\omega$).  In addition, we performed two extra modelings by either fitting  $e$ and $\omega$ directly or assuming $e$=0 and $\omega=90^{\circ}$. The different models were also compared with the fit performed using \texttt{Allesfiter}.
We found that, in general, all the fitted parameters were consistent well within 1$\sigma$. The largest differences are found in the eccentricities as illustrated in Fig.~\ref{fig:ecc_comparison}. We notice that, for planets $b$, $c$ and $d$ the eccentricities derived using (f$_s$,f$_c$) parametrization are systematically large in comparison with the other fits. We found that in this case, eccentricities are correlated with the impact parameters, which were also discrepant with the values obtained with the other fits (see right panels in Fig.~\ref{fig:ecc_comparison}).  These differences likely raise because \texttt{juliet} does not fit for $b_{\mathrm{circ}}$ \citep[the impact parameter of the circular orbit, see e.g.,][]{Gillon2009}, but for the eccentric $b'$, hence producing correlations between $b'$ and the resulting $e$ as:
\begin{equation}
    b' = b_{\mathrm{circ}}(1 + e~sin(\omega))/(1 - e^{2}).
\end{equation}
\texttt{Allesfitter}, instead, assumes (R$_p+$R$_s$)/$a$ and $cos(i)$ as priors, whose combination essentially gives $b_{\mathrm{circ}}$.  Thus, with \texttt{Allesfitter}, we avoided $b'$ versus $e$ correlations; we also obtained non-zero eccentricities, but only 1.5$\sigma$ away from the zero (although \texttt{Allesfitter}'s uncertainties are systematically larger). 
It is also known that fitting directly $e$ and $\omega$ biases towards low eccentricities \citep[e.g][]{Anderson2011,Eastman2013}, behavior that we also confirm.  
Therefore, we attribute the larger eccentricities reported in Table~\ref{tab:results} to the parametrization used by \texttt{juliet}. 
We recall the difficulty in obtaining a robust estimate of the orbital eccentricities with photometry of primary transits alone and thus a detailed RV analysis is required for achieving  a final conclusion in this regards.

\section{Ruling out the putative planet $g$}
\label{sec:stats_planet_g}
Confirming the five-planet architecture of the system or preferring the six-planet scenario is an example of model selection problem. Before acquiring the last \cheops{} light curve of the putative planet $g$ (Fig.~\ref{fig:planet_g_cheops}, bottom panel), we compared the two scenarios by performing several pairs of comprehensive analyses, which involved all the other \cheops{} light curves (where planet $g$ is expected to transit twice) and the two \tess{} sectors, 10 and 11, (where planet $g$ is expected to transit five times). To address the model selection problem, the analyses were performed through the dynamic nested sampling technique \citep[see e.g.,][]{feroz08,feroz19} implemented in the \texttt{dynesty} package \citep{Speagle2020} and incorporated in \texttt{Allesfitter}. In fact, \texttt{dynesty} outputs the Bayesian evidence $\mathcal{Z}$ of the adopted model, so that it is then straightforward to decide whether to prefer a more complicated model (e.g., $\mathcal{M}_1$) against a simpler one ($\mathcal{M}_0$) by computing the Bayes factor $\mathcal{B}_{10}$ as defined in \citet{kass95}:
\begin{equation}
 \mathcal{B}_{10}=\frac{\mathcal{Z}_1}{\mathcal{Z}_0},
 \label{eq:bayesFactor}
\end{equation}
where $\mathcal{Z}_i$ ($i=0,1$) is the Bayesian evidence of the $i^{\mathrm{th}}$ model. As reported by \citet{kass95}, we recall that a very strong evidence against the null hypothesis (i.e. the more complicated $\mathcal{M}_1$ model is accepted) occurs when $\ln{\mathcal{B}}_{10}>5$.

Following a first analysis, where we included only planets from $b$ to $f$ (model $\mathcal{M}_5$), we added the putative planet $g$ to our input file, adopting uniform priors in agreement with the guess values from \citet{Daylan2021} (model $\mathcal{M}_6$). The code was actually able to model also the sixth planet, but we obtained a transit depth $\delta=127\pm21$ ppm, which is half the value suggested by \citet{Daylan2021} and $\sim$\,6$\sigma$ away. Moreover, the posterior distributions of P and T$_0$ have heavy tails, with the uncertainty on T$_0$ that is one order of magnitude larger than what we found for the other planets. This indicates a poor convergence, probably due to the code attempts of modeling transit-like wiggles. 
We also looked at the rms of the residuals coming from the five-planet and six-planet analyses (rms$_5$ and rms$_6$, respectively), considering whether rms$_6$ values were statistically lower than rms$_5$. To this end we applied the Wilcoxon test \citep{wilcoxon45}, whose null hypothesis is that the two samples comes from a single population (i.e., they are statistically equal) obtaining $p$-values\,>\,0.79, which provides no evidence for preferring the six-planet scenario.
On the other hand, we computed the Bayes factor $\mathcal{B}_{65}$ comparing $\mathcal{M}_6$ vs. $\mathcal{M}_5$ obtaining $\ln{\mathcal{B}}_{65}=+8.9$, which would strongly favor the existence of planet $g$.

We then decided to test whether the Bayes factor on its own is actually a reliable indicator in predicting the existence of shallow transit-like signals. To this end, we repeated the analysis of all the light curves mentioned above and replaced the sixth planet candidate from \citet{Daylan2021} with a fake planet having the same characteristics, except for its T$_0$. The fake transits were placed at those timings where we could see a residual noise compatible with weak transit-like features. For the first stress test, we set T$_0=2458959.08$ BJD simulating a transit during \cheops{} observation 2 before the transit of planet $b$ (Fig.~\ref{fig:cheops_lcs1}, right-hand side panel of the first row): model $\mathcal{M}_{\mathrm{f2}}$. After the \texttt{Allesfitter} run, the Bayes factor comparing $\mathcal{M}_{\mathrm{f2}}$ with $\mathcal{M}_5$ gave $\ln{\mathcal{B}_{\mathrm{f25}}}=-0.3$, which states that the existence of this sixth fake planet is not supported. We then performed another stress test, setting T$_0=2458968.0$\,BJD so to mimic a transit during \cheops{} observation 3 (Fig.~\ref{fig:cheops_lcs1}, left-hand side panel of the second row) before the transit of planet $d$: model $\mathcal{M}_{\mathrm{f3}}$. This time the comparison with $\mathcal{M}_5$ yielded to $\ln{\mathcal{B}_{\mathrm{f35}}}=+4.5$, hence favoring the six-planet scenario containing a fake planet though. This is quite surprising, as it means that the transit-like wiggles of the order of 100\,ppm occurring at the transit timings of this fake planet drive the Bayes factor in supporting $\mathcal{M}_{\mathrm{f3}}$. As a double check, we repeated the latter analysis, but looking for a deeper transit ($\delta=500$ ppm) and the new proposed model ($\mathcal{M}_{\mathrm{fd3}}$) was finally rejected by the Bayes factor criterion as expected ($\ln{\mathcal{B}_{\mathrm{fd35}}}=-0.3$).

As the answer from the Bayes factor is not definitive, we decided to schedule one last \cheops{} observation of the putative planet $g$ (observation 15), after ensuring its transit would not overlap with any of the other planets. The detrended light curve is shown in Fig.~\ref{fig:planet_g_cheops} (bottom planet) and our global fitting did not detect any relevant transit feature.

\begin{figure}
 \centering
    \includegraphics[width=0.45\textwidth]{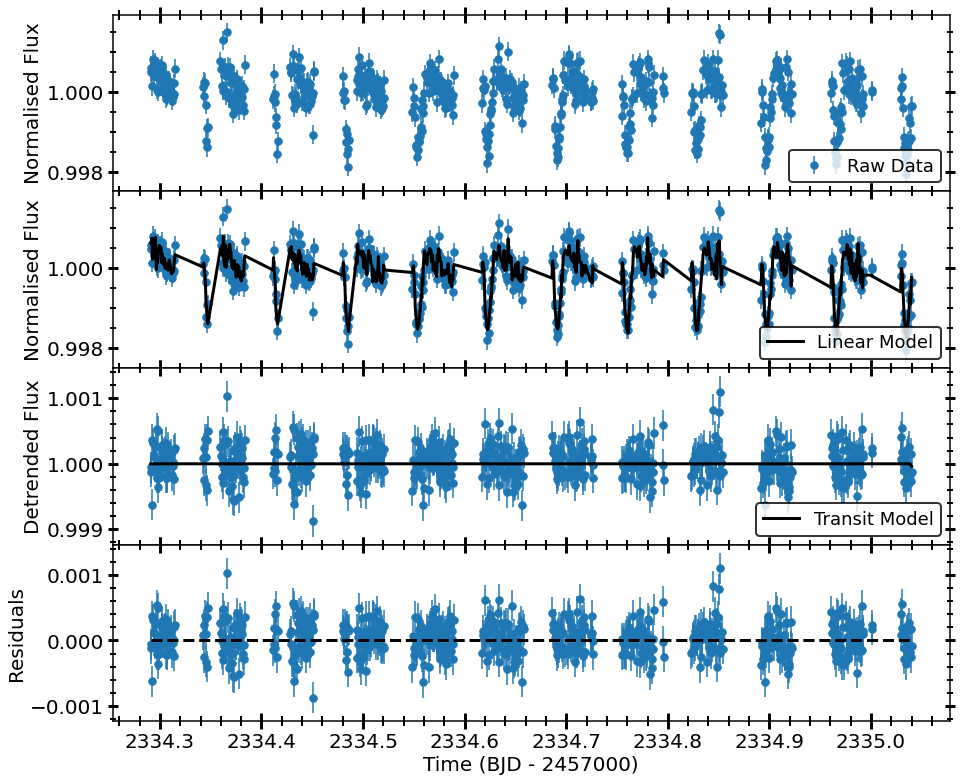} 
     \caption{\cheops{} light curve of the final visit of HD 108236; ({\it top panel}) the DEFAULT DRP fluxes in blue, ({\it second panel}) the raw data in blue plus the linear model built from the PCA PSF method in black, ({\it third panel}) detrended fluxes in blue and the best fit transit model in black, and ({\it bottom panel}) the residuals to the fit in blue.}
    \label{fig:psfpca_detrending}
\end{figure}

To further assess the likelihood of a transiting body within this \cheops{} observation we conduct a more statistically rigorous analysis of this visit. Firstly, we noticed that this dataset suffers from the so-called ramp effect at the beginning of the observations that likely comes from changes in telescope temperature \citep{Maxted2022} that manifests as point-spread function (PSF) shape changes. Moreover, there appears to be roll angle trends that may result as PSF shape changes due to the rotating field of view. To combat these effects, we utilize a novel PSF-based detrending method recently reported in \cite{Wilson2022} to remove these effects in \cheops{} observations of TOI-1064. This tool performs a principal component analysis (PCA) on the auto-correlation function of the \cheops{} subarray images in order to assess subtle changes in the PSF shape. The significant components to be used for decorrelation are then selected via a leave-one-out-cross-validation method. The result of the detrending using these vectors is shown in Fig.~\ref{fig:psfpca_detrending} with the constructed linear model presented in the second panel. Satisfied with the correction of the systematic noise sources, we subsequently computed the true and false inclusion probabilities (TIP and FIP; \citealt{Hara2022}) of the presence of a transit in this dataset by fitting 0 and 1 planet models combined with the linear model determined via the PSF PCA method mentioned above and calculating the TIP and FIP values via comparison of the Bayesian evidence and posterior distributions of the fits. For all transit T$_0$ values within the observing window of the light curve, we find FIP$\sim$1 that statistically confirms that there is no transit within this window and rules out the presence of a planet $g$ on this period. 
A definitive response may likely come from future RV measurements; in any case, the available data ultimately do not support the existence of planet $g$.

\FloatBarrier

\section{Transit times}

The results of the timing analysis of the transits as described in Sect.~\ref{ssec:timing_analysis} are shown in Tables~\ref{tab:transit_times1} and~\ref{tab:transit_times2}.  These transit times are the result of the light curve fits for each planet separately, using as priors the results of the global photometric fit described in Sect.~\ref{sec:analysis}.

\begin{table}
\caption{Central time of each of the transits of planet $b$ and $c$ analyzed in this work.  }
\label{tab:transit_times1}
\centering

\begin{tabular}{ccccc}
\hline
\hline
\multicolumn{5}{c}{ Planet $b$}  \\
 Epoch & T$_{\text{c}}$           & $+\sigma$   &  $-\sigma$  &  Source  \\
\hline

0 & 8572.1080 & 0.0038 & 0.0032 & \tess   \\
1 & 8575.9000 & 0.0055 & 0.0060 & \tess   \\
2 & 8579.7025 & 0.0069 & 0.0082 & \tess  \\
4 & 8587.2939 & 0.0025 & 0.0039 & \tess  \\
5 & 8591.0970 & 0.0078 & 0.0043 & \tess  \\
6 & 8594.8917 & 0.0039 & 0.0038 & \tess  \\
7 & 8598.6734 & 0.0039 & 0.0028 & \tess  \\
8 & 8602.4686 & 0.0052 & 0.0040 & \tess  \\
9 & 8606.2730 & 0.0036 & 0.0050 & \tess  \\
11 & 8613.8544 & 0.0068 & 0.0044 & \tess  \\
12 & 8617.6598 & 0.0021 & 0.0021 & \tess  \\
13 & 8621.4498 & 0.0027 & 0.0042 & \tess  \\
102 & 8959.3020 & 0.0020 & 0.0019 & \cheops  \\
105 & 8970.6821 & 0.0028 & 0.0027 & \cheops  \\
110 & 8989.6601 & 0.0028 & 0.0024 & \cheops  \\
112 & 8997.2494 & 0.0011 & 0.0013 & \cheops  \\
113 & 9001.0425 & 0.0026 & 0.0052 & \cheops  \\
194 & 9308.5086 & 0.0044 & 0.0037 & \tess  \\
195 & 9312.3224 & 0.0042 & 0.0043 & \tess  \\
196 & 9316.0992 & 0.0052 & 0.0035 & \tess  \\
198 & 9323.6931 & 0.0026 & 0.0028 & \tess  \\
199 & 9327.4805 & 0.0040 & 0.0078 & \tess  \\
200 & 9331.2915 & 0.0059 & 0.0079 & \tess  \\

\hline
 \multicolumn{5}{c}{Planet $c$}         \\
Epoch & T$_{\text{c}}$ & $+\sigma$   &  $-\sigma$ & Source   \\
\hline
0 & 8572.3921 & 0.0015 & 0.0013 & \tess  \\
1 & 8578.5998 & 0.0018 & 0.0013 & \tess  \\
2 & 8584.7979 & 0.0038 & 0.0038 & \tess   \\
3 & 8591.0124 & 0.0034 & 0.0036 & \tess  \\
5 & 8603.4095 & 0.0025 & 0.0019 & \tess  \\
6 & 8609.6151 & 0.0022 & 0.0012 & \tess  \\
7 & 8615.8166 & 0.0041 & 0.0047 & \tess  \\
8 & 8622.0309 & 0.0013 & 0.0015 & \tess  \\
56 & 8919.7989 & 0.0013 & 0.0016 & \cheops  \\
64 & 8969.4285 & 0.0020 & 0.0022 & \cheops  \\
65 & 8975.6328 & 0.0005 & 0.0005 & \cheops  \\
66 & 8981.8337 & 0.0021 & 0.0019 & \cheops  \\
70 & 9006.6546 & 0.0008 & 0.0008 & \cheops \\
119 & 9310.6308 & 0.0024 & 0.0038 & \tess  \\
120 & 9316.8357 & 0.0014 & 0.0013 & \tess \\
121 & 9323.0385 & 0.0015 & 0.0014 & \tess \\
122 & 9329.2407 & 0.0014 & 0.0013 & \tess \\

\hline
\end{tabular}
\tablefoot{The transit times are expressed in BJD$_{\text{TDB}}$-2\,450\,000.}
\end{table}

\begin{table}
\caption{Central time of each of the transits of planet $d$, $e$ and $f$ analyzed in this work.   }
\label{tab:transit_times2}
\centering
\begin{tabular}{ccccc}
\hline
\hline
 \multicolumn{5}{c}{Planet $d$}         \\
Epoch & T$_{\text{c}}$ & $+\sigma$   &  $-\sigma$  & Source  \\
\hline

0 & 8571.3355 & 0.0016 & 0.0014 & \tess  \\
1 & 8585.5137 & 0.0019 & 0.0014 & \tess  \\
2 & 8599.6872 & 0.0015 & 0.0011 & \tess  \\
3 & 8613.8654 & 0.0034 & 0.0036 & \tess \\
28 & 8968.2577 & 0.0011 & 0.0009 & \cheops  \\
30 & 8996.6104 & 0.0049 & 0.0043 & \cheops  \\
52 & 9308.4794 & 0.0028 & 0.0025 & \tess \\
53 & 9322.6590 & 0.0022 & 0.0023 & \tess \\

\hline
 \multicolumn{5}{c}{Planet $e$}         \\
Epoch & T$_{\text{c}}$ & $+\sigma$   &  $-\sigma$  & Source  \\
 \hline
0 & 8586.5685 & 0.0014 & 0.0013 & \tess  \\
1 & 8606.1591 & 0.0012 & 0.0011 & \tess  \\
17 & 8919.5979 & 0.0006 & 0.0006 & \cheops \\
21 & 8997.9600 & 0.0008 & 0.0008 & \cheops \\
37 & 9311.4060 & 0.0011 & 0.0013 & \tess  \\
38 & 9330.9942 & 0.0023 & 0.0020 & \tess \\

\hline
\multicolumn{5}{c}{ Planet $f$ }         \\
Epoch &  T$_{\text{c}}$ & $+\sigma$   &  $-\sigma$ & Source   \\
 \hline
-1 & 8586.4885 & 0.0044 & 0.0038 & \tess \\
0 & 8616.0337 & 0.0059 & 0.0050 & \tess \\
12 & 8970.5272 & 0.0015 & 0.0018 & \cheops \\
24 & 9324.9808 & 0.0054 & 0.0037 & \cheops \\
24 & 9324.9857 & 0.0088 & 0.0087 & \tess \\
\hline
\end{tabular}
\tablefoot{ The transit times are expressed in BJD$_{\text{TDB}}$-2\,450\,000.}
\end{table}

\FloatBarrier

\section{Radial velocities fit}

\begin{figure}[h!]
   
    \centering
    \includegraphics[scale=0.4]{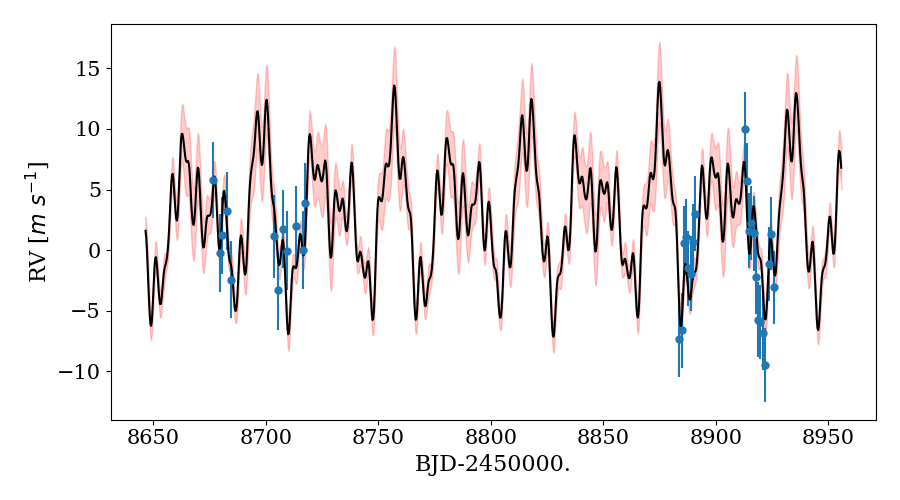} \\
    \includegraphics[scale=0.47]{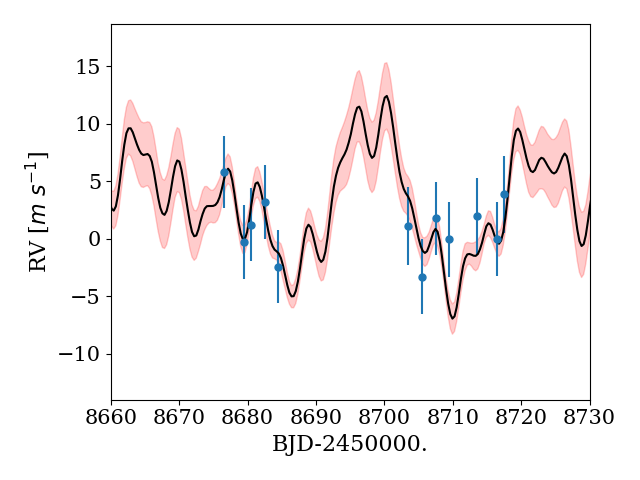} \\
    \includegraphics[scale=0.47]{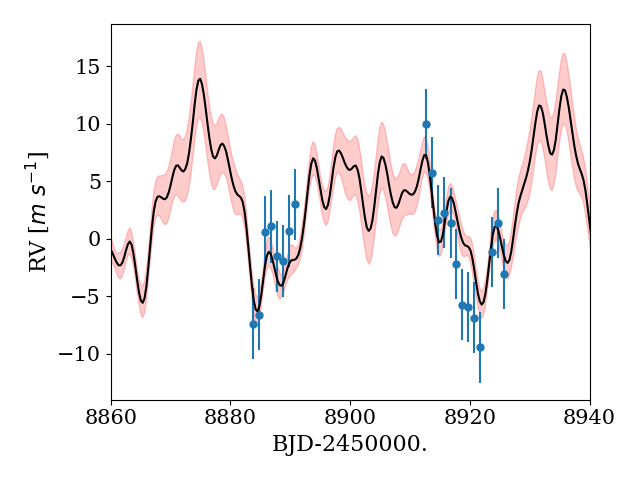} \\
    \caption{Radial velocity measurements from \cite{Teske2021} (blue symbols) with the best fitted model (black solid curve) and its uncertainties (red regions). The bottom panels are closer views of the top panel for the relevant time coverage of the observations. The RV error bars include the estimated jitter. }
    \label{fig:rvs_figures1}
\end{figure}

\begin{figure}
 \centering
    \includegraphics[width=0.45\textwidth]{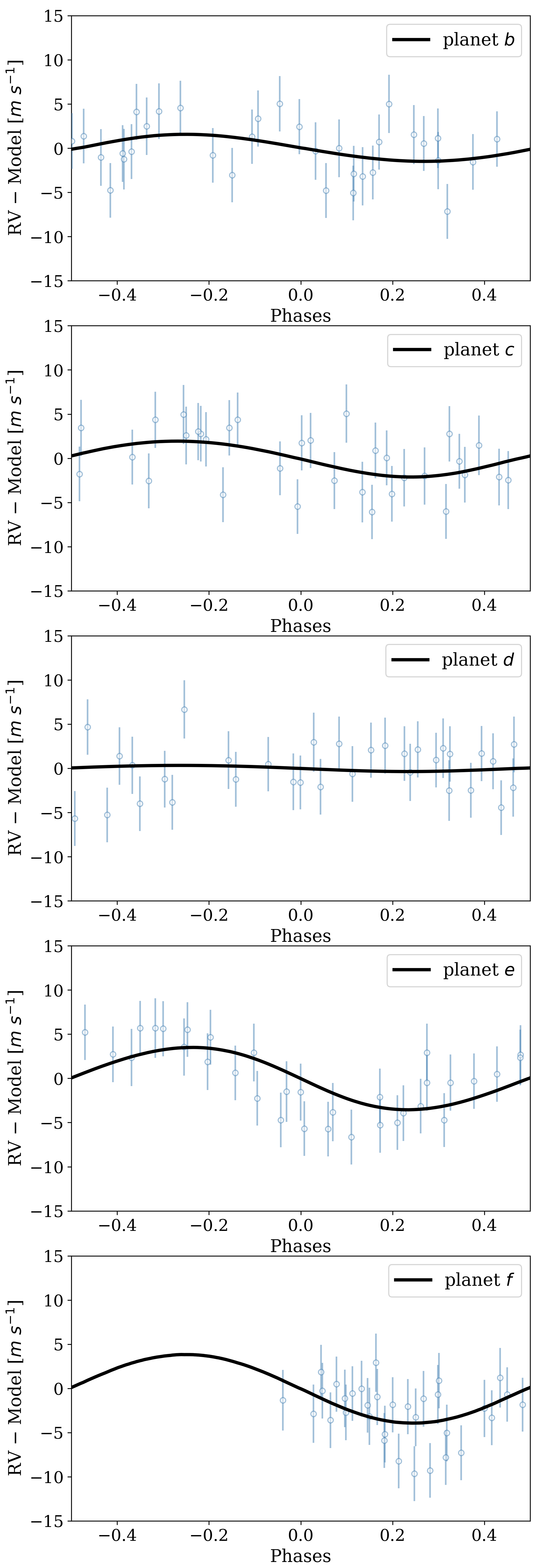} 
     \caption{Phased radial velocities with the full model removed except for the component of the plotted planet (solid curve). }
    \label{fig:rvs_figures2}
\end{figure}

\begin{figure*}
    \centering
    \includegraphics[scale=.45]{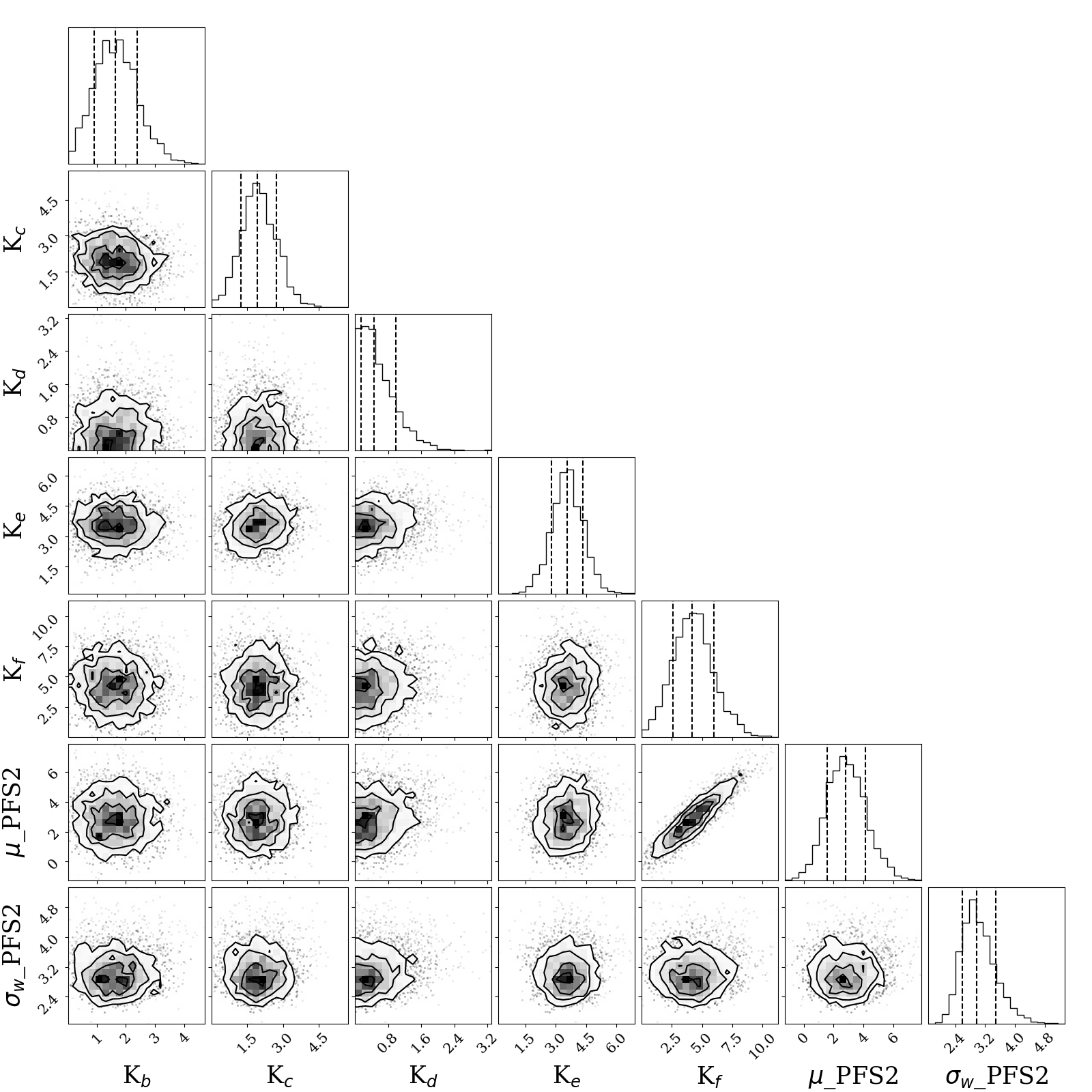}
    \caption{Corner plot of the posterior distributions of the parameters estimated by the modeling of the RV measurements of \starname. }
    \label{fig:corner_plot_rv}
\end{figure*}

The radial velocity measurements reported by \cite{Teske2021} were fitted using \texttt{juliet} package.  The priors of the orbital parameters were defined as normal distributions using the fitted values and uncertainties of the photometric fit (Sect.~\ref{sec:analysis}) reported in Table~\ref{tab:results}.
The full RV model as a function of time, and the phased model for each planet are shown in Figures \ref{fig:rvs_figures1} and \ref{fig:rvs_figures2}, respectively. The corner plot of the posterior distributions of the fitted parameters are shown in Fig.~\ref{fig:corner_plot_rv}.

\end{appendix}

\end{document}